\documentstyle[seceq,supplement]{ptptex}
\begin{document}
\begin{center}
{\bf\large\bf Chapter 5}
\end{center}
\vskip 4mm
\begin{center}
{\large\bf Location of the innermost stable circular orbit of binary
  neutron stars in the post Newtonian approximations of 
general relativity}
\end{center}
\vskip 5mm
\begin{center}
Masaru Shibata, Keisuke Taniguchi$^*$ and Takashi Nakamura$^\dagger$ \\
\vspace{5mm}
{\em Department of Earth and Space Science, Graduate School of 
Science, \\
Osaka University, Toyonaka, Osaka 560, Japan \\
$^*$ Department of Physics, Kyoto University, Kyoto 606-01, Japan \\
$^\dagger$ Yukawa Institute for Theoretical Physics, Kyoto University, 
Kyoto 606-01, Japan}
\end{center}
\vskip 5mm
\begin{abstract}
In this paper, 
we present results obtained from our recent studies on the
location of the innermost stable circular orbit (ISCO) for binary neutron
stars (BNSs) in several levels of post Newtonian (PN) approximations.
We reach the following conclusion at present: (1) 
even in the Newtonian case, there exists the ISCO for 
binary of sufficiently stiff equation of state (EOS). 
If the mass and the radius of each 
star are fixed, the angular velocity at the ISCO 
$\Omega_{\rm ISCO}$ is larger for softer EOS: (2) 
when we include the first PN correction, 
there appear roughly two kinds of effects. One is the 
effect to the self-gravity of each star of binary and the other is 
to the gravity acting between two stars. Due to the former one, 
each star of binary becomes compact and the tidal effect is less 
effective. As a result, $\Omega_{\rm ISCO}$ tends to be 
increased. On the other hand, the latter one has the property to 
destabilize the binary orbit, and $\Omega_{\rm ISCO}$ tends to be 
decreased. If we take into account both effects, however, 
the former effect is stronger than the latter one, and $\Omega_{\rm ISCO}$ 
becomes large with increase of the 1PN correction: 
(3) the feature mentioned above is more remarkable 
for softer EOS if the mass and radius are fixed. This is because 
for softer EOS, each star has the  
larger central density and is susceptible to the GR correction: 
(4) there has been no self consistent calculation including all the 2PN 
effects and only exist studies in which one merely includes the effect of 
the 2PN gravity acting between two stars. In this case, 
the effect has the property to destabilize the binary orbit, so that 
$\Omega_{\rm ISCO}$ is always smaller than that for the Newtonian case. 
If we include the PN effect of the self-gravity to each star, 
$\Omega_{\rm ISCO}$ will increase.
\end{abstract}

\def\beqa{\begin{eqnarray}}
\def\eeqa{\end{eqnarray}}
\def\lsim{\mathrel{\mathpalette\Oversim<}}
\def\gsim{\mathrel{\mathpalette\Oversim>}}
\def\Oversim#1#2{\lower0.5ex\vbox{\baselineskip0pt\lineskip0pt%
            \lineskiplimit0pt\ialign{%
          $\mathsurround0pt #1\hfil##\hfil$\crcr#2\crcr\sim\crcr}}}
\def\alt{\lsim}
\def\agt{\gsim}
\newcommand{\beq}{\begin{equation}}
\newcommand{\eeq}{\end{equation}}
\newcommand{\beqn}{\begin{eqnarray}}
\newcommand{\eeqn}{\end{eqnarray}}
\newcommand{\pa}{\partial}
\def\bI{\hbox{$\,I\!\!\!\!-$}}
\def\a{\alpha}
\def\b{\beta}
\def\p{\partial}
\def\e{\epsilon}
\def\ve{\varepsilon}
\def\r{\rho}
\def\O{\Omega}
\def\t{\tilde}
\def\ra{\rightarrow}
\def\lsim{\mathrel{\mathpalette\Oversim<}}
\def\gsim{\mathrel{\mathpalette\Oversim>}}

\section{Introduction}

 BNSs like PSR1913+16 are believed to be formed after the second
supernova explosion in the evolution of massive binary stars. At the
formation, orbits of BNSs will be highly eccentric because a large
amount of the mass of the system is ejected in the supernova explosion.
If the semi-minor axis of BNSs is less than about a few times of solar
radius, they will coalesce within the age of the universe $\sim
10^{10}$yr due to the emission of gravitational waves (GWs). In the
evolution of the binary the orbital radius as well as the orbital
eccentricity decreases so that the orbit becomes almost circular when
the orbital radius is about $10^2$ times of the neutron star (NS) radius
where the frequency of GWs is $\sim 10$Hz, i.e., in the frequency band
sensitive to the kilometer size laser interferometric GW detector such
as LIGO,\cite{LIGO} VIRGO,\cite{VIRGO} GEO600,\cite{GEO} and
TAMA.\cite{TAMA} After that, evolution of BNSs is divided into two
phases: One is the so called inspiraling phase where the orbital radius
of BNSs is sufficiently larger than the radius of each NS. In this
phase, the hydrodynamic effect of NS is not so important for the orbital
evolution that each star of binary is approximately regarded as a point
mass. The emission time scale of gravitational waves (GWs),
$t_{\rm GW}$, is still much longer than the orbital period and as a
result, the circular orbit of BNSs is stable, so that BNSs evolve
adiabatically in the inspiraling phase. The second phase is the post
inspiraling phase in which the orbital radius becomes a few times of the
NS radius and the hydrodynamic effect as well as general
relativistic (GR) effects between two stars become very important. Just
after BNSs enter this phase, i.e., the orbital radius of BNSs becomes
smaller than the innermost stable circular orbit (ISCO), the circular
orbit becomes unstable due to the hydrodynamic and GR effects and 
BNSs come into a final plunging and merging phase.

 GWs in the inspiraling phase have a characteristic feature; the Fourier
spectrum of the signal behaves as $\propto f^{-7/6}$.\cite{KIP}
However, when the BNSs enter the post inspiraling phase, their wave
form will change drastically because the motion of the BNSs changes
from circular orbit to plunging one. In particular, the ISCO is the
characteristic orbit of such transition. The transition
point will be sensitively determined by mass-radius relation of
NS\cite{cutler}\cite{joan} and the EOS of NSs may be determined
from the mass-radius relation.\cite{lindblom} Hence, the signal around the
ISCO will have important information on the equation of state (EOS), or
in other words, the internal structure of NS. This means that if we
detect a signal from BNSs at the ISCO, we will be able to constrain the
EOS by comparing the signal with theoretical templates. For this
reason, theoretical studies for the ISCO are urgent.

 In Fig.1, we schematically show the location of the ISCO in each level
of approximation which has been determined from recent studies: In the
Newtonian case, there does not exist the ISCO if we regard each star of
binary as point masses, but it appears for BNSs of sufficiently stiff
EOS if we take into account the hydrodynamic effect on
BNSs.\cite{LRS}\cite{LRSLRS} This is because each star is tidally
deformed by the gravity from the companion star and the tidal
deformation induces the attractive force which overcomes the centrifugal
force at small radius. For soft EOS, the degree of the central
condensation is high, so that the effect of tidal force is not so
important compared with the stiff EOS case.

 When we include the first PN (1PN) correction of general relativity, not
only the gravity between two stars, but also the self-gravity in each
star become stronger than those in the Newtonian case. If the former
effect is stronger than the latter one, the orbital radius of the
ISCO ($R_{\rm ISCO}$) becomes larger than that in the Newtonian case,
because the former PN effect has the property to destabilize circular
orbits as we will show later (\S 3 and \S 4). 
On the other hand, if the opposite
relation holds, $R_{\rm ISCO}$ becomes smaller than that in the
Newtonian case because each star of binary is forced to be compact due
to the PN correction of the self-gravity and the tidal effect is
less important. Recently, investigating equilibrium sequences of
corotating binary, Shibata and Taniguchi found that the latter effect is
stronger than the former one\cite{shibapn}\cite{tania}\cite{tanib}, and
$R_{\rm ISCO}$ becomes small due to the 1PN correction. As a result,
the angular frequency at the ISCO increases.

 Another approach has been performed by Taniguchi and
Nakamura (TN)\cite{TN} as well as Lai and Wiseman (LW)\cite{LW} 
and Ogawaguchi and Kojima.\cite{kojima} Note
that the methods adopted by three groups are essentially similar. In
their approaches, they include the second PN (2PN) corrections and some
parts of the higher relativistic corrections to the gravity between two
stars phenomenologically, although they do not take into account the PN
terms to the self-gravity of each star using the Newtonian equilibrium.
In this case, the PN attractive force between two stars tends to
destabilize circular orbits and as a result, $R_{\rm ISCO}$ becomes
larger while the angular velocity becomes smaller compared with the
point particle cases.

 Considering these recent developments for understanding the mechanism
which determines the ISCO, we organize this paper as follows. We
analyze equilibrium sequences of binary stars assuming that $t_{\rm GW}$
is sufficiently larger than the orbital period, and argue the stability
of binary along the sequence because the ISCO is just the critical point
of stability. In \S 2, after we briefly mention why the ISCO exists
even in the Newtonian case, we describe how the location of the ISCO for
fluid stars is determined paying particular attention to the
incompressible star. In the latter part of \S 2, we also argue the
effect of the EOS analyzing equilibrium states of corotating binary of
compressible stars. In \S 3 and \S 4, analyzing equilibrium states of
corotating binaries, we argue how the location of the ISCO changes due
to the 1PN correction for incompressible and compressible fluid cases,
respectively. In the incompressible case, it is possible to obtain a
fairly accurate analytic solution of corotating binaries even in the 1PN
case, so that in \S 3, we present the method to derive it in detail. On
the other hand, in the compressible case, we cannot obtain the analytic
solution, so that we perform numerical computations and show the results
in \S 4. In \S 5, we review the method and results by TN in which the
2PN and some parts of the higher PN corrections are taken into account
only for the gravity acting between two stars. As mentioned above, the
results they (and LW) obtained seem to apparently disagree with those by
the full 1PN study. One of the reasons for this difference is the
starting point of the problem. In the 1PN case, the zeroth order solution
is the Newtonian equilibrium and then 1PN effect is taken into account as 
the correction. While in treatment of TN, 
a semi-relativistic equation of motion (EOM) for point particle with 
ISCO is the zeroth order equation 
and then the finite size effect of the star is taken into
account. Another reason is the consistency of the treatment. In the 1PN
case, 1PN effects of the gravity are consistently taken into account to
both the self-gravity of each star and the gravity between two stars.
While in TN, only 2PN and the some part of the higher order effects of
gravity between two stars are considered phenomenologically. To
understand the effect of the self-gravity more clearly, in the last
subsection of \S 5, we compute 1PN ISCO removing the 1PN self-gravity
force in \S3. From the results, we suggest that if we include the 1PN
effect of the self-gravity in TN, $R_{\rm ISCO}$ will decrease. We
briefly review recent semi-GR calculations in \S 6. Section 7 is devoted to
summary.

Throughout this paper, we use $G$ and $c$ as the gravitational constant
and the speed of light. We define the Newtonian mass $M$ and the
multipole moment $I_{ijk\cdots}$ as
\beq
 M=\int \rho d^3x \hskip 1cm {\rm and } \hskip 1cm 
 I_{ijk \cdots}=\int \rho x_i x_j x_k \cdots d^3x,
\eeq
where $\rho$ denotes the baryon density of star and the integral is
performed inside each star of binary. We also define the tracefree part
of the quadrupole moment as
\beq
 \bI_{ij}=I_{ij}- {1 \over 3} \delta_{ij}\sum_{k=1}^3 I_{kk},
\eeq
where $\delta_{ij}$ denotes the Kroneker's $\delta$. 

In the following sections, we take the axis for the orbital motion of
binary as $x_3$, and two stars have equal Newtonian mass $M$ and the
same structure for simplicity. In \S 2.1, \S3 and \S5, we put the
center of the star 1\footnote{Here, ``the center of mass'' indicates
that defined in the Newtonian order. We should be careful to define
it in the PN order (see \S 3).} at the origin and center of the star 2
at $(-R,0,0)$ (i.e., the orbital separation is $R$ and $x_1$ axis is the
semimajor axis; see Fig.2), and consider the hydrostatic equilibrium for
star 1. Orbital rotation is assumed to be counterclockwise 
in all sections.

%%%%%%%%%%%%%%%%%%%%%%%%%%%%%%%%%%%%%%%%%%%%%%%%%%%%%%%%%%%%%%%
\section{ISCO for binary neutron stars in the Newtonian theory}
%%%%%%%%%%%%%%%%%%%%%%%%%%%%%%%%%%%%%%%%%%%%%%%%%%%%%%%%%%%%%%%

For binary of two point particles in the Newtonian theory, it is well
known that there is no ISCO: In this case, the energy conservation
equation can be written as\cite{gold}
\beq
 {m \over 2}\biggl({dr \over dt}\biggr)^2=E-V(r);~~~
 V(r)={m\ell^2 \over 2r^2}-{GM_t m \over r},
\eeq
where $m$, $M_t$, $\ell$, $E$, $r$ and $t$ denote the reduced mass,
total mass of the binary, the specific angular momentum, the energy, the
orbital separation, and the coordinate time, respectively. The circular
orbit of its radius $r_0=\ell^2/GM_t$ is always stable up to the contact
of binary because $d^2V/dr^2$ is always positive at $r=r_0$.

We can also explain the stability of the circular orbit in the following
way: The energy of the circular orbit is
\beq
 E=-{GM_t m \over 2r_0}. 
\eeq
This expression means that if $E<0$, the circular orbit always exists up
to $r_0=0$. This also shows the existence of the stable circular orbit
for $0< r_0 <\infty$.  Note that this conclusion follows from the fact
that the centrifugal force is always stronger than the gravitational
force at small radii ($r< r_0$).

This is not necessarily the case for NS case if we remind the finite
extension of the star and take into account the hydrodynamic effect on
each NS. Due to the tidal field from the companion star, each NS of
BNSs will be tidally deformed. In such a case, $V(r)$ is approximately
written as
\beq
V(r)={m\ell^2 \over 2r^2}-{GM_tm \over r}-{3GM_t \over 2r^3}\bI_{11}(t), 
\label{vvv}
\eeq
where we assume that the center of each star is located on the $x_1$
axis and $\bI_{11}$ is $11$ component of the tracefree part of the
quadrupole moment of each star. Here, we only take into account the
lowest order multipole among all the multipole moments 
which are generated by the tidal deformation.  We
note that $\bI_{11}$ changes approximately as $O(r^{-3})$ if the star
does not have an intrinsic spin. So that the third term in Eq.
(\ref{vvv}) behaves as $O(r^{-6})$ in reality. Thus, the effect of the
tidal deformation is very small for large $r$, but it cannot be
negligible in the case of small $r$. If the tidal deformation is
sufficiently large, it may overcome the centrifugal potential for a
sufficiently small $r$, and in such a case, the orbit becomes unstable
before each star of binary comes into contact; i.e., for $E < E_{\rm
 crit}$ where $E_{\rm crit}$ is a critical value of the energy, no
circular orbit exists. This is essentially what Lai, Rasio, and
Shapiro (LRS) pointed out\cite{LRS}\cite{LRSLRS} already.

%%%%%%%%%%%%%%%%%%%%%%%%%%%%%%%%
\subsection{Incompressible case}
%%%%%%%%%%%%%%%%%%%%%%%%%%%%%%%%

We explain the qualitative fact mentioned above by showing sequences of
equilibrium states of incompressible binary stars in more details. In
the Newtonian case, the hydrostatic equation in the rotating frame of
the angular velocity $\Omega$ is\cite{EFE}
\beq
 \rho \sum_{j=1}^3 u_j {\pa u_i \over \pa x_j}=-{\pa P\over \pa x_i}+
 \rho{\pa \over \pa x_i}\biggl[U+{\Omega^2 \over 2}
 \biggl\{\biggl(x_1 + {R \over 2}\biggr)^2+x_2^2 \biggr\}\biggr]
 +2\rho\Omega \sum_{l=1}^3 \epsilon_{il3}u_l,\label{hysteq}
\eeq
where $u_i$, $P$, and $\epsilon_{ijk}$ denote the velocity in the
rotating frame, the pressure, and the completely antisymmetric tensor,
respectively.  $U$ is the Newtonian potential which satisfies
\beq
 \Delta U=-4\pi G\rho, 
\eeq
and it can be split into two parts; one is the self-gravity part, and
the other is the contribution from the companion star. We separate $U$
as
\beq
 U=U^{1 \rightarrow 1}+U^{2 \rightarrow 1}, 
\eeq
where $U^{1 \rightarrow 1}$ denotes the self-gravity part and 
$U^{2 \rightarrow 1}$ denotes the contribution from the companion. 
These are written, respectively, as
\beqn
 U^{1 \rightarrow 1}&&=\pi G\rho\Bigl(A_0-\sum_{i=1}^3 A_ix_i^2\Bigr),
 \label{u11}\\
 U^{2 \rightarrow 1}&&={GM \over R}\biggl(1-{x_1 \over R}+
 {2x_1^2-x_2^2-x_3^2 \over 2R^2}+{-2x_1^3+3x_1(x_2^2+x_3^2) \over 2R^3}
 +O(R^{-4})\biggr)\nonumber\\
 &&~~~~+{3G\bI_{11} \over 2R^3}\biggl(1-{3x_1 \over R}+O(R^{-2})\biggr),
\eeqn
where we only include the contribution from the quadrupole moment of
each star, and neglect the higher multipoles. In this case, we may
assume that each star of binary is an ellipsoid of its axial length
$a_1$, $a_2$ and $a_3$.  Index symbol $A_i$ is that defined in the
textbook of Chandrasekhar\cite{EFE} and $A_0=\sum_iA_ia_i^2$ is
calculated from
\beq
 A_0=a_1^2\alpha_2\alpha_3\int^{\infty}_0 {dt \over \sqrt{(1+t)
 (\alpha_2^2+t) (\alpha_3^2+t) } } \equiv a_1^2 \tilde A_0
 (\alpha_2,\alpha_3),
\eeq
where $\alpha_2=a_2/a_1$ and $\alpha_3=a_3/a_1$. 
Note that $A_i$ is the function of $\alpha_2$ and $\alpha_3$\cite{EFE}. 

For ellipsoids, the pressure $P$ is given by 
\beq
 P=P_0 \biggl( 1-\sum_{i=1}^3 {x_i^2 \over a_i^2} \biggr). \label{pressure}
\eeq
Hereafter, we assume for simplicity that the internal motion of one star 
is the same as that of the other star, and we restrict its 
form given as 
\beq
 u_1={a_1 \over a_2}\Lambda x_2 ~~~~~~~~~u_2=-{a_2 \over a_1}\Lambda x_1. 
\label{intvelo}
\eeq
In this case, the vorticity $\omega=2\Omega+
\sum_{j,k}\epsilon_{3jk}\pa_j u_k$ 
in the inertial frame is spatially constant everywhere inside the star as
\beq
 \omega=2\Omega-\biggl( {a_2 \over a_1}+{a_1 \over a_2}\biggr)\Lambda
 \equiv \Omega (2+f_{R});~~~f_R\equiv -{a_1^2+a_2^2 \over a_1a_2}
 {\Lambda \over \Omega}. 
\eeq
Thus, the circulation in the equatorial plane becomes
\beq
 C=\pi a_1a_2 \Omega (2+f_R). 
\eeq

 {}From the first tensor virial (TV) relation, we obtain the 
orbital angular velocity as
\beq
 \Omega^2={2GM \over R^3}+{18G\bI_{11} \over R^5}. 
\eeq
 {}From the second TV relation, we obtain the following three 
equations;
\beqn
 -\Lambda^2 &&={2P_0 \over \rho a_1^2}-2\pi G\rho A_1+{2GM \over R^3}
 +\Omega^2-2{a_2 \over a_1}\Lambda\Omega,\nonumber \\
 -\Lambda^2 &&={2P_0 \over \rho a_2^2}-2\pi G\rho A_2 - {GM \over R^3}
 +\Omega^2-2{a_1 \over a_2}\Lambda \Omega,\nonumber \\
 0 &&={2P_0 \over \rho a_3^2}-2\pi G\rho A_3-{GM \over R^3}.\label{tvtv}
\eeqn
Note that relations among $R/a_1$, $\alpha_2$ and $\alpha_3$ are 
derived from these three equations. 

The energy and the angular momentum are calculated as
\beqn
 E&&={M \over 5}\biggl[\biggl(\Omega-{a_2 \over a_1}\Lambda\biggr)^2a_1^2
 +\biggl(\Omega-{a_1 \over a_2}\Lambda\biggr)^2 a_2^2\biggr]
 +{MR^2 \over 4}\Omega^2 \nonumber \\
 &&~~~-{4 \over 5}\pi G \rho MA_0-{GM \over R}\biggl(M+{3\bI_{11} \over R^2}
 \biggr),\label{Nenergy}\\
 J &&={2M \over 5}(a_1^2+a_2^2)\Omega-{4M \over 5}a_1a_2\Lambda
 +{MR^2 \over 2}\Omega.
\eeqn
In Eq. (\ref{Nenergy}), the terms in brackets $[\cdot\cdot\cdot]$ 
denote the spin kinetic energy of each star, 
the next two terms denote the kinetic energy of the 
orbital motion and the binding energy by the self-gravity, and 
the final two terms denote the binding energy between two stars. 
The sum of the kinetic energy of the orbital motion and 
the binding energy between two stars mainly 
concerns the stability of the orbit and it becomes 
\beq
-{GM^2 \over 2R}+{3M\bI_{11} \over 2R^3}. 
\eeq
Since $a_1>a_2,a_3$ and as a result $\bI_{11}>0$, 
the second term is positive definite. Thus, 
the energy has a minimum critical value $E_{\rm crit}$ at a small 
radius due to the second term. As mentioned above, 
this just indicates that the tidal effect acts as the destabilization. 

Before presenting results on equilibrium states, we mention the possible
state of internal motion of star.  If the viscosity of the fluid is so
small that the dissipation time scale of the circulation by the
viscosity ($t_{\rm vis}$) is much longer than $t_{\rm gw}$, the
circulation conserves throughout the whole evolution. On the other hand,
if the viscosity is so large that $t_{\rm vis}$ is shorter than $t_{\rm
 gw}$, BNSs will settle down to the corotating configuration.
Kochanek, and Bildsten and Cutler showed that 
in order to achieve corotating
binary dissipating the circulation, the viscosity must
satisfy\cite{irre}
\beq
\eta \agt 10^{29} \biggl({R \over a_0}\biggr)^2 
\biggl({M \over 1.4M_{\odot}}\biggr)^4
\biggl({10{\rm km} \over a_0}\biggr)^5{\rm g/cm/s}, 
\eeq
where $a_0=(3M/4\pi\rho)^{1/3}$ and $M_{\odot}$ denotes the solar mass. 
Since this value is much larger than the microscopic viscosity,\cite{ito}
we should consider sequences of equilibrium states of constant $C$ 
as realistic BNSs. 

By the way, 
for large $R\rightarrow \infty$, $\Omega \rightarrow 0$, so that 
$C\rightarrow -\pi(a_1^2+a_2^2)\Lambda$. This implies that 
$-\Lambda$ at $R\rightarrow \infty$ is regarded as the 
intrinsic spin angular velocity of each star (we denote it as 
$\Omega_{\rm s}$ following LRS\cite{LRSLRS}). 
Since $|\Omega_{\rm s}|$ for real NS's will be 
less than $\sim 0.6\Omega_0$, where 
$\Omega_0\equiv (GM/a_0^3)^{1/2}$,\cite{RNS}\cite{COMPACT}
$|C|$ should be less than $\sim 4(GMa_0)^{1/2}$. 
Note that for pulsars we have ever known,\cite{taylor}
$|\Omega_{\rm s}| \alt 0.2 \Omega_0$. 

In Figs.3, we show the energy and angular momentum as functions of
$\Omega/\Omega_0$ for $C/(G M a_0)^{1/2}=-2 \sim 2$.  We also show them
for corotating binary for comparison.  We see that there exists a
critical angular velocity where the energy and angular momentum are
simultaneously minima.  As LRS showed,\cite{LRS} the minima in the
energy and angular momentum along the constant-$C$ sequence indicate the
onset of dynamical instability. Thus, the minimum corresponds to the
ISCO. As mentioned above, this minimum occurs due 
to the tidal interaction between two stars at small separation. Each
star deforms and the quadrupole dependent term in the effective
interaction potential between two stars (i.e., the third term in Eq.
(\ref{vvv})) becomes sufficiently large.

Figures 3 also show some important features for the ISCO.  First, the
effect of the spin of each star is not so important in determining the
ISCO as long as it is not very large, i.e., $|C|/(GMa_0)^{1/2} < 1$.
Second, even for corotating sequence where the circulation does not
conserve, there exists a critical angular velocity where the energy and
angular momentum are simultaneously minima, and it locates near the
minima for constant-$C$ sequence.  Although the minima for corotating
sequence do not mean the onset of the dynamical instability, but of the
secular instability\cite{LRS} (so we call it the innermost stable
corotating circular orbit(ISCCO)), its existence is useful for analysis
of the ISCO. This is because the angular velocity at the minima for
corotating sequence is only slightly smaller than that for
constant-$C$ (small $C$) sequences and as a result, we can approximately
find the location of the ISCO for constant-$C$ sequences by
investigating corotating sequence instead.

%%%%%%%%%%%%%%%%%%%%%%%%%%%%%%
\subsection{Compressible case}
%%%%%%%%%%%%%%%%%%%%%%%%%%%%%%

Next, we argue the effect of EOS.  Since the location of the ISCO for
binary stars is determined by the degree of the tidal deformation, it
depends on the EOS of NS. NS becomes more centrally concentrated as the
EOS is softer. Thus, the tidal effect is less important for softer EOS.
This means that the effect of EOS is quantitatively very important in
determining the location of the ISCO.  Actually, using an approximate
analytical model, LRS have shown that if we assume the polytropic EOS as
\beq
P=K\rho^{\Gamma}; ~~~\Gamma=1+{1 \over n}, 
\eeq
where $K$ and $n$ are 
the polytropic constant and the polytropic index, 
$R_{\rm ISCO}$ is smaller for larger $n$, and 
for sufficiently large $n$ ($ \agt 1$), the ISCO disappears. 

Although LRS presented approximate locations of the ISCO for various
$n$, we do not know the precise ones. Furthermore, their 
approximation is good only for small $n$, so that quantitative
dependence on the EOS may not be understood well for realistic NS whose
EOS has $0.5 \alt n \alt 1$.\cite{COMPACT}  To get precise locations of
the ISCO, we need to perform numerical calculation, but we have not yet
established an accurate numerical method of obtaining equilibrium states
including arbitrary circulation. As we mentioned above, however,
corotating sequence may be regarded as an approximate sequence of
constant circulation if the spin of each star is not so large, and we
have accurate methods of obtaining corotating sequences.\cite{hachisu}
Thus, we here see the effect of the EOS by investigating corotating
sequences.

Equilibrium states of corotating binary stars are obtained by 
consistently solving the 
Newtonian potential $U$ and the integrated Euler equation as 
\beq
K(n+1)\rho^{1/n}=U+{\varpi^2 \over 2}\Omega^2; ~~
\varpi^2=\biggl(x_1+{R \over 2}\biggr)^2+x_2^2. 
\eeq
In Figs.4, we show the energy and the angular momentum as 
functions of the angular velocity for $n=0.25,~0.5$, 0.75 and
1 ($\Gamma=5,~3,~7/3$ and 2). 
The energy, angular momentum, and angular velocity are 
shown in units of $GM^2/a_0$, $(GM^3a_0)^{1/2}$, and 
$\Omega_0\equiv (GM/a_0^3)^{1/2}$, respectively, where $a_0$ denotes the 
radius of the spherical polytropic star. In each figure, 
the circle of the largest $\Omega$ denotes the 
configuration at contact of two stars. 
It is found from figs.4 that 
with increase of $n$, $\Omega$ at the energy 
and angular momentum minima increases and 
the minima tend to disappear for given $a_0$ and $M$.  
The angular velocity at the minima, $\Omega_{\rm N:ISCCO}$,
is $\simeq (0.252-0.256)\Omega_0$ for
$n=0.25$, $\simeq (0.265-0.269)\Omega_0$ for $n=0.5$,
$\simeq (0.279-0.283)\Omega_0$ for $n=0.75$ and 
$\simeq (0.296-0.298)\Omega_0$ for $n=1$.\footnote{
Note that in the analysis by LRS,\cite{LRS}
they somewhat overestimate $\Omega_{\rm N:ISCCO}$ for compressible star.
For example, $\Omega_{\rm N:ISCCO}=0.279\Omega_0$ for $n=0.5$
and $\Omega_{\rm N:ISCCO}=0.314\Omega_0$ for $n=1$ in
their analysis.} Note that for
incompressible case, it is $\simeq 0.248\Omega_0$. From these numerical 
results, the angular velocity at the minima for compressible star 
of $n=0 - 1$ is found to be approximately expressed as (see Fig.5) 
\beq
\Omega_{\rm N:ISCCO} \simeq 0.248(1+ 0.2n^{\alpha})\Omega_0,
\label{ISCCO}
\eeq
where $\alpha \sim 1.5$. Thus, the frequency of GWs 
corresponding to $\Omega_{\rm N:ISCCO}$ is calculated from 
$f_{\rm N:ISCCO}\equiv\Omega_{\rm N:ISCCO}/\pi$, which is 
\beq
f_{\rm N:ISCCO} \simeq 590 (1+ 0.2n^{1.5} )
\biggl({M \over 1.4M_{\odot}}\biggr)^{1/2}
\biggl({15 {\rm km} \over a_0}\biggr)^{3/2}{\rm Hz}. 
\eeq

The above results concerning the dependence of the energy and angular
momentum minima on the EOS are for corotating binaries, but we may
expect that similar dependence holds for constant-$C$ sequences if the
spin of each star is not so large. This is because change of the EOS
mainly affects the density profile of each star almost independent of
the spin and circulation.  Actually, from the results by an approximate
study of LRS,\cite{LRS} it is found that the following quantity is
almost independent of the EOS;
\beq
{\Omega_{\rm N:ISCO}~ {\rm for}~C=0~{\rm sequence} \over \Omega_{\rm N:ISCCO} 
~{\rm for~ corotating~ sequence}}. 
\eeq
Thus, we are allowed to conclude that even for constant-$C$ sequences 
the angular frequency
at the ISCO for compressible star 
will increase approximately with $n^{1.5}$ as Eq. (\ref{ISCCO}). 

%%%%%%%%%%%%%%%%%%%%%%%%%%%%%%%%%%%%%%%%%%%%%%%%%%%%%%%%%%%%%%%%%%%%%%%%%%%
\section{First post Newtonian study: Incompressible case}\label{section1PN}
%%%%%%%%%%%%%%%%%%%%%%%%%%%%%%%%%%%%%%%%%%%%%%%%%%%%%%%%%%%%%%%%%%%%%%%%%%%

In \S 2, we briefly mention that the ISCO exists even in the Newtonian
case if the EOS of NS is sufficiently stiff.  In this section, we see
how the location of the ISCO obtained in the Newtonian theory is
affected by the 1PN correction of general relativity.  We again make use
of sequences of corotating binary stars.  In this section, as a first
step, we investigate equilibrium states for incompressible binary stars
for which fairly accurate approximate solutions are obtained in an
analytic manner (see details in Ref. 12)).

%%%%%%%%%%%%%%%%%%%%%%%%
\subsection{Formulation}
%%%%%%%%%%%%%%%%%%%%%%%%

Non-axisymmetric equilibrium configurations of uniformly rotating
incompressible fluid in the 1PN approximation are obtained by solving
the integrated form of the Euler equation and the Poisson equations for
gravitational potentials consistently.  The integrated form of the Euler
equation was derived by Chandrasekhar\cite{PNeom} and it can be written
as\cite{shibapn}\cite{asada}\footnote{Note that we use the standard PN
  gauge.}
\beqa
 & &\int {dP \over \rho} -\frac{1}{c^2} \int
 \biggl(\varepsilon+{P \over \rho}\biggr)
 {dP \over \rho}  \nonumber \\
 &=& U-\frac{X_0}{c^2} + \left\{ \frac{\varpi^2}{2} + \frac{1}{c^2}
  \left( 2\varpi^2 U-X_{\O} + \hat{\beta}_{\varphi} \right) \right\}
 \O^2 + \frac{\varpi^4}{4 c^2} \O^4 + {\rm const}, \label{ftve}
\eeqa
where $X_0$, $X_{\O}$, and $\hat \beta_{\varphi}$ are the gravitational
potentials in the 1PN order which are obtained by solving the following
Poisson equations:
\beqn
&&\Delta \hat P_1=-4\pi G\rho \biggl(x_1+{R \over 2}\biggr)~, \nonumber\\
&&\Delta \hat P_2=-4\pi G\rho x_2~, \nonumber \\
&&\Delta X_0=4\pi G\rho \Bigl(\varepsilon+2U+{3P \over \rho}\Bigr),
\nonumber \\
&&\Delta X_{\Omega}=8\pi G\rho \varpi^2 ,\label{PNpot}
\eeqn
and
\beqn
 \hat \beta_{\varphi}=-\biggl[ &&{7 \over 2}\biggl\{
 \Bigl(x_1+{R \over 2}\Bigr)\hat P_1+x_2\hat P_2 \biggr\} \nonumber \\
 +&&{1 \over 2} \biggl\{\Bigl(x_1+{R \over 2}\Bigr)^2 \hat P_{2,2}
 +x_2^2\hat P_{1,1}-\Bigl(x_1+{R \over 2}\Bigr)x_2(\hat P_{1,2}+\hat
 P_{2,1}) \biggr\}\biggr].
\eeqn
Note that in the incompressible case,
$\varepsilon=0$ and $\rho=$constant.  Thus, the left hand-side of Eq.
(\ref{ftve}) is
\beq
 {P \over \rho}-{1 \over 2c^2}\biggl({P \over \rho} \biggr)^2.
\eeq

In the 1PN approximation, we have conserved quantities such as the 
conserved mass $M_*$, the energy $E$, and the angular momentum $J$ 
which are written as\cite{chand}\cite{shibapn}
\beqn
M_*&=&
\int \rho_* d^3x=\int \rho\Bigl\{1+{1 \over c^2}\Bigl({v^2 \over 2}+3U
\Bigr) \Bigr\}d^3x,\label{masseq}\\
E&=&\int \rho\Bigl\{\varepsilon+{v^2 \over 2}-{1 \over 2}U \nonumber \\
&& \hskip 0.3cm +{1 \over c^2}\Bigl({5 \over 8}v^4+{5 \over 2}v^2U+
{1 \over 2}\hat \beta_{\varphi}\Omega^2
+\varepsilon v^2+{P \over \rho}v^2+2\varepsilon U-{5 \over 2}U^2 \Bigr)
\Bigr\}d^3x,\label{energyeq}\\
J&=&\int \rho \biggl[ v_{\varphi}\Bigl\{1
+{1 \over c^2}\Bigl(v^2 +6U
+\varepsilon+{P \over \rho}\Bigr) \Bigr\}
+{\hat \beta_{\varphi}\Omega \over c^2}\biggr]d^3x.\label{anguleq}
\eeqn
$M_*$ conserves throughout the whole evolution of the system even if
there exists a dissipation process such as the emission of GWs. Thus,
in the case when we consider a sequence of a constant $M_*$, it may be
regarded as an evolution sequence of the system.

Our purpose is to calculate the 1PN correction of the 
angular velocity, the energy and the angular momentum for 
corotating binary.  To get the solution in the Newtonian order, 
we solve Eq. (\ref{tvtv}) with $\Lambda=0$ using 
the angular velocity 
\beq
\Omega^2={2 GM \over R^3}+{18G\bI_{11} \over R^5}\equiv \Omega^2_{\rm N}. 
\label{omen}
\eeq
In the 1PN approximation, we can expect that the 
following types of quantities will be the main terms of the angular 
velocity in the 1PN order:
\beq
\sim {GM \over R^3}\times {GM \over a_0c^2},~~~~
\sim {GM \over R^3}\times {GM \over R  c^2},~~~~
\sim {GMa_0^2 \over R^5}\times {GM \over a_0c^2}~~~{\rm and}~~~
\sim {GMa_0^2 \over R^5}\times {GM \over Rc^2},
\eeq
where $a_0$ is a typical radius of the star and we use the relation 
$\bI_{11} \sim Ma_0^2$. We will derive the details of four types of 
terms shown above in  the correction of the energy and the angular momentum 
 neglecting the effect of higher multipole terms. 
 
As shown in Ref. 12), the neglection of the higher multipole terms
brings us a great advantage so that we can regard each star of the
binary as an ellipsoid whose shape is the same as that in the Newtonian
order.  This is because the deformation is induced by higher order
effect of $R^{-1}$. Hence, we hereafter perform calculation setting
density profile of each star as an ellipsoid of its axial length $a_1$,
$a_2$ and $a_3$.

%%%%%%%%%%%%%%%%%%%%%%%%%%%%%%%%%%%%%
\subsection{Gravitational potentials}
%%%%%%%%%%%%%%%%%%%%%%%%%%%%%%%%%%%%%

As we did for the Newtonian potential $U$, we decompose PN gravitational
potentials into two parts; one is the contribution from star 1 and the
other is from star 2. In the following, we denote the former part as
$\phi^{1 \rightarrow 1}$, and the latter one as $\phi^{2 \rightarrow
  1}$, where $\phi$ denotes one of the potentials.  We also define
$\phi^{1 \rightarrow 2}$ and $\phi^{2 \rightarrow 2}$ as the
contribution from star 1 to 2 and star 2 itself, respectively.

%%%%%%%%%%%%%%%%%%%%%
\subsubsection{$X_0$}
%%%%%%%%%%%%%%%%%%%%%

As we mentioned above, the 1PN potentials 
are divided into two parts as $X_0=X_0^{1\ra 1} +X_0^{2 \ra 1}$, 
and we consider them separately. 
\begin{itemize}
\item Contribution from star 1

$X_0^{1 \ra 1}$ is derived from the Poisson equation\cite{shibapn}
\beqa
  \Delta X_0^{1 \ra 1} = 4\pi G\r 
 \biggl[ 2 \pi G\r \biggl(A_0 -\sum_l A_l x_l^2
    \biggr) +\frac{3P_0}{\r} \biggl( 1- \sum_l \frac{x_l^2}{a_l^2} \biggr) 
    +2 U^{2 \ra 1} \biggr],\nonumber \\
\eeqa
and the solution becomes
\beqn
  X_0^{1 \ra 1} &=& -\a_0 U^{1 \ra 1} +\a_1 D_1 +\sum_l
  \eta_l D_{ll} 
 -\frac{M}{R^3} \Bigl(2D_{11} -D_{22} -D_{33}\Bigr)\nonumber \\
&&    -\frac{M}{R^4} \Bigl(-2D_{111} +3D_{122} +3D_{133}\Bigr),
\eeqn
where
\beqa
  \a_0 &=& 2 \pi G\r A_0 + \frac{3P_0}{\r} +\frac{2GM}{R} +\frac{3G
    \bI_{11}}{R^3}, \\
  \a_1 &=& \frac{2GM}{R^2} +\frac{9 G\bI_{11}}{R^4}, \\
  \eta_l &=& 2\pi G\r A_l +\frac{3P_0}{\r a_l^2}.
\eeqa
$D_i$, $D_{ii}$, and $D_{1ii}$ are the solutions of equations
\beqa
  \Delta D_i &=& -4 \pi G\r x_i, \\
  \Delta D_{ii} &=& -4 \pi G\r x_i^2, \\
  \Delta D_{1ii} &=& -4 \pi G\r x_1 x_i^2,
\eeqa
and the solutions at star 1 are\cite{EFE}
\beqa
  D_i &=& \pi G\r a_i^2 \Bigl( A_i -\sum_l A_{il} x_l^2 \Bigr) x_i , \\
  D_{ii} &=& \pi G\r \biggl[ a_i^4 \Bigl( A_{ii} -\sum_l A_{iil} x_l^2
  \Bigr) x_i^2 \nonumber \\
  & &\hspace{20pt} + \frac{1}{4} a_i^2 \Bigl( B_i -2\sum_l B_{il}
  x_l^2 + \sum_l \sum_m B_{ilm} x_l^2 x_m^2 \Bigr) \biggr], \\
  D_{111} &=& \pi G\r \biggl[ a_1^6 \Bigl( A_{111} -\sum_l A_{111l} x_l^2
  \Bigr) x_1^3 \nonumber \\
  & &\hspace{20pt} + \frac{3}{4} a_1^4 \Bigl( B_{11} -2 \sum_l
  B_{11l} x_l^2 +
  \sum_l \sum_m B_{11lm} x_l^2 x_m^2 \Bigr) x_1 \biggr], \\
  D_{1ii} &=& \pi G\r \biggl[ a_1^2 a_i^4 \Bigl( A_{1ii} -\sum_l A_{1iil}
  x_l^2 \Bigr) x_1 x_i^2 \nonumber \\
  & &\hspace{20pt} +\frac{1}{4} a_1^2 a_i^2 \Bigl( B_{1i} -2\sum_l
  B_{1il} x_l^2 + \sum_l \sum_m B_{1ilm} x_l^2 x_m^2 \Bigr) x_1 \biggr],
  \nonumber \\
\eeqa
where $B_{ijk \cdots}$ are index symbols defined by
Chandrasekhar.\cite{EFE}

\item Contribution from star 2

The equation for $X_0^{2 \ra 1}$ is
\beqa
  \Delta X_0^{2 \ra 1} = 4\pi G\r \biggl[ 2 \pi G\r 
  \biggl(A_0 -\sum_l A_l y_l^2 \biggr) 
  +\frac{3P_0}{\r} \biggl( 1- \sum_l \frac{y_l^2}{a_l^2} \biggr) 
    +2 U^{1 \ra 2} \biggr],\nonumber \\
\eeqa
where $y_1 =-(x_1+R)$, $y_2=x_2$ and $y_3=x_3$. The solution is expressed as
\beq
  X_0^{2 \ra 1} = -\a_0 U^{2 \ra 1} -\a_1 D_1^{2 \ra 1}
      +\sum_l \eta_l D_{ll}^{2 \ra 1} 
  -\frac{M}{R^3} \Bigl(2D_{11}^{2 \ra 1} - D_{22}^{2 \ra 1}
  -D_{33}^{2 \ra 1}\Bigr), \label{X_0^(2)}
\eeq
where $D_{ij \cdots}^{2 \ra 1}$ are calculated from the same equations
as the case of $D_{ij \cdots}$,\cite{EFE} i.e.,
\beqa
  \Delta D_{ij \cdots}^{2 \ra 1} =-4 \pi G\r y_i y_j \cdots.
\eeqa
The solutions of $D_{ij \cdots}^{2 \ra 1}$ are 
\beqa
  D_1^{2 \ra 1} &=& \frac{GI_{11}}{R^2} \biggl(1-\frac{2x_1}{R}
    +\frac{6x_1^2-3x_2^2-3x_3^2}{2R^2} + O(R^{-3}) \biggr), \\
  D_2^{2 \ra 1} &=& \frac{GI_{22}}{R^2} \biggl( \frac{x_2}{R}
    -\frac{3x_1x_2}{R^2} + O(R^{-3}) \biggr), \\
  D_{ii}^{2 \ra 1} &=& \frac{GI_{ii}}{R} \biggl( 1-\frac{x_1}{R}
    +\frac{2x_1^2 -x_2^2 -x_3^2}{2R^2}
    +\frac{-2x_1^3+3x_1(x_2^2+x_3^2)}{2R^3} + O(R^{-4}) \biggr)
    \nonumber \\
   & &+\frac{3 G\bI_{ii11}}{2R^3} 
   \biggl( 1-\frac{3x_1}{R} + O(R^{-2}) \biggr),
\eeqa
where
\beqa
  \bI_{ii11} = I_{ii11} -\frac{1}{3} \sum_lI_{iill}.
\eeqa

\end{itemize}

%%%%%%%%%%%%%%%%%%%%%%%%
\subsubsection{$X_{\O}$}
%%%%%%%%%%%%%%%%%%%%%%%%

\begin{itemize}

\item Contribution from star 1

The equation for $X_{\O}^{1 \ra 1}$ is 
\beqa
  \Delta X_{\O}^{1 \ra 1} =8 \pi G\r \biggl( x_1^2 +x_2^2 +Rx_1
    +\frac{R^2}{4} \biggr).
\eeqa
Then the solution is
\beqa
  X_{\O}^{1 \ra 1} =-2 \biggl( D_{11} +D_{22} +RD_1 +\frac{R^2}{4} U^{1
      \ra 1} \biggr).
\eeqa

\item Contribution from star 2

The equation $X_{\O}^{2 \ra 1}$ may be written as 
\beqa
  \Delta X_{\O}^{2 \ra 1} =8 \pi G\r \biggl( y_1^2 +y_2^2 -Ry_1
    +\frac{R^2}{4} \biggr).
\eeqa
Then, using $D_{ij \cdots}^{2 \ra 1}$, the solution is easily 
derived as
\beqa
  X_{\O}^{2 \ra 1} &=&-2 \biggl( D_{11}^{2 \ra 1} +D_{22}^{2 \ra 1}
    -RD_1^{2 \ra 1} +\frac{R^2}{4} U^{2 \ra 1} \biggr), \\
  &=& -\frac{R^2}{2} U^{2 \ra 1} -\frac{2GI_{11}}{R^2} x_1 -
  \frac{2GI_{22}}{R} \Bigl( 1- \frac{x_1}{R} \Bigr).
\eeqa

\end{itemize}

%%%%%%%%%%%%%%%%%%%%%%%%%%%%%%%%%%%%%%%
\subsubsection{$\hat{\beta}_{\varphi}$}
%%%%%%%%%%%%%%%%%%%%%%%%%%%%%%%%%%%%%%%

$\hat{\beta}_{\varphi}$ is obtained from 
$\hat P_1$ and $\hat P_2$ which satisfy, 
\beqa
  \Delta \hat P_1 &=&-4 \pi G\r \left( x_1 +\frac{R}{2} \right),
  \\
  \Delta \hat P_2 &=&-4 \pi G\r x_2.
\eeqa
$\hat P_i$ is also written as $\hat P_i^{1 \ra 1} +\hat P_i^{2 \ra 1}$, where
\beqa
\hat P_1^{1 \ra 1} &=& D_1 +\frac{R}{2} U^{1 \ra 1}, \\
\hat P_2^{1 \ra 1} &=& D_2
\eeqa
and
\beqa
\hat P_1^{2 \ra 1} &=& D_1^{2 \ra 1} -\frac{R}{2} U^{2 \ra 1}, \label{P_1^2}\\
\hat P_2^{2 \ra 1} &=& D_2^{2 \ra 1}. \label{P_2^2}
\eeqa

%%%%%%%%%%%%%%%%%%%%%%%%%%%%%%%%%%%%%%%%%%%%%%%%%%%%%%%%%%%%%%%%%%%%%%%
\subsection{The Post-Newtonian angular velocity and definition of mass 
and center of mass}
%%%%%%%%%%%%%%%%%%%%%%%%%%%%%%%%%%%%%%%%%%%%%%%%%%%%%%%%%%%%%%%%%%%%%%%

As in the Newtonian case, the orbital angular velocity in the 1PN order is 
also derived using the first TV equation as 
\beqn
0=\int {\pa P \over \pa x_1} d^3x
&=&\int \rho U_{,1} d^3x+{R \over 2}M\Omega^2
+{1 \over c^2}\biggl[-\int \rho X_{0,1} d^3x \nonumber \\
&& +\Omega^2_{\rm N}
\int\rho \Bigl(2\varpi^2U_{,1}+4x_1U +2RU-X_{\Omega,1}+ \beta_{\varphi,1}
\Bigr)d^3x \nonumber \\
&& ~~~
+\Omega^4_{\rm N}
\biggl({R \over 2}(3I_{11}+I_{22})+{MR^3 \over 8}\biggr)\biggr]
,\label{tveq}
\eeqn
where ``$,k$'' denotes the partial derivative with respect to $x_k$. 
Since all the gravitational potentials can be written as the polynomial 
form of $x_k$, we can immediately perform the integral shown above and 
obtain $\O^2$ as 
\beqa
  \O^2 &=& \frac{2GM}{R^3} \left[ 1+ \frac{G}{c^2} \left\{ 2\pi \r A_0
   -\frac{9M}{4R} -\frac{M}{10R^3} (28a_1^2 -14a_2^2 -9a_3^2)
   +O(R^{-4}) \right\} \right] \nonumber \\
   & &+ \frac{18 G\bI_{11}}{R^5} \left( 1+\frac{28}{15 c^2} \pi G\r A_0
     +O(R^{-2}) \right),  \label{PNoav1}
\eeqa
where we make use of the relations $A_0 =\sum_l A_l a_l^2$ and 
Eqs. (\ref{tvtv}) with $\Lambda=0$ to simplify the expression. 

Next, we discuss the mass of the system and 
definition of the center of mass for each star. 
This is because there are several definitions of them in the 1PN 
approximation, and we should clarify the difference between the 
similar ones. 

First, we consider the conserved mass which is defined as
\beqa
  M_{\ast} &=& \int d^3 x \r \biggl[ 1+\frac{1}{c^2} \biggl( \frac{v^2}{2} 
      +3U \biggr) \biggr] \nonumber \\
  &=& M \biggl[ 1+\frac{G}{c^2} \biggl( \frac{13M}{4R} +\frac{12 \pi \r
        A_0}{5} +\frac{M}{20R^3} (34a_1^2 -11a_2^2 -15a_3^2) +O(R^{-5})
    \biggr) \biggr].\nonumber \\
\label{conmas}
\eeqa
Using $M_{\ast}$, $\O^2$ becomes
\beqa
  \O^2 &=& \frac{2GM_{\ast}}{R^3} \left[ 1+
    \frac{G}{c^2} \left\{ -\frac{2\pi \r A_0}{5}
      -\frac{11M_{\ast}}{2R} 
  -\frac{M_{\ast}}{20R^3} (90a_1^2 -39a_2^2 -33a_3^2) \right\}
  +O(R^{-4}) \right] \nonumber \\
  & &+ \frac{18G(\bI_{11})_{\ast}}{R^5} \left[ 1+\frac{G}{c^2} \left\{
      -\frac{8}{15} \pi \r A_0 -\frac{13 M_{\ast}}{4R} +O(R^{-2})
      \right\} \right],
\eeqa
where $(\bI_{11})_{\ast}=M_*\bI_{11}/M$. 
Thus, $\O^2$ looks as if it depends on the internal structure of the
star even in the limit of $a_i/R \ra 0$. Since we believe that in the 
EOM for the point particle, the quantities
depending on the internal structure does not appear, $M_{\ast}$ is not
desirable to describe the EOM for the point particle. Instead, in the case
when the EOM is derived, one usually adopts the PPN mass\cite{Will}
defined as
\beqa
  M_{{\rm PPN}} &=& \int d^3 x \r \biggl[ 1+ \frac{1}{c^2} \biggl(
     \frac{v^2}{2} +3U -\frac{1}{2} U_{{\rm self}}
    +\frac{v_{{\rm self}}^2}{2} \biggr) \biggr] \nonumber \\
  &=& M \biggl[ 1+\frac{G}{c^2} \biggl( \frac{13M}{4R} +2\pi \r A_0
   +\frac{M}{20R^3} (38a_1^2 -7a_2^2 -15a_3^2) +O(R^{-5}) \biggr) \biggr],
\nonumber \\
\eeqa
where $U_{\rm self}$ and $v_{\rm self}$ are the 
self-gravity part of the Newtonian potential and the spin velocity of 
each star. 
When we rewrite Eq. (\ref{PNoav1}) using the PPN mass, the orbital
angular velocity does not depend on the internal structure of the star
and it agrees with that for the point particle\cite{BDWW} in the limit
of $a_i/R \ra 0$ as
\beqa
  \O^2 &=& \frac{2GM_{{\rm PPN}}}{R^3} \left[ 1
    +\frac{G}{c^2} \left\{ -\frac{11M_{{\rm PPN}}}{2R}
   -\frac{M_{{\rm PPN}}}{20 R^3}
   (94 a_1^2 -35a_2^2 -33a_3^2)
 +O(R^{-4}) \right\} \right] \nonumber \\
  & &+\frac{18G(\bI_{11})_{{\rm PPN}}}{R^5} \left[ 1+\frac{G}{c^2}
    \left\{ -\frac{2}{15} \pi \r A_0 -\frac{13M_{{\rm PPN}}}{4R}
      +O(R^{-2}) \right\} \right], \label{PNoav3}
\eeqa
where $(\bI_{11})_{\rm PPN}=M_{\rm PPN}\bI_{11}/M$.
Thus, when we compare the present results with the point 
particle calculations, we should use the PPN mass. 
In the present case, however,
$M_{{\rm PPN}}$ is not a conserved quantity although $M_{\ast}$ is. When 
we consider a sequence of equilibrium configurations as an
evolutionary sequence, we should fix $M_{\ast}$.

Next, we consider the definition of the center of mass for each star. 
In the PPN formalism, it is defined as\cite{Will}
\beqa
  x^i_{{\rm PPN}} =\frac{1}{M_{{\rm PPN}}}\int d^3 x \r x^i \left[ 1+
    \frac{1}{c^2} \left(
    \frac{v^2}{2} +3U -\frac{1}{2} U_{{\rm self}}
    +\frac{v_{{\rm self}}^2}{2} \right) \right].
\eeqa
$x_1$ coordinate of the center of mass for star 1 deviates from
0 to
\beqa
  \frac{G}{c^2} \left( -\frac{2M a_1^2}{5R^2} +O(R^{-4}) \right).
\eeqa
Thus, in the PPN formalism, the following orbital separation should be used;
\beqa
  R_{{\rm PPN}} =R \left[ 1+ \frac{G}{c^2} \left\{ -\frac{4M
        a_1^2}{5R^3} +O(R^{-5}) \right\} \right].
\eeqa

It is worth noting that 
when we define the center of mass by the conserved mass as
\beqa
  x^i_{\ast} =\frac{1}{M_{\ast}} \int d^3 x \r_{\ast} x^i,
\eeqa
the result is the same up to $O(R^{-4})$. 
Thus, in this paper, we do not have to distinguish $R_*$ from $R_{\rm PPN}$. 
Even in the general case, the difference between 
$R_*$ and $R_{\rm PPN}$ is expected to be small. 

Using $R_*$ and/or $R_{\rm PPN}$, $\O^2$ is rewritten as 
\beqa
  \O^2 &=& \frac{2GM_{\ast}}{R_{\ast}^3} \Biggl[ 1+
    \frac{G}{c^2} \biggl\{ -\frac{2\pi \r A_0}{5}
      -\frac{11M_{\ast}}{2R_{\ast}} \nonumber  \\
& & \hskip 3cm  -\frac{M_{\ast}}{20R_{\ast}^3} (138a_1^2 -39a_2^2 -33a_3^2)
  +O(R_{\ast}^{-4})\biggr\} \Biggr] \nonumber \\
& &+ \frac{18G(\bI_{11})_{\ast}}{R_{\ast}^5} \left[ 1+\frac{G}{c^2}
    \left\{ -\frac{8}{15} \pi \r A_0 -\frac{13 M_{\ast}}{4R_{\ast}}
      +O(R_{\ast}^{-2}) \right\} \right],
\eeqa
or
\beqa
  \O^2 &=& \frac{2GM_{{\rm PPN}}}{R_{{\rm PPN}}^3} \Biggl[ 1
    +\frac{G}{c^2} \biggl\{ -\frac{11M_{{\rm PPN}}}{2R_{{\rm PPN}}}
\nonumber \\
& & \hskip 3cm   -\frac{M_{{\rm PPN}}}{20 R_{{\rm PPN}}^3}
   (142 a_1^2 -35a_2^2 -33a_3^2)
 +O(R_{{\rm PPN}}^{-4}) \biggr\} \Biggr] \nonumber \\
  & &+\frac{18G(\bI_{11})_{{\rm PPN}}}{R_{{\rm PPN}}^5} \left[
   1+\frac{G}{c^2} \left\{ -\frac{2}{15} \pi \r A_0
   -\frac{13M_{{\rm PPN}}}{4R_{{\rm PPN}}} 
      +O(R_{{\rm PPN}}^{-2}) \right\} \right].\label{omepnp}
\eeqa
Here, we should note that the effect of the spin-orbit coupling terms 
will appear in $\O^2$ from $O(R^{-6}_{\rm PPN})$.\cite{Kidder}
According to the 1PN study, it becomes 
\beq
\O^2 = {2GM_{\rm PPN} \over R_{\rm PPN}^3 }
\left[ 1-{2 GM_{\rm PPN} \over  c^2R_{\rm PPN}^3}(a_1^2+a_2^2)\right],
\label{ooooo}
\eeq
where we omit other terms which are not related to this discussion. 
Eq. (\ref{ooooo}) shows that the terms of $O(R^{-6}_{\rm PPN})$ in 
Eq. (\ref{omepnp}) cannot be explained only by the spin-orbit 
coupling term. This will mean that there appears a new effect, say the 1PN 
quadrupole one, in Eq. (\ref{omepnp}). \footnote{In this paper, we use
  the PPN mass in order to compare the angular velocity with that of the 
  point particle case. However, the PPN mass is useful only for the
  point particle case because it does not conserve if we take into
  account the tidal forces on each star of binary. In the context of
  this paper, we have to use the mass which is valid even if the tidal
  forces exist. Unfortunately, no one has proposed such a mass as far as 
  we know.}

%%%%%%%%%%%%%%%%%%%%%%%%%%%%%%%%%%%%%%%%%%%%%%%%
\subsection{The energy and the angular momentum}
%%%%%%%%%%%%%%%%%%%%%%%%%%%%%%%%%%%%%%%%%%%%%%%%

%%%%%%%%%%%%%%%%%%%%%%%%%%%%%%%%
\subsubsection{The Total Energy}
%%%%%%%%%%%%%%%%%%%%%%%%%%%%%%%%

The 1PN total energy is calculated from $E=E_{{\rm N}}^{{\rm
    def}}+E_{{\rm PN}}^{{\rm def}}/c^2$, where\cite{chand}\cite{shibapn}
\beqa
  E_{{\rm N}}^{{\rm def}} &=& \int \r \left( \frac{1}{2} v^2
    -\frac{1}{2} U \right) d^3 x, \\
  E_{{\rm PN}}^{{\rm def}} &=& \int \r \left( \frac{5}{8} v^4
    +\frac{5}{2} v^2 U +\frac{P}{\r} v^2 -\frac{5}{2} U^2 +\frac{1}{2}
    \hat{\beta}_{\varphi} \O^2 \right) d^3 x.
\eeqa
For the 1PN corotating binary, they become
\beqa
  E_{{\rm N}}^{{\rm def}} &=& M \left[ -\frac{4\pi G\r A_0}{5}
    +\frac{\O^2}{5} (a_1^2 +a_2^2) +\frac{R^2 \O^2}{4} -\frac{GM}{R}
    -\frac{3 G\bI_{11}}{R^3} +O(R^{-5}) \right], \nonumber \\
  E_{{\rm PN}}^{{\rm def}} &=& 2G^2M \Biggl[ -\frac{34}{21} (\pi \r A_0)^2
    -\frac{11M \pi \r A_0}{3R} -\frac{7M^2}{32R^2} \nonumber \\
  & & \hspace{50pt} +\frac{M \pi \r
    A_0}{R^3} \biggl\{ \frac{68}{105} (a_1^2+a_2^2) -\frac{61}{7}
    \frac{\bI_{11}}{M} \biggr\}  \nonumber \\
  & &\hspace{50pt}  +\frac{M^2}{240R^4} (302a_1^2 +59a_2^2
    -209a_3^2) +O(R^{-5}) \Biggr].\label{eneeq}
\eeqa
If we substitute $\O^2$ into the above formulas, $E$
may be rewritten as $E_{{\rm N}} + E_{{\rm PN}}/c^2$ where
\beqa
  E_{{\rm N}} &=& M \left[ -\frac{4\pi G\r A_0}{5}
    -\frac{GM}{2R}
    +\frac{3 G\bI_{11}}{2R^3}
    +\frac{\O_{{\rm N}}^2}{5} (a_1^2 +a_2^2)
    +O(R^{-5}) \right], \\
  E_{{\rm PN}} &=& 2G^2M \Biggl[ -\frac{34}{21} (\pi \r A_0)^2
    -\frac{19M \pi \r A_0}{6R} -\frac{25M^2}{32R^2} \nonumber \\ 
  & & \hspace{50pt}+\frac{M \pi \r
    A_0}{R^3} \left\{ \frac{22}{21} (a_1^2+a_2^2) -\frac{158}{35}
    \frac{\bI_{11}}{M} \right\}  \nonumber \\
  & &\hspace{50pt}  +\frac{M^2}{240R^4} (26a_1^2 +35a_2^2
    -155a_3^2) +O(R^{-5}) \Biggr].
\eeqa

%%%%%%%%%%%%%%%%%%%%%%%%%%%%%%%%%%%%%%%%%%
\subsubsection{The Total Angular Momentum}
%%%%%%%%%%%%%%%%%%%%%%%%%%%%%%%%%%%%%%%%%%

We can calculate the 1PN total angular momentum from 
$J=J_{{\rm N}}^{{\rm def}} +J_{{\rm PN}}^{{\rm def}}/c^2$, 
where\cite{chand}\cite{shibapn}
\beqa
  J_{{\rm N}}^{{\rm def}} &=& \int \r v_{\varphi} d^3 x, \\
  J_{{\rm PN}}^{{\rm def}} &=& \int \r \left[ v_{\varphi} \left( v^2 +6U 
    +\frac{P}{\r} \right) + \hat{\b_{\varphi}} \O \right] d^3 x, 
\eeqa
and $v_{\varphi} = \O \varpi^2$.
For the PN corotating binary, they become
\beqa
  J_{{\rm N}}^{{\rm def}} &=& 2M \O \left( \frac{R^2}{4} + \frac{a_1^2
    +a_2^2}{5} \right), \nonumber \\
  J_{{\rm PN}}^{{\rm def}} &=& GM \O_{{\rm N}} \left[ R^2 \pi \r A_0 +5 R 
    M + \frac{164}{105} \pi \r A_0 (a_1^2+a_2^2) \right. \nonumber \\
  & &\hspace{50pt} \left. +\frac{M}{10R} (20a_1^2 
    +15a_2^2 -11a_3^2) +O(R^{-2}) \right]. \label{angeq}
\eeqa
If we substitute $\O^2$ into the above formulas, $J$ may be rewritten as 
$J_{{\rm N}}+ J_{{\rm PN}}/c^2$ where
\beqa
  J_{{\rm N}} &=& 2M \O_{{\rm N}} \left( \frac{R^2}{4} + \frac{a_1^2
    +a_2^2}{5} \right), \\
  J_{{\rm PN}} &=& GM \O_{{\rm N}} \left[ \frac{3}{2} R^2 \pi \r A_0
    +\frac{71}{16} R M +\frac{\pi \r A_0}{1050} (2018 a_1^2 +2081 a_2^2
    +21a_3^2) \right. \nonumber \\
  & &\hspace{50pt} \left.+\frac{M}{80 R} (122a_1^2 +85 a_2^2 -97 a_3^2)
    +O(R^{-2}) \right].
\eeqa

%%%%%%%%%%%%%%%%%%%%%%%%%%%%%%%%%%%%%%%%%%%%%%%%%%
\subsection{Construction of equilibrium sequences} 
%%%%%%%%%%%%%%%%%%%%%%%%%%%%%%%%%%%%%%%%%%%%%%%%%%

To construct equilibrium sequences of 1PN corotating binary stars 
fixing $M_*$ and $\rho$, we use the following method: 

\noindent
Step (1): Using Eqs. (\ref{tvtv}) with $\Lambda=0$,
we numerically calculate equilibrium sequences in the Newtonian order. 
Up to this stage, $\alpha_2$, $\alpha_3$, and $\tilde R=R/a_1$ are 
determined. 

\noindent
Step (2): $a_1$ is determined from the condition $M_*=$constant using 
Eq. (\ref{conmas}):
\beq
a_1^3={3 M_* \over 4\pi\rho\alpha_2\alpha_3}
\biggl[1-{\pi G \rho \over c^2}\biggl({3M_* \over 4\pi\rho\alpha_2\alpha_3}
\biggr)^{2/3}\biggl({12 \tilde A_0 \over 5}
+{13 \alpha_2\alpha_3 \over 3\tilde R}
+{\alpha_2\alpha_3 \over 15 \tilde R^3}(34-11\alpha_2^2-15\alpha_3^2)\biggr)
\biggr].\label{aaeq}
\eeq

\noindent
Step (3): After substituting Eq. (\ref{aaeq}) into Eqs. (\ref{PNoav1}),
(\ref{eneeq}), and (\ref{angeq}), we rewrite the 1PN expressions for the 
orbital angular velocity, the energy and the angular momentum as
\beqa
  \tilde{\O}^2 &\equiv& \frac{\O^2}{\O_*^2}
  = \tilde{\O}_{\rm N}^2 +\frac{GM_{\ast}}{c^2 a_{\ast}}
  \tilde{\O}_{\rm PN}^2, \\
  \tilde{E} &\equiv& \frac{E}{(GM_{\ast}^2/a_{\ast})}
  = \tilde{E}_{\rm N} +\frac{GM_{\ast}}{c^2 a_{\ast}}
  \tilde{E}_{\rm PN},\\
  \tilde{J} &\equiv& \frac{J}{(GM_{\ast}^3 a_{\ast})^{1/2}}
  = \tilde{J}_{\rm N} +\frac{GM_{\ast}}{c^2 a_{\ast}}
  \tilde{J}_{\rm PN},
\eeqa
where
\beqa
  a_{\ast} &=&\left( \frac{3M_{\ast}}{4 \pi \r} \right)^{1/3},\\
  \tilde{\O}_{\rm N}^2 &=&  \a_2 \a_3 \left[
   \frac{2}{\tilde{R}^3} +\frac{6}{5\tilde{R}^5} (2-\a_2^2-\a_3^2)
 \right], \\
 \tilde{\O}_*^2 &=&{GM_* \over a_*^3},\\
  \tilde{\O}_{\rm PN}^2 &=& (\a_2 \a_3)^{4/3} \left[
   \frac{3}{\tilde{R}^3} \frac{\tilde{A_0}}{\a_2 \a_3}
   -\frac{9}{2\tilde{R}^4} +\frac{42}{25\tilde{R}^5}
   \frac{\tilde{A_0}}{\a_2 \a_3} (2-\a_2^2-\a_3^2) \right. \nonumber \\
  & &\left. \hspace{70pt}-\frac{1}{5\tilde{R}^6}(28-14\a_2^2-9\a_3^2)
  \right],\label{omega1PN} \\
%%%%%%%%%
  \tilde{E}_{\rm N} &=& (\a_2\a_3)^{1/3}\left[ -\frac{3}{5}
    \frac{\tilde{A_0}}{\a_2 \a_3}
  -\frac{1}{2 \tilde{R}} +\frac{1}{10 \tilde{R}^3}(2-\a_2^2-\a_3^2)
  \right. \nonumber \\
  & &\left. \hspace{60pt}+
  \frac{\tilde \Omega_{\rm N}^2}{5\alpha_2\alpha_3} (1+\a_2^2) \right], \\
%%%%%%%%%
  \tilde{E}_{\rm PN} &=& (\a_2\a_3)^{2/3} \biggl[
  -\frac{3}{140} \left( \frac{\tilde{A_0}}{\a_2
      \a_3} \right)^2 +\frac{55}{48 \tilde{R}^2} +\frac{1}{700
    \tilde{R}^3} \frac{\tilde{A_0}}{\a_2 \a_3} (398+401\a_2^2+\a_3^2)
  \nonumber \\
  & &\hspace{60pt} -\frac{1}{120
      \tilde{R}^4}(194+215\a_2^2+165\a_3^2) \biggr], \label{energy1PN}\\
%%%%%%%%%%%
  \tilde{J}_{\rm N} &=& 2 (\a_2\a_3)^{-1/6} \left[
   \frac{\tilde \Omega_{\rm N}^2}{\alpha_2\alpha_3} 
 \right]^{1/2} \left( \frac{\tilde{R}^2}{4}+\frac{1+\a_2^2}{5} \right),\\
%%%%%%%%%%%
  \tilde{J}_{\rm PN} &=& (\a_2\a_3)^{1/6} \left[
   \frac{\tilde \Omega_{\rm N}^2}{\alpha_2\alpha_3}
 \right]^{1/2} 
  \biggl[ -\frac{3}{8} \tilde{R}^2 \frac{\tilde{A_0}}{\a_2
     \a_3} +\frac{83}{48} \tilde{R} \nonumber \\
  & & +\frac{1}{1400}
   \frac{\tilde{A_0}}{\a_2 \a_3} (338+401\a_2^2+21\a_3^2)
 -\frac{1}{240\tilde{R}} (494+155\a_2^2+141\a_3^2) \biggr]. \nonumber \\
\eeqa
Then, using the numerical value of $\alpha_2$, $\alpha_3$, and $\tilde R$ 
determined at step (1), we calculate the sequence of the angular velocity, 
the energy and the angular momentum as functions of the
orbital separation. 

We repeat this procedure changing the mean radius of each star $a_*$. 
Once a sequence is obtained, we search the minimum point of the energy. 
If we find it, we call it the ISCCO. 
The 1PN approximation is valid only for small 
$C_{\rm s}\equiv GM_{\ast}/c^2 a_{\ast}$, 
i.e., the characteristic value of the compactness of each star. 
Thus, we hereafter show results only for small $C_{\rm s}$. 

In Figs.6(a) and (b), we show $\tilde{E} =E/(GM_{\ast}^2/a_{\ast})$ and
$\tilde{J}= J/(GM_{\ast}^3 a_{\ast})^{1/2}$ as functions of 
$\Omega/\Omega_*$. The figures show the important fact that 
the angular velocity at the ISCCO increases approximately 
in proportion to $C_{\rm s}$. 

In Fig.7, we show $\tilde{\O}=\O/\O_*$ at the ISCCO as a function of
$C_{\rm s}$. We show $\tilde{\O}$ at the energy minimum as well as 
that at the angular momentum minimum. The figure indicates that 
two minima are almost coincident, but slightly different. 
We guess that this disagreement is due to the 
neglect of the effect from the higher multipole deformation. 
In any case, we may expect that the ISCCO locates near two minima. 
Fig.7 clearly shows that the orbital angular velocity at the ISCCO 
increases almost linearly with increase of $C_{\rm s}$ as 
\beq
\Omega_{\rm ISCCO}=\Omega_{\rm N:ISCCO}\bigl(1+C_{\rm PN}(n)
C_{\rm s}\bigr),
\eeq 
where $C_{\rm PN}(n)$ is a constant which depends on the 
EOS (see next section) and for the incompressible case, 
$C_{\rm PN}(0) \simeq 0.5$. 

We briefly argue the meaning of 
the result for $\Omega_{\rm ISCCO}$ using 
Eq. (\ref{energy1PN}). In Eq. (\ref{energy1PN}), there appear 
mainly two kinds of effects in the first and second terms, 
respectively: The first term 
concerns the internal structure of each star and denotes 
that each star is forced to be compact due to the 1PN gravity 
because of its negative definite character. Thus, 
the first term acts as the stabilization because 
the tidal effect is less important in the case when the star 
becomes compact. On the other hand, the second term concerns the gravity 
acting between two stars because it exists in the limit 
$a_1, a_2, a_3 \rightarrow 0$. This term tends to act as 
destabilization of circular orbits because of its dependence of 
$\tilde R~(O(\tilde R^{-2})$) and its positive definite 
character. In this way, there exist two opposite effects in the 1PN 
corrections, and the present result shows that the former effect 
dominates over the latter one. 

%%%%%%%%%%%%%%%%%%%%%%%%%%%%%%%%%%%%%%%%%%%%%%%%%%%%%%%
\section{First post Newtonian study: Compressible case}
%%%%%%%%%%%%%%%%%%%%%%%%%%%%%%%%%%%%%%%%%%%%%%%%%%%%%%%

For compressible stars, we can no  longer 
solve the integrated form of the Euler equation analytically 
although we make approximate solutions assuming 
density profile of each star appropriately.\cite{tanib}\cite{baum}
To obtain accurate equilibrium states, we need numerical calculation. 
In this section, we show numerical results for 
sequences of 1PN corotating 
equilibrium binaries obtained by solving the integrated form of 
the Euler equation as well as the Poisson equations for the 
gravitational potentials. We use a numerical method 
which is developed in previous papers.\cite{shibapn}

In the compressible case, the integrated form of the 
Euler equation for uniformly rotating fluid is 
\beqn
&& K(n+1) \rho^{1/n}
-{1 \over 2c^2}\Bigl(K(n+1)\rho^{1/n}\Bigr)^2 \nonumber \\
&&=U-{X_0 \over c^2}
+\biggl\{{\varpi^2 \over 2}+{1 \over c^2}\Bigl(2\varpi^2U-X_{\Omega}
+\hat \beta_{\varphi}\Bigr)\biggr\}\Omega^2
+{\varpi^4 \over 4c^2}\Omega^4+{\rm const},
\eeqn
where we assume the polytropic EOS, i.e., 
\beq
P=K\rho^{1+1/n}~~{\rm and}~~\varepsilon = {nP \over \rho}. 
\eeq
Hereafter, we set $n=0.5$, $0.75$ and 1 because real NSs 
will have $n=0.5 \sim 1$.\cite{COMPACT}
The gravitational potential in the 1PN order 
are obtained by solving Eqs. (\ref{PNpot}). 
The conserved mass, the energy, and the angular momentum 
are defined as Eqs. (\ref{masseq})-(\ref{anguleq}). 
Following a previous section, we hereafter show 
the angular velocity, the energy and 
the angular momentum in units of $\Omega_*\equiv(GM_*/a_*^3)^{1/2}$, 
$GM_*^2/a_*$ and $(GM_*^3a_*)^{1/2}$, 
where 
\beqn
a_*&=&\biggl({ K M_*\over 2.524 G}\biggr)^{1/5}\hskip 1.1cm {\rm for}~n=0.5,
\nonumber \\
&=&\biggl({ K M_*^{1/3}\over 0.9960 G}\biggr)^{1/3}
\hskip 0.9cm {\rm for}~n=0.75,
\nonumber \\
&=& \biggl({\pi K \over 2G}\biggr)^{1/2}\hskip 1.7cm {\rm for}~n=1,
\nonumber \\
\eeqn
We also define the center of mass for each star as 
\beq
x_{1*}={1 \over M_*}\int \rho_* x_1 d^3x,
\eeq
and as a result, the orbital separation is defined to be $R_*=2x_{1*}$. 
In numerical computation, sequences of equilibrium states are obtained 
fixing $M_*$ and $K$. 

First, we demonstrate that our numerical results are reliable 
by showing the relation between the orbital separation and the 
angular velocity. 
If we neglect the contribution from the quadrupole deformation of 
each star, the angular velocity in the 1PN 
approximation can be approximately expressed as\cite{Will}\cite{shibapn} 
\beq
\Omega=\sqrt{{GM_* \over R_*^3}}\biggl[1+{G \over c^2}\biggl(
-{3-n \over 2(5-n)}{M_* \over a_*}-{11M_* \over 8R_*}\biggr)\biggr].
\label{analy}
\eeq
Numerical results should agree well with this analytic formula. 
In Fig.8, we show the relation 
between $R_*/a_*$ and $\Omega/\Omega_*$ for $n=0.5$
and $C_{\rm s}\equiv GM_*/a_*c^2=0$ (open circles), 
0.0327 (filled circles) and 
0.0654 (open squares) as examples. 
Dotted lines are drawn using Eq. (\ref{analy}) 
as counter parts of the numerical results. The figure shows that 
numerical results agree with the analytic formula well except for 
slight deviation caused by the deformation of each star. 

In Figs.9, we show the energy and angular momentum as functions of 
$\Omega/\Omega_*$ for sequences of $n=0.5$(a), 0.75(b) and 1(c). 
In Fig.9(a), open circles, filled circles, and open squares denotes 
sequences for $C_{\rm s}=0$, 0.0327 and 
0.0654, in Fig.9(b), they denote sequences for $C_{\rm s}=0$, 0.02 and 
0.04, and in Fig.9(c), they denote 
sequences for $C_{\rm s}=0$, 0.0167 and 0.0333, respectively. 
It is shown that the minima of the energy and the angular momentum 
almost coincide, and $\Omega/\Omega_*$ 
at those minima (i.e., ISCCO) increases with increase 
of $C_{\rm s}$ as in the incompressible case. 

In figs.10, we show the orbital angular velocity at the ISCCO as a 
function of $C_{\rm s}$ for $n=0.5$(a), 0.75(b) and 1(c). 
It is found that as in the case $n=0$, $\Omega_{\rm ISCCO}$ almost 
linearly increases with increase of $C_{\rm s}$ as
\beq
\Omega_{\rm ISCCO}=\Omega_{\rm N:ISCCO}
\bigl(1+C_{\rm PN}(n)C_{\rm s}\bigr), 
\eeq
where 
$C_{\rm PN}(0.5) \sim 1.1$, $C_{\rm PN}(0.75) \sim 1.8$, 
and $C_{\rm PN}(1)\sim 2.5$. 
Thus, not only $\Omega_{\rm N:ISCCO}$, but also $C_{\rm PN}$ changes 
with the EOS, and for softer EOS, 
$\Omega_{\rm ISCCO}$ sensitively increases with $C_{\rm s}$ 
because the 1PN effect of the self gravity is larger for the
softer EOS. 

The summary of \S 3 and \S 4 is as follows.
\begin{enumerate}
\item $\Omega_{\rm ISCCO}$ increases with the increase of the 
1PN GR correction. \\
This is because each star of the binary becomes compact due to the strong 
GR self-gravity and as a result, the tidal effect becomes less effective. 
\item The feature mentioned at 1 is more remarkable for larger $n$, 
i.e., for softer EOS. \\
The reason for this behavior is that for softer EOS,
each star has the 
larger central density and is susceptible to the GR correction.
\end{enumerate}

\noindent
These conclusions are obtained by analyzing equilibrium sequences 
of corotating binary. However, it seems that 
these features may not be very sensitive to the spin and circulation 
of each star because the above properties mainly originate from 
the change of the density profile due to the GR correction. 
Therefore, we may expect that the similar trend will hold 
for constant-$C$ sequences although more investigation is needed
for this point.

%%%%%%%%% taniguchi part %%%%%%%%%%%%%%%%%

%%%%%%%%%%%%%%%%%%%%%%%%%%%%%%%%%%%%%%%%%%%%%%%%%%%%%%%%%%%%%%%%%%%%%%%%
\section{GR orbital effects: Pseudo-Newtonian potential 
approach}\label{pseudo}
%%%%%%%%%%%%%%%%%%%%%%%%%%%%%%%%%%%%%%%%%%%%%%%%%%%%%%%%%%%%%%%%%%%%%%%%

In previous two sections, we investigated the stability of corotating 
binary in the 1PN approximation, and showed $\Omega_{\rm ISCCO}$ 
 increases with the increase of the GR correction. 
This may be viewed from another point. 
In the 1PN approximation ISCO does not exist in the point 
particle limit.\cite{KWW} Therefore with the increase of relativity 
parameter $C_{\rm s}$, $\Omega_{\rm ISCO}$ should increase since the
increase of the relativity parameter means the way toward the point 
particle limit where the circular orbit exists for any radius. 
On the other hand, if we include the 
GR effect to the gravity acting between two stars in which ISCO
exists even in the point particle limit   
but we do not include the self-gravity of each star,
$\Omega_{\rm ISCO}$ decreases compared with $\Omega_{\rm ISCO}$ in the
point particle limit.
This seems to apparently disagree with the results in previous sections. 
In the following, we first review the result of the work by 
TN,\cite{TN} which is one of such studies. (We note that 
there is the other work by LW,\cite{LW}
in which the qualitative features of the results are essentially 
the same as those by TN.) 
In the last subsection, we clarify the reason for this apparent disagreement. 
We will show that this disagreement simply comes from the 
difference of two treatments. We also suggest that if 1PN effect
of the self-gravity is included in the formalism of TN, $\Omega_{\rm ISCO}$
will increase from the results without the 1PN self-gravity effect.

%%%%%%%%%%%%%%%%%%%%%%%%%%%%%%%%%%%%%%%%%%%%%%%%%
\subsection{Outline of method}
%%%%%%%%%%%%%%%%%%%%%%%%%%%%%%%%%%%%%%%%%%%%%%%%%

Here, we consider the so called 
Roche-Riemann problem\cite{EFE} as a model of a binary that 
consists of a finite size star (star 1) and a point-like gravity
source (star 2). 
To mimic the GR potential between two stars in which 
circular orbits are unstable for small radii, we use 
the pseudo-Newtonian potential proposed 
by Paczy\'nsky and Wiita.\cite{PW}
Details of the method by TN are as follows. 

First, we regard star 2 as a point-like star of mass $m_2$ and
denote the gravitational potential by it as $V_2(r)$. 
To include the effects of general relativity phenomenologically, 
we adopt pseudo-Newtonian potential as $V_2(r)$ (see below). 
Then, as star 1, we adopt the 
incompressible and homogeneous ellipsoid of its axial length 
$a_1$, $a_2$ and $a_3$, mass $m_1$ and the density $\rho_1$. 
The structure of star 1 is determined by the Newtonian 
hydrostatic equation. Thus, the self-gravity part of the 
gravitational potential $V_1$ is the same as that in Eq. (\ref{u11}).  

%%%%%%%%%%%%%%%%%%%%%%%%%%%%
\subsection{Basic equations}
%%%%%%%%%%%%%%%%%%%%%%%%%%%%

We use the tensor virial method\cite{EFE} to derive the equations
necessary for determining the orbital angular velocity and 
for constructing equilibrium configurations of star 1. 
We choose the coordinate system such that the origin is at the center 
of mass of star 1 and 
the $x_1$-axis points to the center of mass of star 2 which is 
located at $(-R,0,0)$. 
The $x_3$-axis is chosen to be the axis of the orbital rotation. 
In the frame of reference rotating with $\Omega$, 
the hydrostatic equation of star 1 is
written as
\begin{eqnarray}
  \rho_1 \sum_{j=1}^3 u_j {\pa u_i \over \pa x_j}&=&-{\pa P\over \pa
  x_i}+\rho_1 {\pa \over \pa x_i}\biggl[ V_1 +V_2 +{\Omega^2 \over 2}
  \biggl\{\biggl( {m_2 R \over m_1+m_2}+ x_1
  \biggr)^2+x_2^2 \biggr\} \biggr] \nonumber \\
  & &+2\rho_1 \Omega \sum_{l=1}^3\epsilon_{il3}u_l, \label{Eulereq}
\end{eqnarray}
where $u_i$, $P$, and $R$ are the internal velocity, the pressure, and
the separation between two stars, respectively. 

Following Chandrasekhar,\cite{EFE}
we expand the interaction potential $V_2$ in the power series of
$x_k$ up to the second order assuming $R \gg a_1, a_2$ and $a_3$. 
We assume that the potential $V_2$ depends only on the distance
$r$ from the center of mass of star 2 as
\begin{equation}
  V_2 =V_2(r),
\end{equation}
where $r$ is given by
\begin{equation}
  r=\bigl\{ (R+x_1)^2 +x_2^2 +x_3^2 \bigr\}^{1/2}.
\end{equation}
The expansion of $V_2(r)$ near the coordinate origin becomes
\begin{eqnarray}
  V_2 =(V_2)_0 +\biggl( {\pa V_2 \over \pa r} \biggr)_0 x_1 +{1 \over 2} 
  \biggl( {\pa^2 V_2 \over \pa r^2} \biggr)_0 x_1^2 +{1 \over 2R}
  \biggl( {\pa V_2 \over \pa r} \biggr)_0 (x_2^2 +x_3^2),
  \label{V2expand}
\end{eqnarray}
where the subscript 0 denotes the derivatives at the origin of the
coordinate. From the force balance at the center, 
we obtain the orbital angular velocity as 
\begin{eqnarray}
  {m_2 R \over m_1+m_2} \Omega^2 = -\biggl( {\p V_2 \over \p r}
  \biggr)_0 (1+ \delta), \label{forcebalance}
\end{eqnarray}
where $\delta$ is the quadrupole term of the interaction
potential\cite{LRS}. If we take the Newtonian potential
\beqa
  V_2(r)={Gm_2 \over r}
\eeqa
as an interaction potential, $\delta$ is written as
\beqa
  \delta={3 \over 10 R^2} (2 a_1^2 -a_2^2 -a_3^2). \label{newtondelta}
\eeqa

Substituting Eqs. (\ref{V2expand}) and (\ref{forcebalance}) into
Eq. (\ref{Eulereq}), we have
\begin{eqnarray}
  \rho_1 \sum_{j=1}^3 u_j {\pa u_i \over \pa x_j}&=&-{\pa P\over \pa
  x_i}+ \rho_1 {\pa \over \pa x_i}\biggl[V_1 - \delta \biggl( {\p V_2
  \over \p r} \biggr)_0 x_1 +{1 \over 2} \Omega^2 (x_1^2 +x_2^2) +{1
  \over 2} \biggl( {\p^2 V_2 \over \p r^2} \biggr)_0 x_1^2 \nonumber \\
  & &+{1 \over 2R} \biggl(
  {\p V_2 \over \p r} \biggr)_0 (x_2^2 +x_3^2) \biggr] +2 \rho_1 \Omega
  \sum_{l=1}^3\epsilon_{il3} u_l. \label{Eulersub}
\end{eqnarray}
Multiplying $x_j$ to Eq. (\ref{Eulersub}) and integrating over the
volume of star 1, we have
\beqa
  0&=& 2T_{ij} +W_{ij} +\biggl\{
  \O^2 +\biggl( {\p^2 V_2 \over \p r^2} \biggr)_0 \biggr\} \delta_{1i}
  I_{1j}
  + \biggl\{ \O^2 +{1 \over R} \biggl( {\p V_2 \over \p r} \biggr)_0
  \biggr\} \delta_{2i} I_{2j} \nonumber \\
  & &+{1 \over R} \biggl( {\p V_2 \over \p r}
  \biggr)_0 \delta_{3i} I_{3j}
  +2 \O \sum_{l=1}^3\epsilon_{il3} \int \r_1 u_l x_j d^3 x + \delta_{ij} \Pi,
  \label{stveq}
\eeqa
where
\beqa
  T_{ij} &\equiv& {1 \over 2} \int \r_1 u_i u_j d^3 x:~{\rm
    Kinetic~energy~tensor}, \\
  W_{ij} &\equiv& \int \r_1 {\p V_1 \over \p x_i} x_j d^3 x:~{\rm
    Potential~energy~tensor}, \\
  I_{ij} &\equiv& \int \r_1 x_i x_j d^3 x:~{\rm
    Moment~of~inertia~tensor}
\eeqa
and
\beqa
  \Pi \equiv \int P d^3 x.
\eeqa
In Eq. (\ref{stveq}) there appear no terms related to $\delta$. Since it is 
possible to take the coordinate system moving with a constant velocity, 
the term proportional to $\delta$ in Eq. (\ref{Eulersub}) can be vanished. 
Equation (\ref{stveq}) is the basic equation to construct 
equilibrium configurations of star 1. Note that in this 
model, the GR effects generated by star 2 are included in the 
orbital angular velocity and the tidal potential at star 1. 

Using Eqs. (\ref{u11}), (\ref{pressure}), and (\ref{intvelo})
for the self-gravity part of the potential $V_1$, 
the pressure $P$, and the internal velocity
$u_i$, respectively, we obtain three equations from Eq. (\ref{stveq});
\beqa
  -\Lambda^2 &=& {2 P_0 \over \r_1 a_1^2} -2 \pi G \r_1 A_1 + \biggl(
  {\p^2 V_2 \over \p r^2} \biggr)_0 +\O^2 -2 {a_2 \over a_1} \Lambda \O, 
  \nonumber \\
  -\Lambda^2 &=& {2 P_0 \over \r_1 a_2^2} -2 \pi G \r_1 A_2 + {1 \over
    R} \biggl( {\p V_2 \over \p r} \biggr)_0  +\O^2 -2 {a_1 \over a_2}
  \Lambda \O, \nonumber \\
  0 &=& {2 P_0 \over \r_1 a_3^2} -2\pi G \r_1 A_3 +{1 \over R} \biggl(
  {\p V_2 \over \p r} \biggr)_0. \label{stveqstveq}
\eeqa
If we take the Newtonian potential as an interaction potential,
Eq. (\ref{stveqstveq}) reduces to Eq. (\ref{tvtv}).
Combining these equations with the angular velocity equation
\beqa
  \O^2 = -{1+p \over R} \biggl( {\p V_2 \over \p r} \biggr)_0 (1 +
  \delta),
\eeqa
where $p\equiv m_1/m_2$, we can construct equilibrium sequences. 

%%%%%%%%%%%%%%%%%%%%%%%%%%%%%%%%%%%%%%%%%%%%%%%%
\subsection{Modified pseudo-Newtonian potential}
%%%%%%%%%%%%%%%%%%%%%%%%%%%%%%%%%%%%%%%%%%%%%%%%

Although there are a wide variety of choices of $V_2(r)$ to incorporate GR 
effects phenomenologically, we use the so called pseudo-Newtonian potential 
modifying the original form proposed by Paczy\'nsky and Wiita.\cite{PW}
This potential fits the effective potential for test particles 
orbiting Schwarzschild black hole quite well, 
as we will show later. We will use
the modified pseudo-Newtonian potential defined by
\beqa
  V_2(r) &=& {G m_2 \over r-r_{\rm pseudo}}, \\
  r_{\rm pseudo} &=& r_s \{1 +g(p) \}, \\
  g(p) &=& {7.49 p \over 6(1+p)^2} -{10.4 p^2 \over 3(1+p)^4} +{29.3 p^3 
  \over 6(1+p)^6}, \\
  r_s &\equiv& {2G M_{\rm tot} \over c^2}, \\
  M_{\rm tot} &=& m_1+m_2,
\eeqa
where $g(p)$ is a correction term to fit the ISCOs obtained from 
the hybrid 2PN EOM of Kidder, Will, and Wiseman\cite{KWW}. For
$p=0$, the modified pseudo-Newtonian potential agrees with the 
original pseudo-Newtonian potential proposed by Paczy\'nsky and Wiita.

Figure 11(a) shows effective potentials (solid lines) and locations of 
circular orbits (dots) in the modified pseudo-Newtonian potential 
($p=0$ and $r_{\rm pseudo} = r_s$) and in the Schwarzschild
metric. Although by this choice of the parameter ($r_{\rm pseudo}=r_s$), 
the locations of the ISCOs in the modified pseudo-Newtonian potential 
agree with those in the Schwarzschild metric, the angular momenta at the 
ISCO are different, that is, the angular momentum in the modified 
pseudo-Newtonian potential ($J_{\rm pseudo}$) for $p=0$ is $(9/8)^{1/2}$ 
times larger than that in the Schwarzschild metric ($J_{\rm Sch}$) at
the ISCO. Therefore in Figs.11(a) and (b) we compare circular orbits
with different angular momenta related as
\beqa
  J_{\rm pseudo} =\biggl( {9 \over 8} \biggr)^{1/2} J_{\rm Sch}.
\eeqa
 From Fig.11(b) we see that the radii of circular orbits of the modified 
pseudo-Newtonian potential agree with those of the effective 
potential around Schwarzschild black hole within $10\%$ accuracy near
the ISCO. This is the reason we believe that the 
modified pseudo-Newtonian potential 
defined here expresses the effect of general relativity
within $10\%$ accuracyor so.

In the pseudo-Newtonian case, the quadrupole term $\delta$ is written as
\beqa
  \delta= {3 \over 10} \biggl[ 2a_1^2 -{(3R-r_{\rm pseudo}) (R-r_{\rm
      pseudo}) \over 3R^2} (a_2^2+a_3^2) \biggr] {1 \over (R-r_{\rm
      pseudo})^2 }. \label{quadterm}
\eeqa
When we take the limit $r_{\rm pseudo} \ra 0$, Eq. (\ref{quadterm})
recovers the Newtonian potential case (Eq. (\ref{newtondelta})).

%%%%%%%%%%%%%%%%%%%%%%%%%%%%%%%%%%%%%%%%%%%%%%%%
\subsection{The energy and the angular momentum}
%%%%%%%%%%%%%%%%%%%%%%%%%%%%%%%%%%%%%%%%%%%%%%%%

In this subsection, we show the total energy and angular momentum. We
regard the minimum point of the total energy or the total angular
momentum as the location of the ISCO. 

The total energy is formally written as
\beqa
  E=T+W+W_i,
\eeqa
where $T,~W$ and $W_i$ denote the kinetic, the self-gravity and the
interaction energy, respectively. In the pseudo-Newtonian
potential approach, $T$ and $W$ have the same form as that in the
``pure'' Newtonian case, but the interaction energy has the 
different form as 
\beqa
  W_i =-{Gm_1 m_2 \over R-r_{\rm pseudo}} -{G m_2 \over 2(R-r_{\rm
        pseudo})^3} \biggl[ 2I_{11}- \biggl( {R-r_{\rm pseudo} \over R}
    \biggr) (I_{22}+I_{33}) \biggr]. \nonumber \\
\eeqa
Thus, the total energy is
\beqa
  E&=&{m_1 \over 10} \biggl[ \biggl( \O -{a_2 \over a_1} \Lambda
  \biggr)^2 a_1^2 + \biggl( \O -{a_1 \over a_2} \Lambda \biggr)^2 a_2^2
  \biggr] +{m_1 R^2 \over 2(1+p)} \O^2 -{2 \over 5} \pi G \r_1 m_1 A_0
  \nonumber \\
  & &-{Gm_1 m_2 \over R-r_{\rm pseudo}} -{G m_2 \over 2(R-r_{\rm
      pseudo})^3} \biggl[ 2I_{11}- \biggl( {R-r_{\rm pseudo} \over R}
  \biggr) (I_{22}+I_{33}) \biggr], \nonumber \\
\eeqa
where the angular velocity is written as
\beqa
  \O^2 &=&{G (m_1+m_2) \over R(R-r_{\rm pseudo})^2} \biggl[ 1+ {3 \over
    10(R-r_{\rm pseudo})^2} \biggl\{ 2a_1^2 \nonumber \\
  & &-{(3R-r_{\rm
      pseudo})(R-r_{\rm pseudo}) \over 3R^2} (a_2^2+a_3^2) \biggr\}
  \biggr].
\eeqa

The form of the total angular momentum in the 
pseudo-Newtonian potential approach is the same as that in the pure Newtonian
case as 
\beqa
  J={m_1 R^2 \over 1+p} \O +{m_1 \over 5} (a_1^2+a_2^2) \O -{2 \over 5}
  m_1 a_1 a_2 \Lambda.
\eeqa
Note that the pseudo-Newtonian effect is included implicitly 
in the angular velocity.

%%%%%%%%%%%%%%%%%%%%
\subsection{Results}
%%%%%%%%%%%%%%%%%%%%

In previous subsections, 
we described equations needed to get equilibrium
sequences and to determine the location of the ISCO. Here, 
we show the results for the equal mass binary ($p=1$) in the 
corotating case ($f_R=0$) and the irrotational case ($f_R=-2$).

In Figs.12 and 13, we show $\tilde{E}$(a) and $\tilde{J}$(b) as
functions of $\tilde{\O}$ in the corotating and irrotational cases,
respectively.  Note that the minima for the energy and angular momentum
denote the ISCCO for $f_R=0$ case, and the ISCO for $f_R=-2$ case.  In
Fig.14, we show $\bar{\O}\equiv\O(Gm_1/c^3)$ at the ISCO and/or ISCCO as
a function of the compactness parameter of star 1 $Gm_1/a_0 c^2$
for $p=1$. Note that this normalization is different from that in Fig.7
and so on. In this normalization, the orbital angular velocity of the
Newtonian order (the value at $Gm_1/a_0 c^2=0$) becomes zero, and it is
not convenient to compare $\bar{\Omega}$ with the Newtonian angular
velocity. However, the approach of TN is based on the point particle
binary case, i.e., the case of $Gm_1/a_0 c^2 \ra \infty$. Then, it is
natural to choose the normalization $\bar{\Omega}$ in this section.

 {}From these figures, we find that due to the finite size effect of the
star, $\bar{\O}$ at the ISCO (ISCCO) is always smaller than the point
particle limit ($\bar{\O}$=0.0402) i.e. the finite size effect of the
star always destabilize the binary.  However as a function of the
relativity parameter, $\bar{\O}$ at the ISCO (ISCCO) increases which is
qualitatively the same result as in 1PN treatment. In this sense there
is no disagreement between 1PN and the result in this section
apparently.

%%%%%%%%%%%%%%%%%%%%%%%%%%%%%%%%%%%%%%%%%%%%%%%%%%%%%%%%%%%%%
\subsection{1PN calculations without 1PN self-gravity effect} 
%%%%%%%%%%%%%%%%%%%%%%%%%%%%%%%%%%%%%%%%%%%%%%%%%%%%%%%%%%%%%

One of the unsatisfactory points in TN is that the 
PN effect of the self-gravity
is not included. While in 1PN calculations in \S 3 and \S4, 1PN effect
of the self-gravity as well as the gravity between two stars are
consistently taken into account. In \S 4 we suggested that 1PN effect of
the self-gravity will have stabilizing effect. To expect PN effect of
the self-gravity in TN in this subsection we will calculate $\bar{\O}$ at
the ISCCO removing the 1PN self-gravity terms in the calculations of \S
3.

%%%%%%%%%%%%%%%%%%%%%%%%%
\subsubsection{Equations}
%%%%%%%%%%%%%%%%%%%%%%%%%

As mentioned above, we only incorporate the PN effect 
only to the gravity acting between two stars in this approximation. 
So that equations we use are quite similar to those in \S 2.1 except for the 
angular velocity which we use
\begin{eqnarray}
  \Omega^2 ={2GM \over R^3} \biggl[ 1+ {3 \over 5R^2} (2a_1^2 -a_2^2
  -a_3^2) \biggr] -{9 G^2 M^2 \over 2 R^4 c^2}. 
  \label{2PNpointomega}
\end{eqnarray}
Here the PN effect appears in the last term. 
(Note that for $a_1, a_2, a_3 \rightarrow 0$, 
the angular velocity agrees with that for two point masses.)
As in \S 2.1, we use Eqs. (\ref{tvtv}) 
to determine $\alpha_2=a_2/a_1$ and $\alpha_3=a_3/a_1$. 

In calculating equilibrium sequences, we must fix the conserved 
mass $M_*$ (see Eq. (\ref{conmas})).
In this section, the conserved mass should be written as
\begin{eqnarray}
  M_{\ast} = M \biggl( 1+ {13GM \over 4R c^2} \biggr),
\end{eqnarray}
where we neglect the stellar structure terms in the 1PN order.

As was done in previous sections, 
we regard the minimum of the energy as 
the location of the ISCO. In this approximation, the energy is written as
\begin{eqnarray}
  E&=& -{3G M^2 \over 5 a_0^3} A_0 -{GM^2 \over 2R} +{GM^2 \over 10 R^3}
  (2a_1^2 -a_2^2 -a_3^2) \nonumber \\
  & &+{M \over 5} \Omega_{\rm N}^2 \Biggl[ (a_1^2
  +a_2^2) \bigg\{ 1+ \biggl( {a_1 a_2 f_R \over a_1^2+a_2^2} \biggr)^2
  \biggr\} +{4 a_1^2 a_2^2 \over a_1^2+a_2^2} f_R \Biggr] \nonumber \\
  & &-{25 G^2 M^3 \over 16 R^2 c^2},
\end{eqnarray}
where $a_0$ denotes the radius of the spherical star in the Newtonian
order.

%%%%%%%%%%%%%%%%%%%%%%%
\subsubsection{Results}
%%%%%%%%%%%%%%%%%%%%%%%

In the following calculation, we normalize the angular velocity and the
energy by $(GM_*/a_*^3)^{1/2}$ and $GM_*^2/a_*$, respectively, where
$a_*$ represents the radius of the PN spherical star defined by the
conserved mass as
\beq
  a_* \equiv \biggl( {3 M_* \over 4 \pi \rho} \biggr)^{1/3} 
  = a_0 \biggl( 1+ {13 GM_* \over 12 R c^2} \biggr).
\eeq
The normalized angular velocity and energy are written as
\beqa
  \tilde{\O}^2 &=&{2 \a_2 \a_3 \over \tilde{R}^3} \biggl[ 1+ {3 \over 5
    \tilde{R}^2} (2-\a_2^2 -\a_3^2) \biggr] -{9 (\a_2\a_3)^{4/3} \over 2
    \tilde{R}^4} C_{\rm s}, \label{1PNorbitvelocity} \\
  \tilde{E} &=& (\a_2 \a_3)^{1/3} \biggl[ -{3 \over 5} {\tilde{A}_0
    \over \a_2 \a_3} -{1 \over 2\tilde{R}} +{1 \over 10 \tilde{R}^3} (2
  -\a_2^2 -\a_3^2) \nonumber \\
  & &+{\tilde{\O}_{\rm N}^2 \over 5 \a_2 \a_3} \biggl\{
  (1+ \a_2^2) \bigl\{1+ \bigl( {\a_2 f_R \over 1+\a_2^2} \bigr)^2
  \bigr\} + {4 \a_2^2 \over 1+\a_2^2} f_R \biggr\} \biggr] \nonumber \\
  & &+ {55(\a_2 \a_3)^{2/3} \over 48 \tilde{R}^2} C_{\rm s},
  \label{1PNorbitenergy}
\eeqa
where $\tilde{R}=R/a_1$ and $C_{\rm s}\equiv G M_{\ast}/c^2 a_{\ast}$.

Figures 15(a) and (b) show the energy as a function of the angular velocity
in the corotating case ($f_R=0$) and the irrotational one ($f_R=-2$),
respectively. The solid, dotted, short dashed, and long dashed lines
represent the cases for $C_{\rm s}=0$, 0.01, 0.03 and 0.05,
respectively.  Note that for $f_R=-2$, the minima correspond to the ISCO
and for $f_R=0$, the minima correspond to the ISCCO.  In Fig.16, we show
$\bar{\O}$ at the ISCCO as a function of $C_{\rm s}$ with (the dotted
line) or without (the solid line) 1PN effect of the self-gravity, where
we use the normalization $\bar{\Omega}$ to present Fig.16 in order to
compare it with Fig.14.  As expected the dotted line is above the solid
line. Therefore from these calculations we can say that the 1PN effect
of the self-gravity will increase $\bar{\O}$ and stabilize the binary.
Note here that the Fig.16 corresponds to the lower left corner of
Fig.14. We checked the numerical value of $\bar{\O}$ in the solid lines
of Fig.14 and Fig.16 for the same relativity parameter and found the
almost similar value. This is not strange.
For small relativity parameter both treatments essentially agree
since the pseudo-Newtonian
potential is almost the same as the Newtonian potential. Therefore this
result suggests that at least for small relativity parameter if we
include 1PN effect of the self-gravity to TN, $\bar{\O}$ at ISCO (ISCCO)
will increase. What will happen for large relativity parameter is not
certain but we expect that the tendency will not change.

%%%%%%%%%%%%%%%%%%%%%%%%%%%%%%%%%%%%%%%%%%%%%%%%%%%%%%%%
\section{Brief review on recent semi-relativistic works}
%%%%%%%%%%%%%%%%%%%%%%%%%%%%%%%%%%%%%%%%%%%%%%%%%%%%%%%%

So far, we have investigated the location of the ISCO in the PN
approximation. Although we have obtained qualitative properties of the
GR correction for the location of the ISCO, the PN study is not
sufficient to know the location accurately because it is 
accurate for small $C_{\rm s}$, but NS is a highly GR star of $C_{\rm s}
\sim 0.2$. Thus, we need calculation including full GR terms or
sufficient GR corrections. In the fully GR case, there is no equilibrium
state of BNSs because of emission of GWs.  Even so, we will be able to
consider quasi-equilibrium states because $t_{\rm GW}$ is still much
longer than the orbital period for orbits outside the ISCO.  
However, we have not known yet how to
define the ``quasi-equilibrium'' in the fully GR case in contrast to the
1PN and 2PN cases in which we can distinguish quantities related to GWs
from others.  Thus, for fully GR calculation, we must begin from the
unresolved first step, where we need to clarify the notion of the
``quasi-equilibrium'' and establish the formulation along the notion.

Instead of fully GR one, two groups have been performing semi-GR
calculation.\cite{wilson}\cite{cornell} The essence of their method is
to assume that the three metric has the conformal flat form
approximately.
In this case, all the geometric variables which appear in the
formalism are determined by solving Poisson equations, and wave
equations do not appear. Thus, effect of GWs is neglected and the
formalism reduces to that similar to the PN one.  Although some fully GR
effects are taken into account in this formalism, the meaning of this
approximation is still not sure. For example, if the orbital separation
of BNSs is sufficiently large (i.e., tidal effects are not important)
and spin angular velocity of each star is negligible, this formalism
yields the exact GR solution. On the other hand, from the PN point of
view, the formalism includes all the 1PN terms consistently, but not the
2PN terms when close BNSs are concerned. Thus, we may regard it as a
formalism in which some GR corrections are merely included in the 1PN
approximation.

For a small compactness $C_{\rm s}$, this formalism will yield the same
answer as that obtained by the 1PN calculation. Actually, this is the
case for numerical results by Baumgarte et al. (BCSST):\cite{cornell}
They obtained equilibrium states of corotating BNSs and their numerical
results agree with those obtained in \S 4 for small $C_{\rm s}$ cases,
and even for large $C_{\rm s}$, they are similar to ours. From this
fact, we may consider that this formalism is essentially very similar to
the 1PN one. On the other hand, Wilson et al. did not obtain equilibrium
states of corotating binary, but of non-uniform internal motion.  Since
the velocity field is different from the uniformly rotating one, the
results may quantitatively disagree with ours and BCSST. However,
qualitative nature also does not at all agree with ours and BCSST; in
the analysis by Wilson et al., (1)$\Omega_{\rm ISCO}$ decreases with
increase of the GR effect and (2) the maximum density increases with
decrease of the orbital separation, while the opposite relation holds in
the analysis of ours and BCSST. There have been several researchers 
who doubt that accuracy of their calculation is not enough and
their results are something
incorrect.\cite{eanna}\cite{cornell}\cite{tanib} However, we cannot
completely deny a possibility that we miss some GR effects we have not
been aware of. We may not reach a certain conclusion even for the case
of this simplest version of semi-GR calculation at 
present. Hence, to clarify the
meaning of this approximation and accuracy of their numerical results,
many works should be done in this area in near future.

\section{Summary}

In this manuscript, we have presented recent results obtained from the 
study on the location of the ISCO in several levels of the PN 
approximation. The following is the brief summary. 

\begin{itemize}
\item Even in the Newtonian case, there exists the ISCO for 
binary of sufficiently stiff EOS. If the mass and the radius of each 
star are fixed, $\Omega_{\rm ISCO}$ is larger for softer EOS. 
\item There exist roughly two kinds of the 1PN corrections. One is the 
correction to the self-gravity of each star of binary, and the other is the 
correction to the gravity acting between two stars. The former one tends to 
increase $\Omega_{\rm ISCO}$, but the latter one tends to 
decrease it. If we take into account both effects, however, 
the former effect is stronger than the latter one, and $\Omega_{\rm ISCO}$ 
becomes large with increase of the 1PN correction. 
\item The feature mentioned above is more remarkable 
for softer EOS when we fix the mass and radius. This is because 
for softer EOS, each star has the 
larger central density and is susceptible to the GR correction. 
\item There has been no self consistent calculation including all the 2PN 
effects. But, there exist studies in which 
one only considers the 2PN gravity acting between two stars. In this case, 
$\Omega_{\rm ISCO}$ is always smaller than that for the Newtonian case. 
If we include the PN effect of the self-gravity, 
$\Omega_{\rm ISCO}$ will increase.

\end{itemize}

\vskip 5mm
\begin{center}
{\bf Acknowledgment}
\end{center}
\vskip 5mm

Numerical computations were mainly performed on FACOM VX4 in
data processing center of National Astronomical Observatory in Japan.
This work was in part supported by a Grant-in-Aid for Science Research
from the Ministry of Education, Culture,
Science and Sports (08NP0801, 08237210 and 09740336).

\vskip 5mm

\begin{center}
  {\large FIGURE CAPTIONS}
\end{center}

\vspace{0.5cm}

\begin{itemize}
\item Fig.1: Location of the ISCO is shown in each level of 
approximation, schematically. 
\item Fig.2: Sketch of corotating binary. The origin of the coordinate 
is located at the center of mass of star 1.
\item Fig.3: The energy(a) and the angular momentum(b) 
as functions of the angular 
velocity for Newtonian incompressible binary stars of 
$C/(GMa_0)^{1/2}=-2$, $-1.5$, $-1$, $-0.5$, 0, 0.5, 1, 1.5, 2 and 
corotating binary. 
\item Fig.4: The energy and the angular momentum as functions of the angular
velocity for Newtonian 
corotating binary of compressible EOS. (a) for $n=0.25$, 
(b) for $n=0.5$, (c) for $n=0.75$ and (d) for $n=1$. 
\item Fig.5: $\Omega_{\rm N:ISCCO}/\Omega_0$ 
as a function of the polytropic index $n$. Solid line denotes 
$0.248(1+0.2n^{1.5})$. Note that $\Omega_{\rm N:ISCCO}$ was determined 
with $\sim \pm 1\%$ uncertainty. 
\item Fig.6: The total energy(a) and angular momentum(b)
of the equilibrium sequence as functions of
$\Omega/\Omega_*$ for 1PN incompressible binary stars.
Solid, dotted, dashed and long dashed lines denote
$C_{\rm s}=GM_{\ast}/c^2a_{\ast}=0$ (the Newtonian case), 
0.01, 0.03 and 0.05, respectively.
\item Fig.7: The orbital angular velocity at the ISCCO as a function of the
compactness parameter $C_{\rm s} \equiv GM_{\ast}/c^2a_{\ast}$ for 
1PN incompressible binary stars. 
\item Fig.8: The relation between $R_*/a_*$ and 
$\Omega/\Omega_*$ for $n=0.5$
and $C_{\rm s}=0$ (open circles), 0.0327 (filled circles) and
0.0654 (filled squares). Dotted lines are drawn using Eq. (\ref{analy}) as 
counter parts of numerical results.
\item Fig.9: The energy and angular momentum as functions of
$\Omega/\Omega_*$ for $n=0.5$(a), 0.75(b) and 1(c) for the 1PN case. 
In Fig.9(a), open circles, filled circles and open squares denotes
sequences for $C_{\rm s}=0$, 0.0327 and 
0.0654, in Fig.9(b), they denote
sequences for $C_{\rm s}=0$, 0.02 and 0.04, and in Fig.9(c), 
they denote sequences for $C_{\rm s}=0$, 0.0167 and 0.0333, respectively.
\item Fig.10: The angular velocity at the ISCCO as a function of $C_{\rm s}$ 
for $n=0.5$(a), 0.75(b) and 1(c). Note that $\Omega_{\rm ISCCO}$ 
was determined with $\sim \pm 1\%$ uncertainty.
%%%%%%%%%%%%% taniguchi part %%%%%%%%%%%%
\item Fig.11: (a)The effective potentials of a test particle in the
  Schwarzschild metric (left) and the pseudo-Newtonian potential (right) 
  as a function of the normalized distance $R/r_s$. The vertical axes
  denote \\
  $\Psi_{\rm Sch}=\sqrt{(1-(2GM/Rc^2))((G^2 J_{\rm Sch}^2/R^2
  c^6)+1)}-1$ and $\Psi_{\rm pseudo}=-(GM/(R-r_s)c^2)+(G^2 J_{\rm
  pseudo}^2/2 R^2 c^6)$, respectively. The dots express the place of
  circular orbits. The value of the effective potential of the
  Schwarzschild black hole at the infinity is shifted to zero. (b)The
  fractional deviation of circular orbits of the pseudo-Newtonian
  potential from those of the Schwarzschild effective potential as a
  function of of the angular momentum. Circular orbits with different
  angular momentum related as $J_{\rm pseudo}=(9/8)^{1/2} J_{\rm Sch}$
  are compared (see text).
\item Fig.12: The total energy(a) and angular momentum(b) of the
  equilibrium sequences as functions of the angular velocity
  in the case of corotating binaries. Solid, dotted, short dashed,
  long dashed, short dash-dotted, and long dash-dotted lines
  denote $Gm_1/a_0 c^2=0.05$, 0.1, 0.15, 0.2 and 0.25, respectively.
\item Fig.13: The total energy(a) and angular momentum(b) for 
  irrotational binaries  as functions of the angular velocity. 
  The conventions are the same as in  Fig.12.
\item Fig.14: The relation between the compactness of star 1
  $(Gm_1/a_0 c^2)$ and the angular velocity of the binary $\bar{\O}=
  \O(Gm_1/c^3)$ at the ISCO. Solid and dashed lines denote the cases of
  the corotating (Roche Ellipsoids) and irrotational
  binaries (Irrotational Roche-Riemann Ellipsoids), respectively. The
  value written at the upper right corner is that of the angular
  velocity in the point particle binary case.
\item Fig.15: The 1PN total energy without self-gravity terms 
as a function of the angular velocity 
  in the corotating case(a) and the irrotational one(b),
  respectively. Solid, dotted, short dashed and long dashed lines 
  represent  the cases for $C_{\rm s}=0, 0.01, 0.03$ and 0.05, respectively.
\item Fig.16: The relation between 
  $C_{\rm s}$ and $\bar{\O}=\O(GM_{\ast}/c^3)$ at the ISCCO. 
  Solid and dotted lines denote the cases of the 1PN corotating binaries
  without and with the 1PN self-gravity, respectively.
\end{itemize}
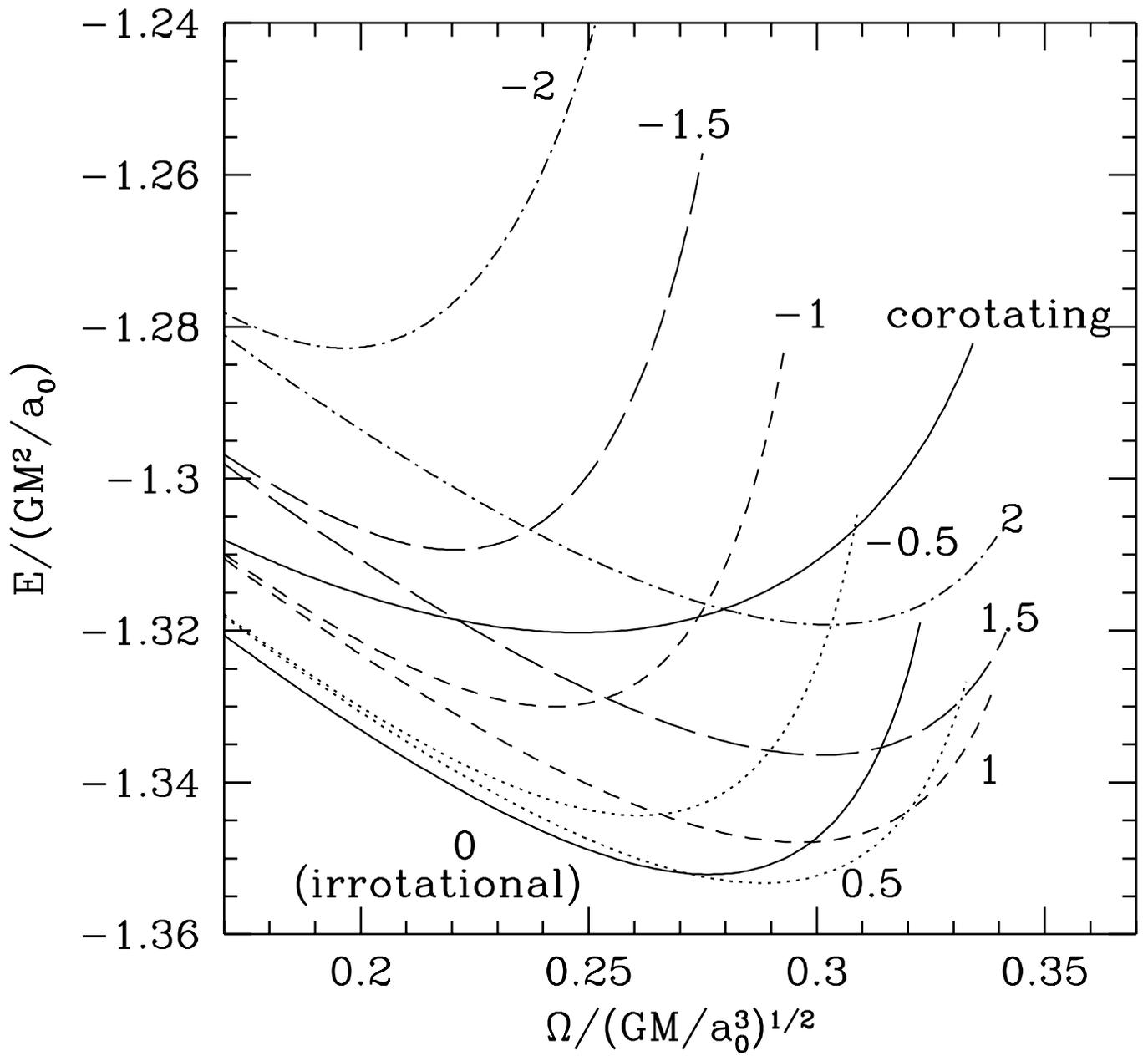
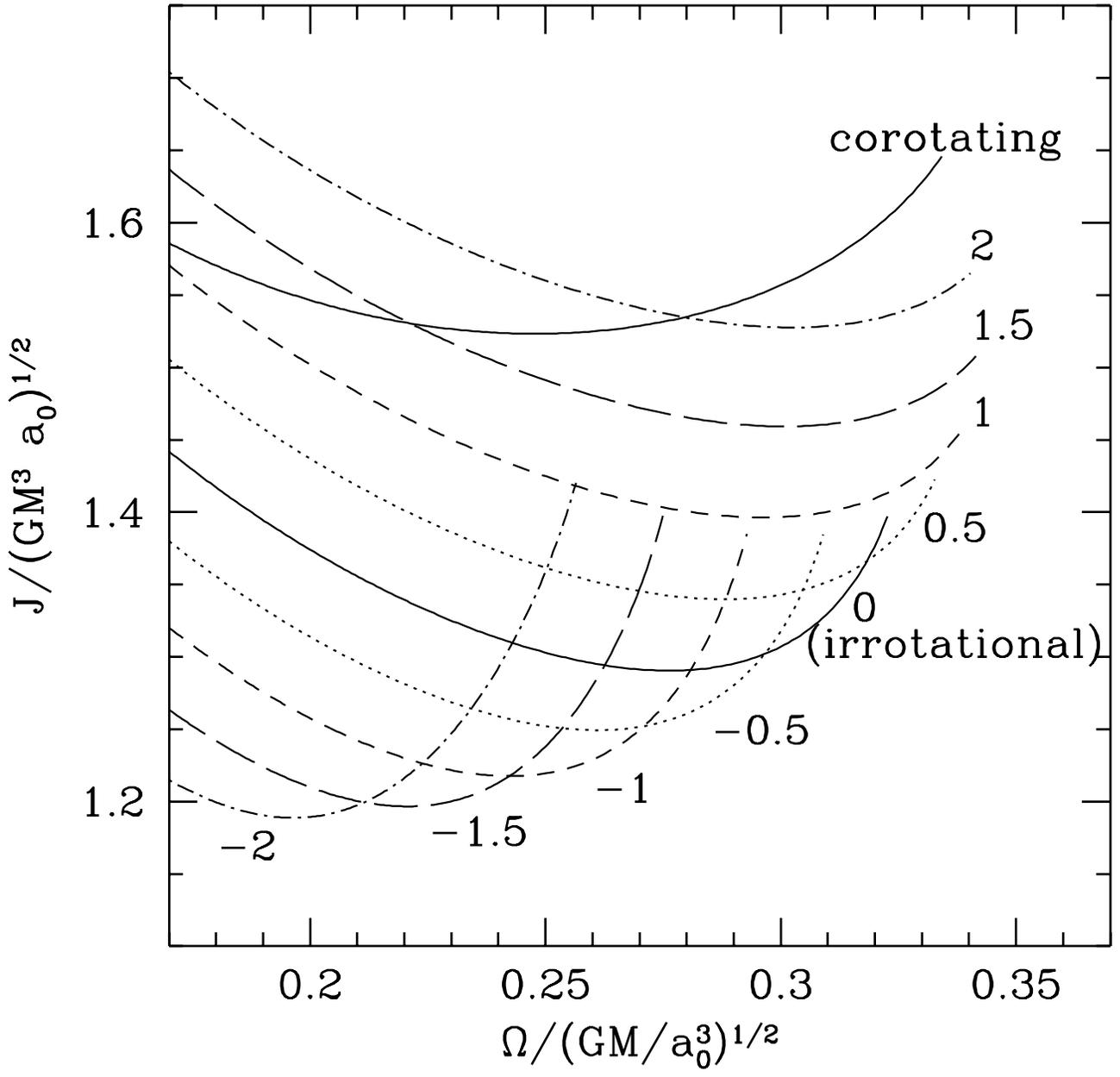
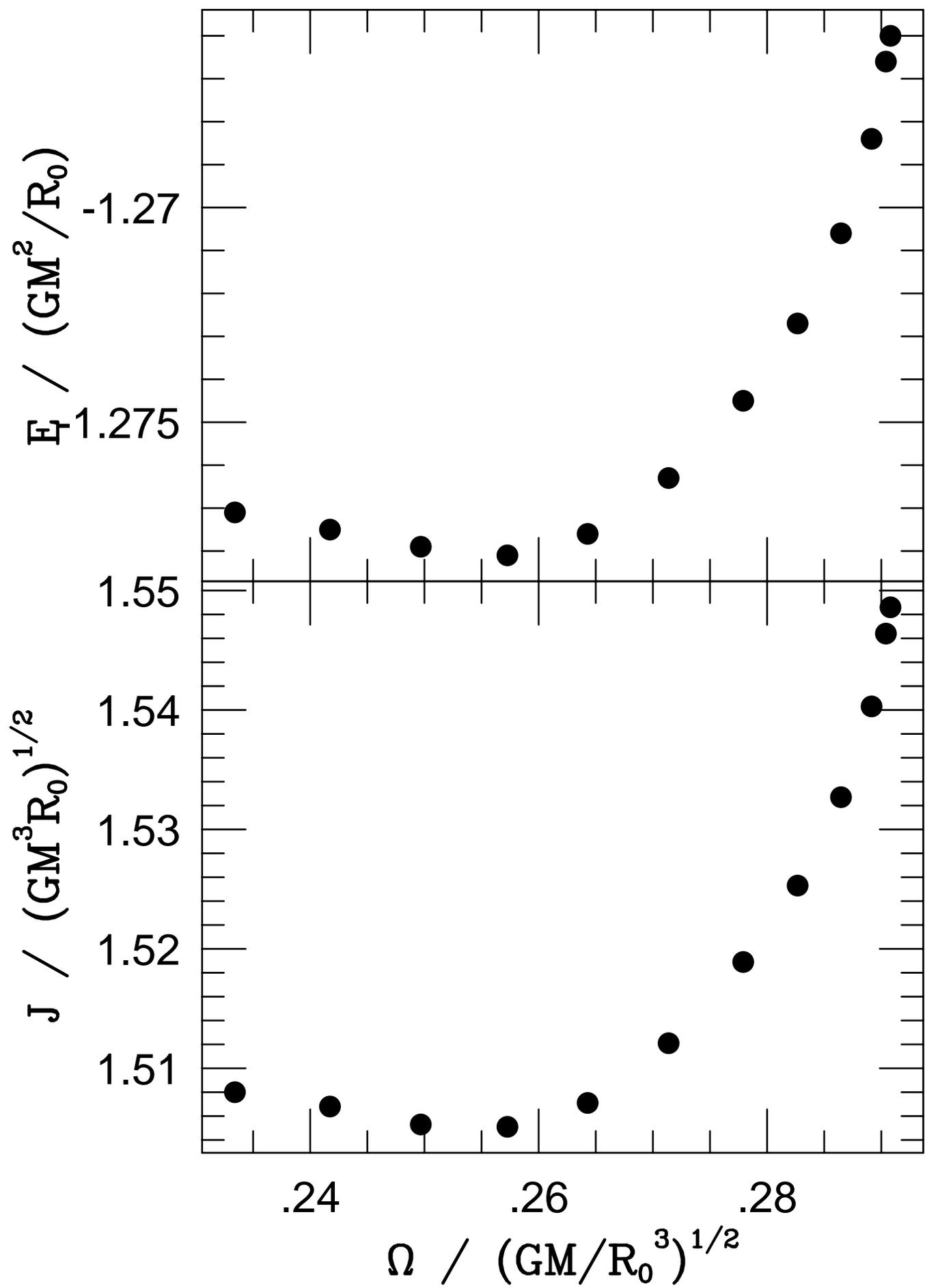
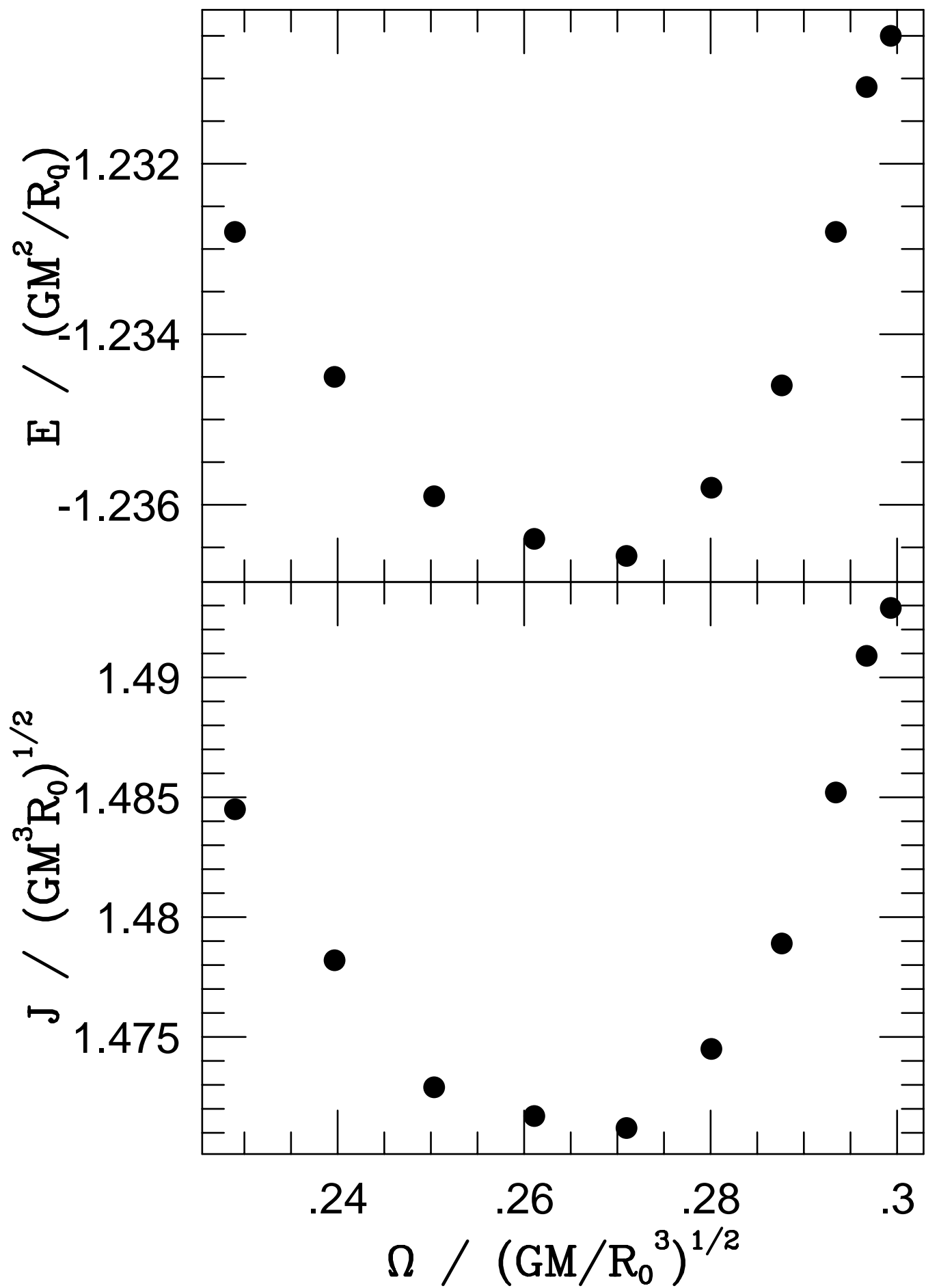
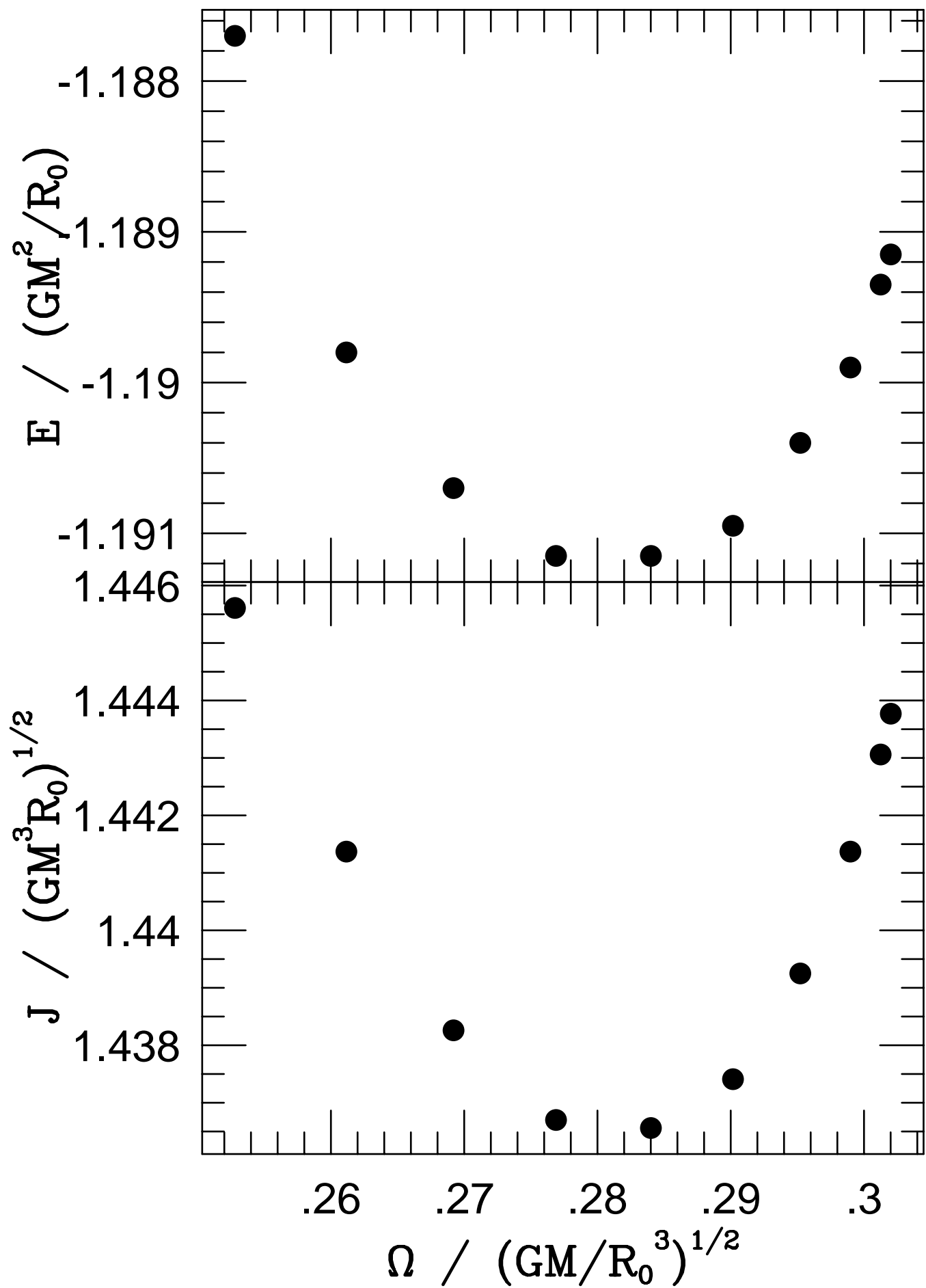
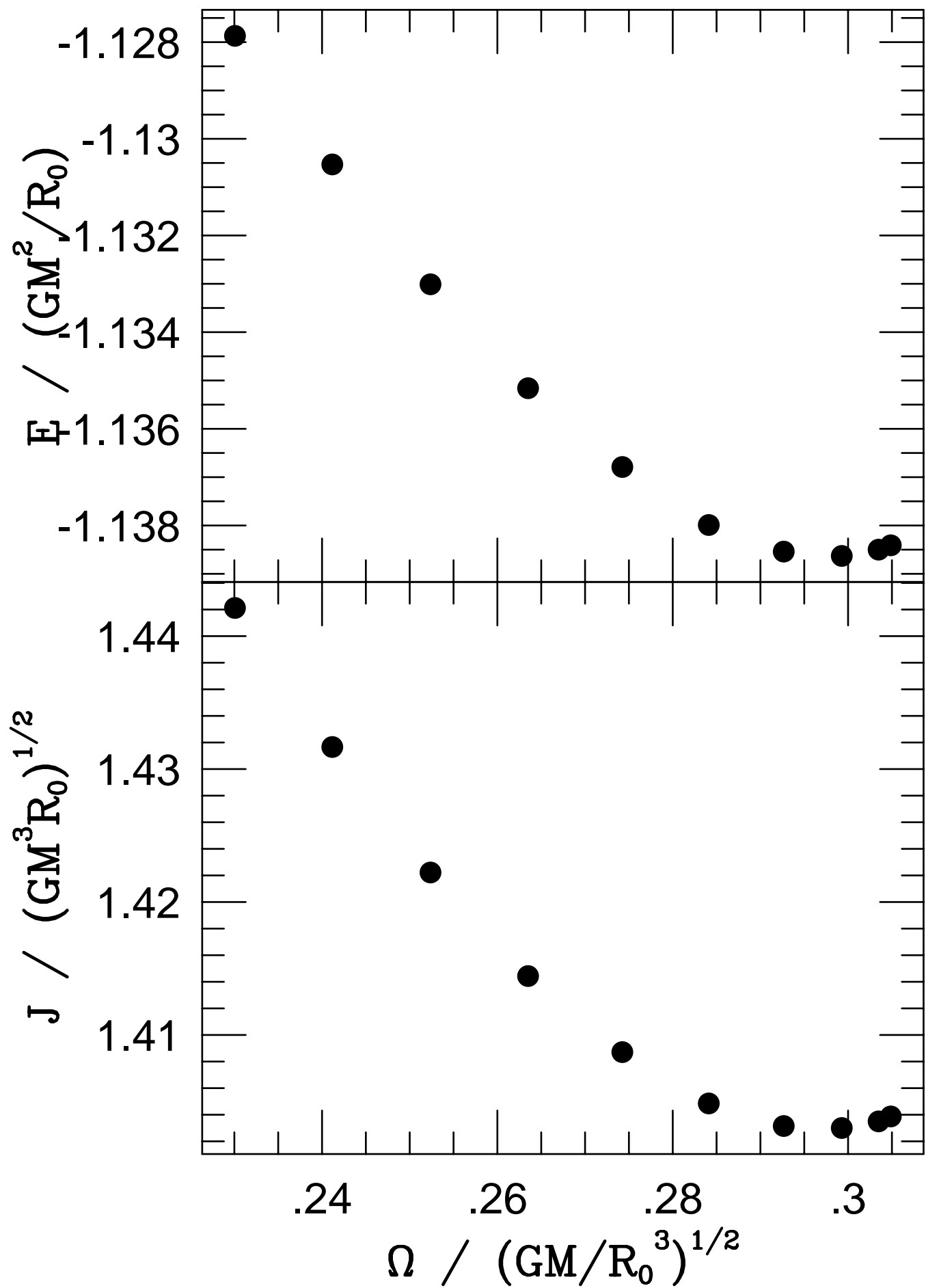
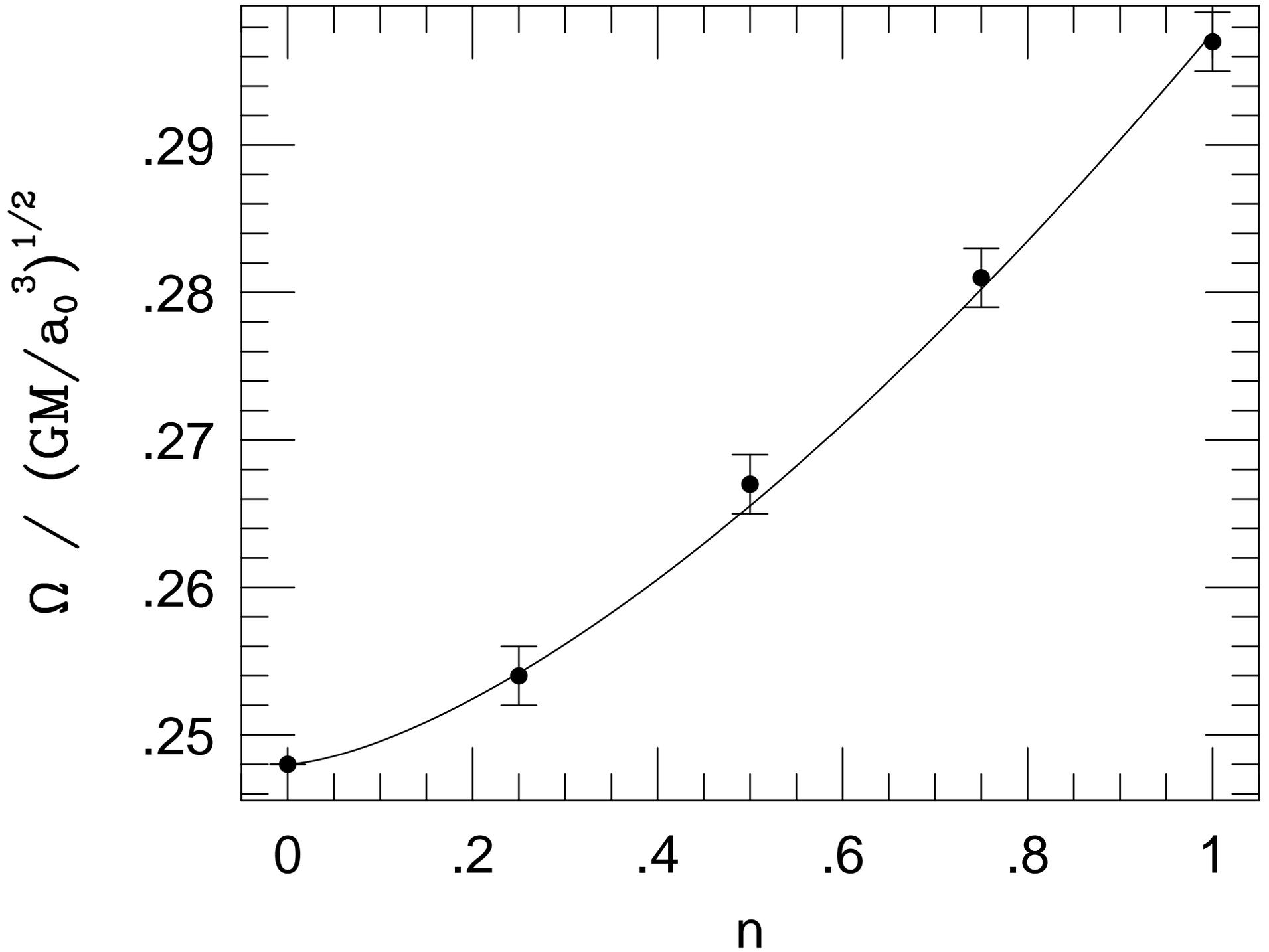
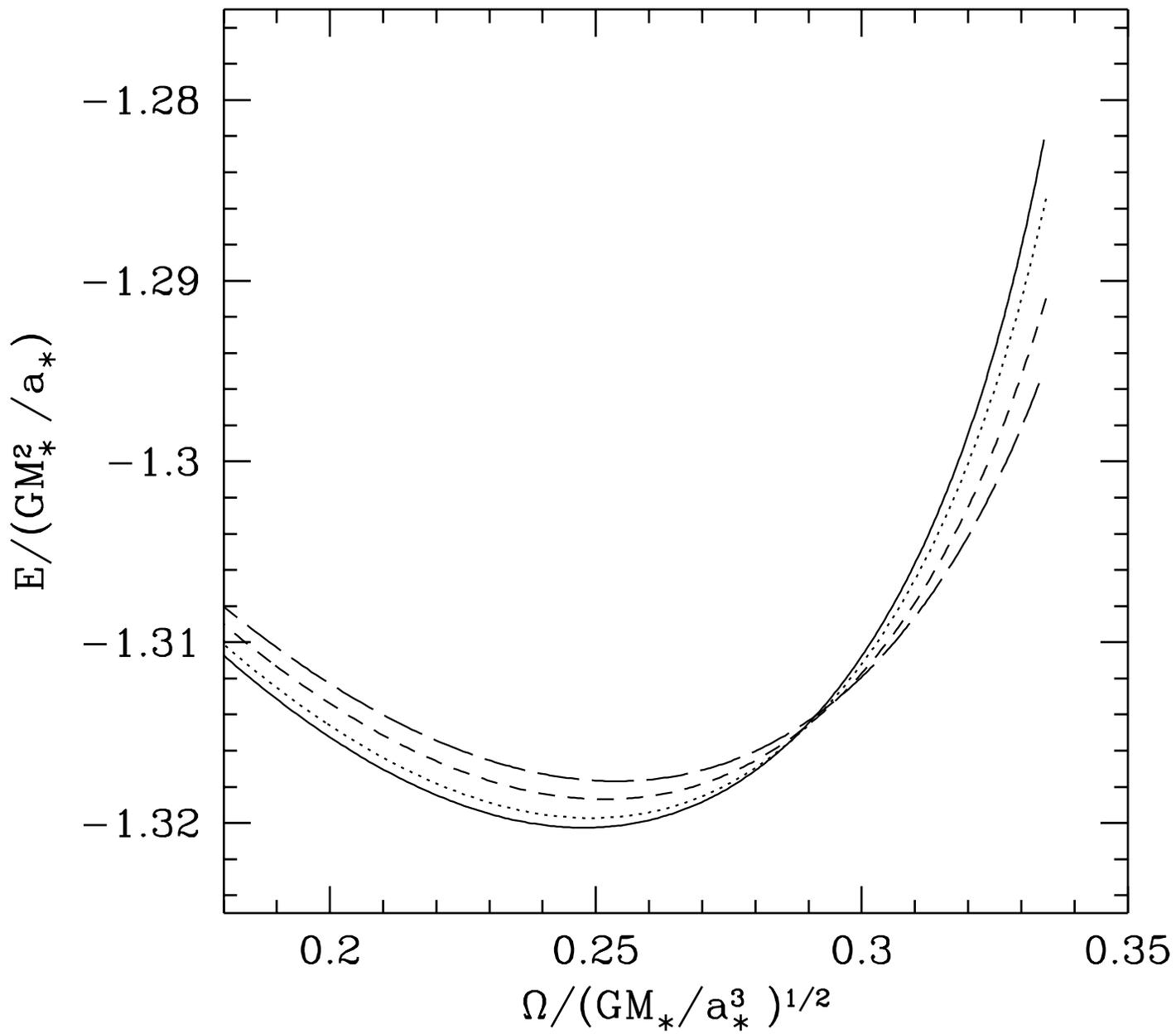
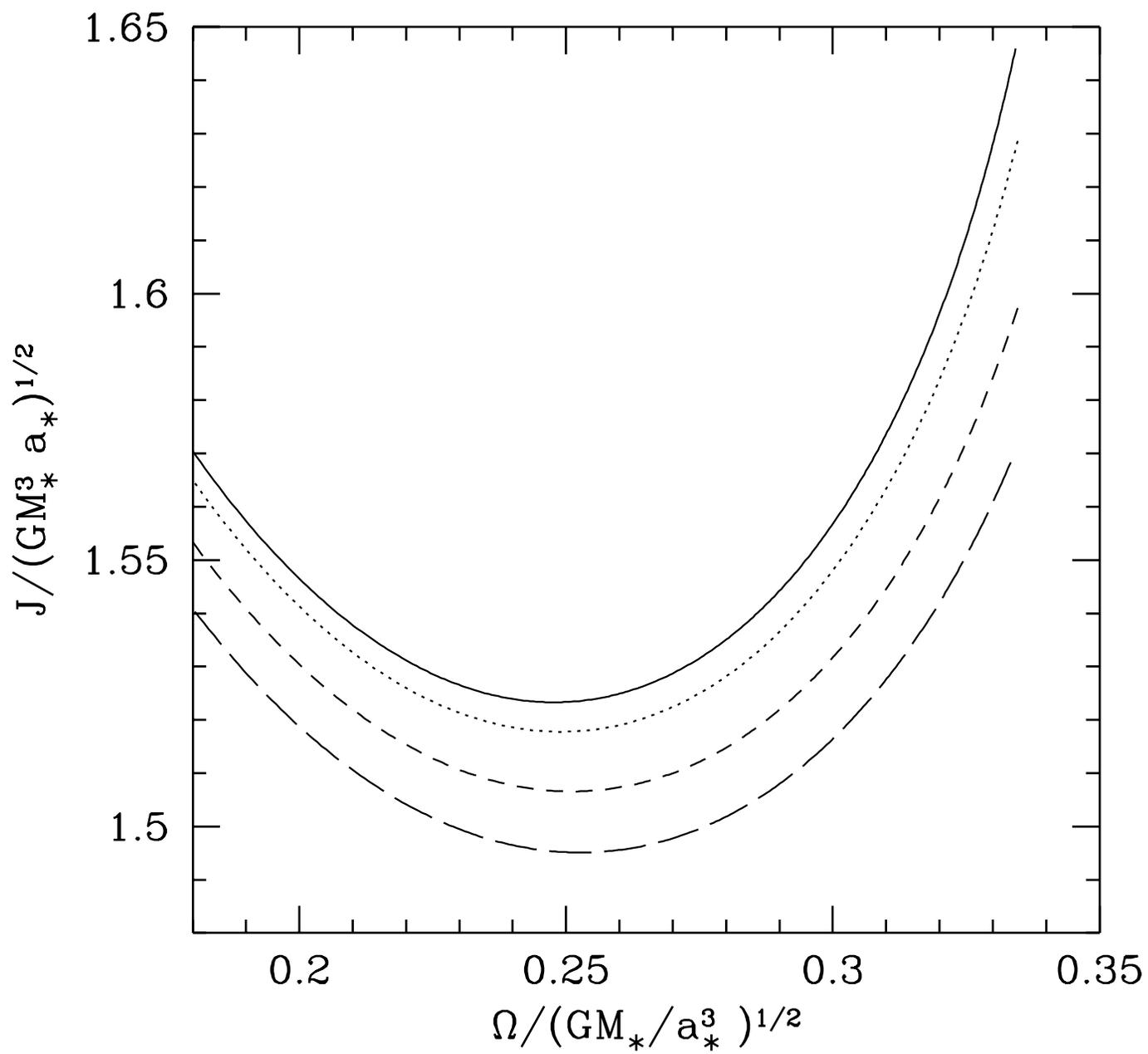
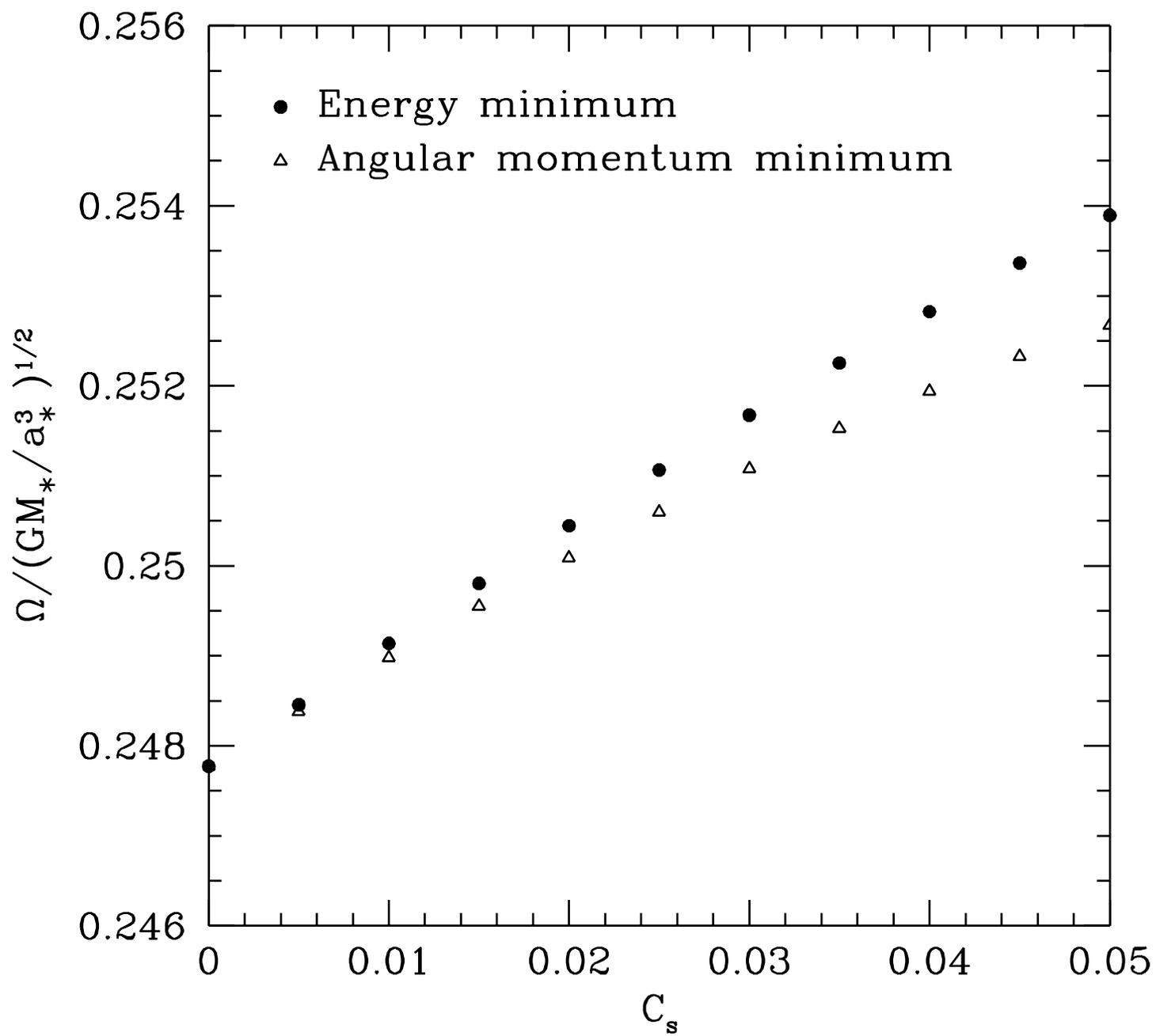
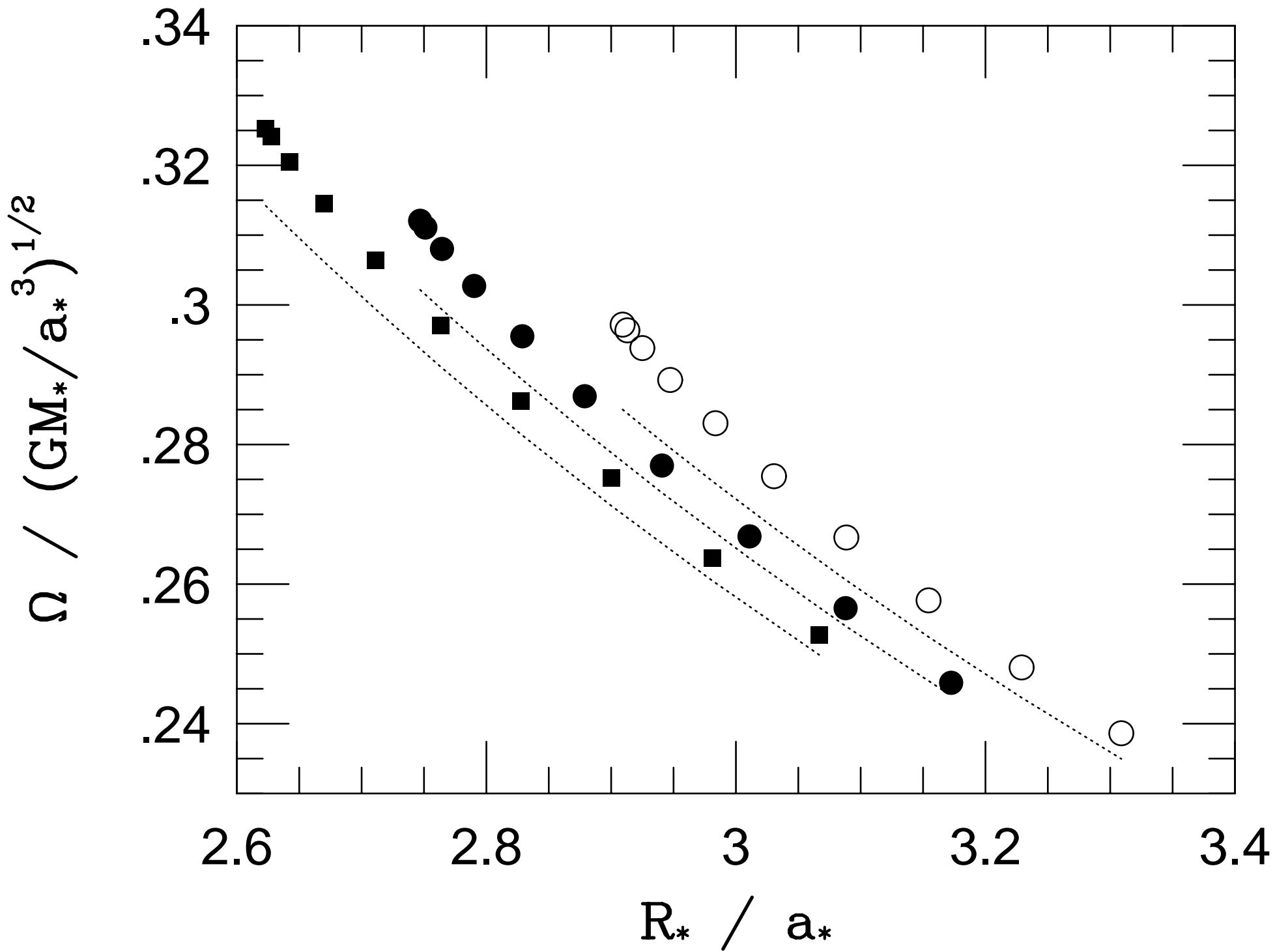
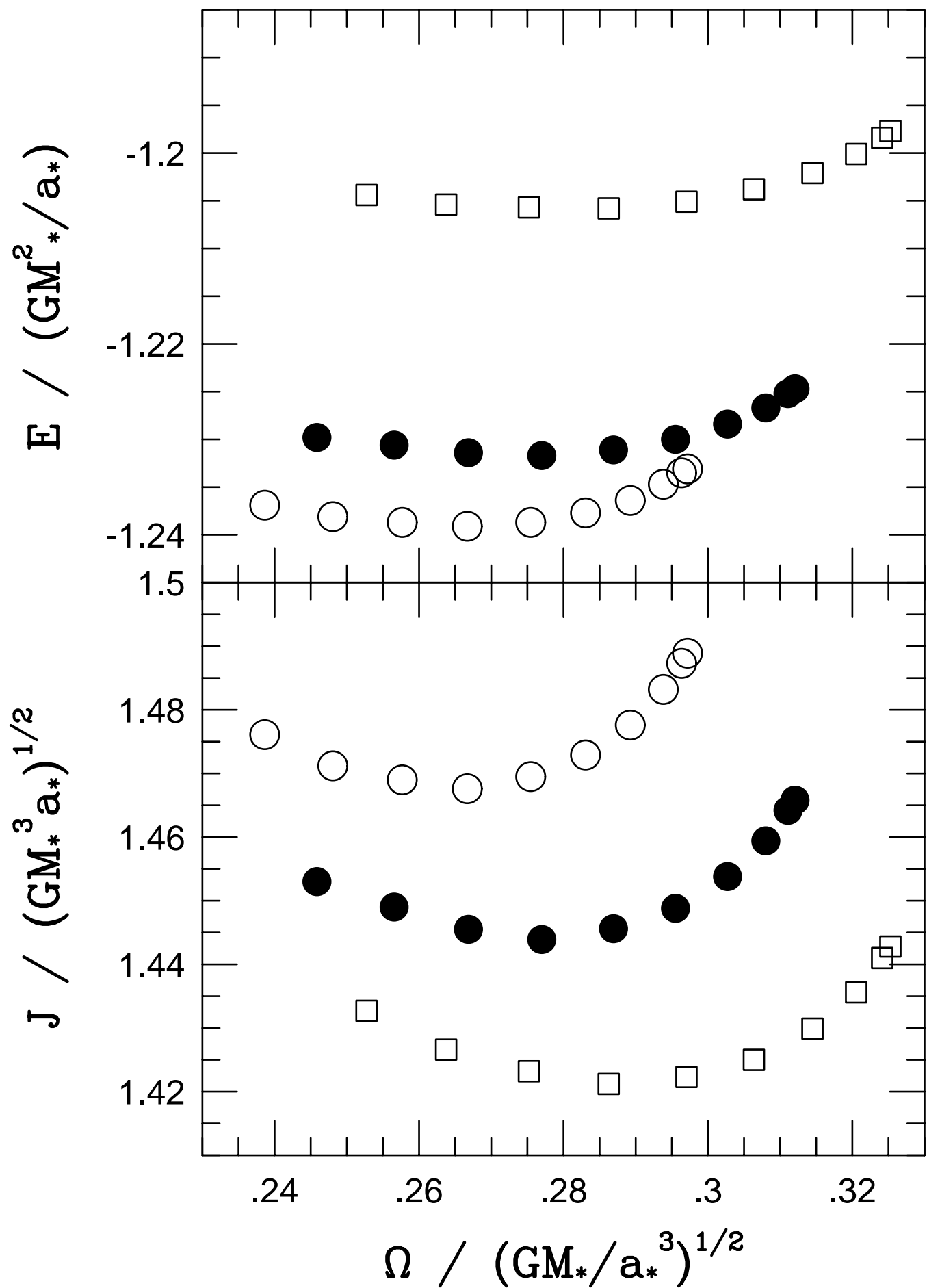
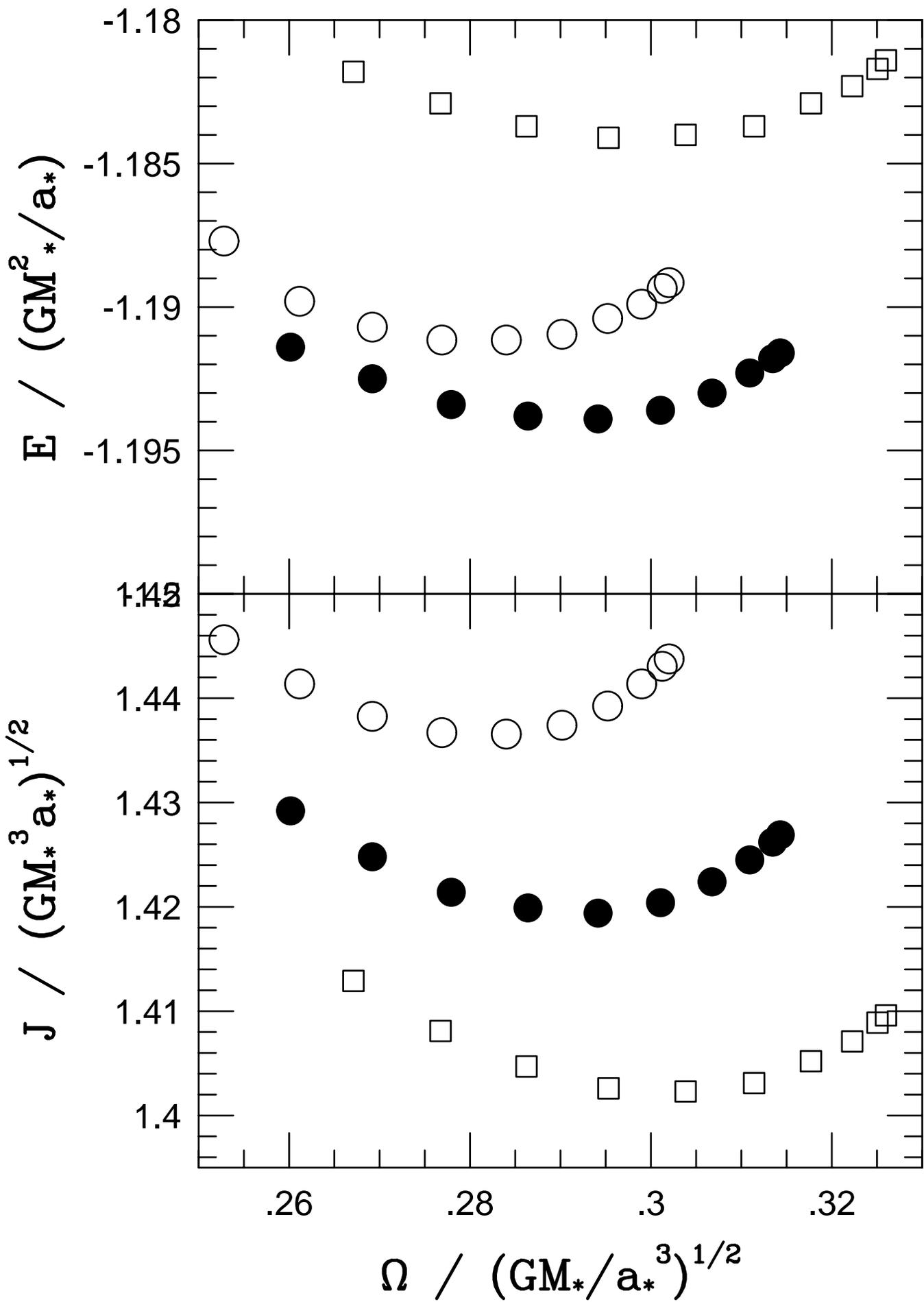
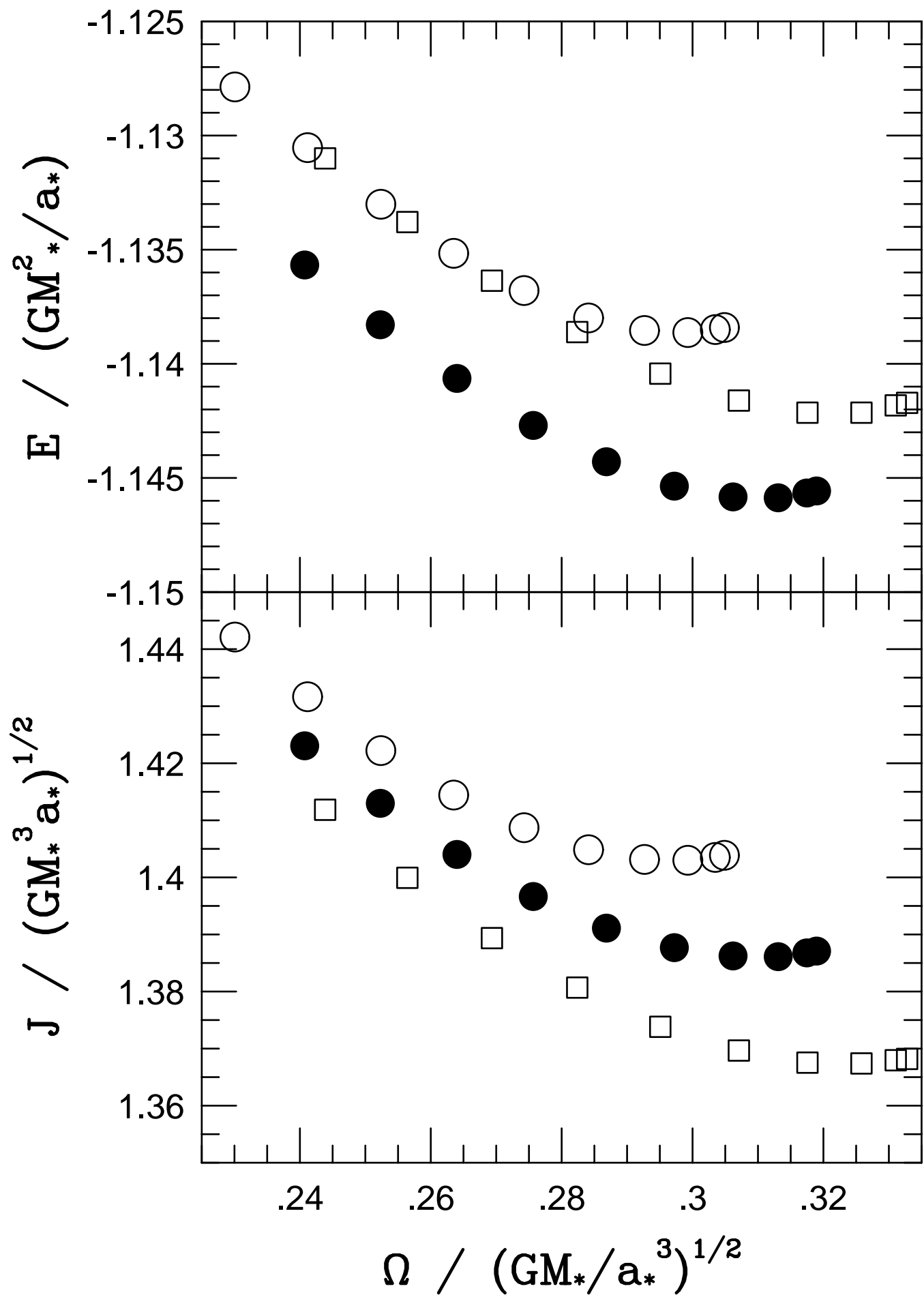
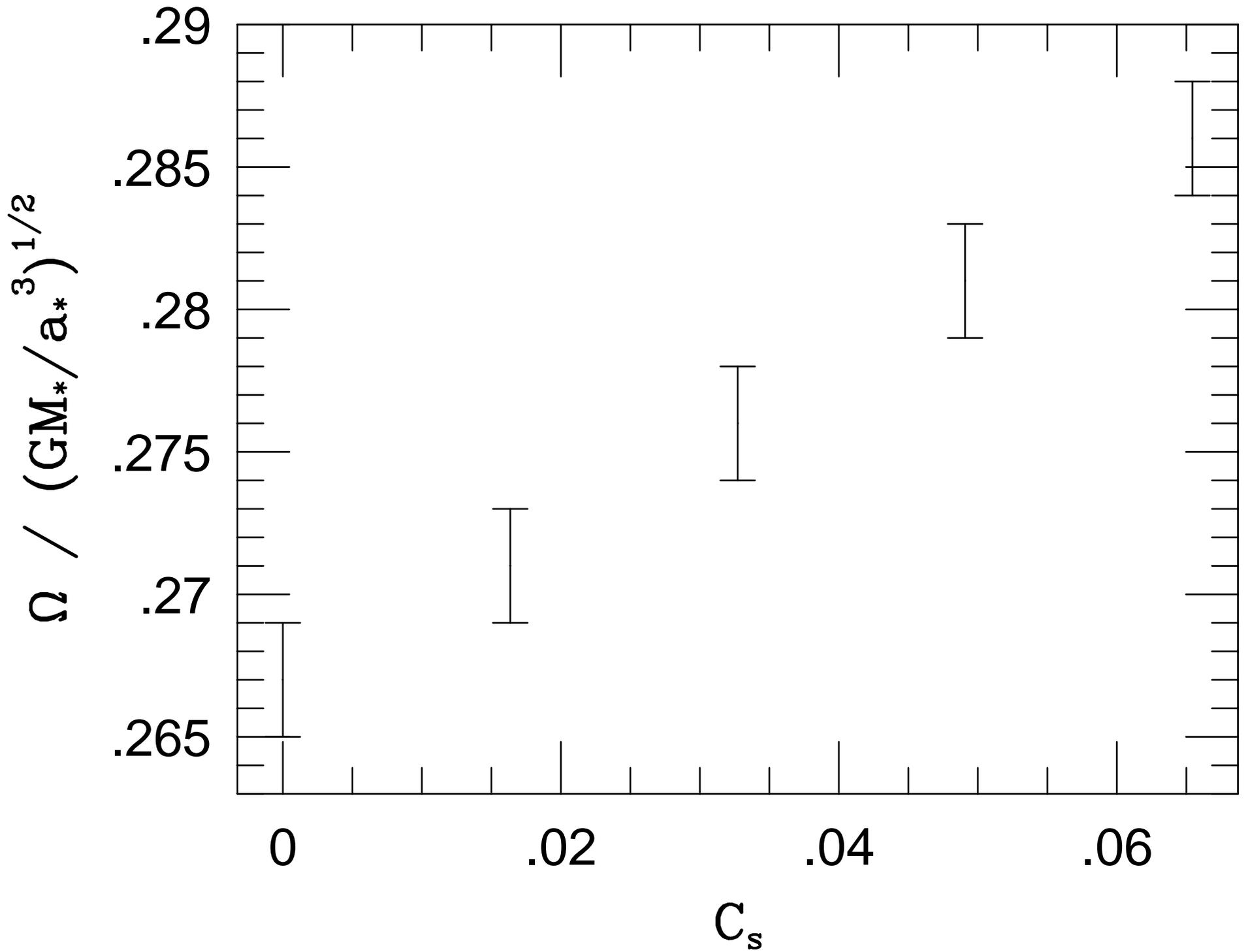
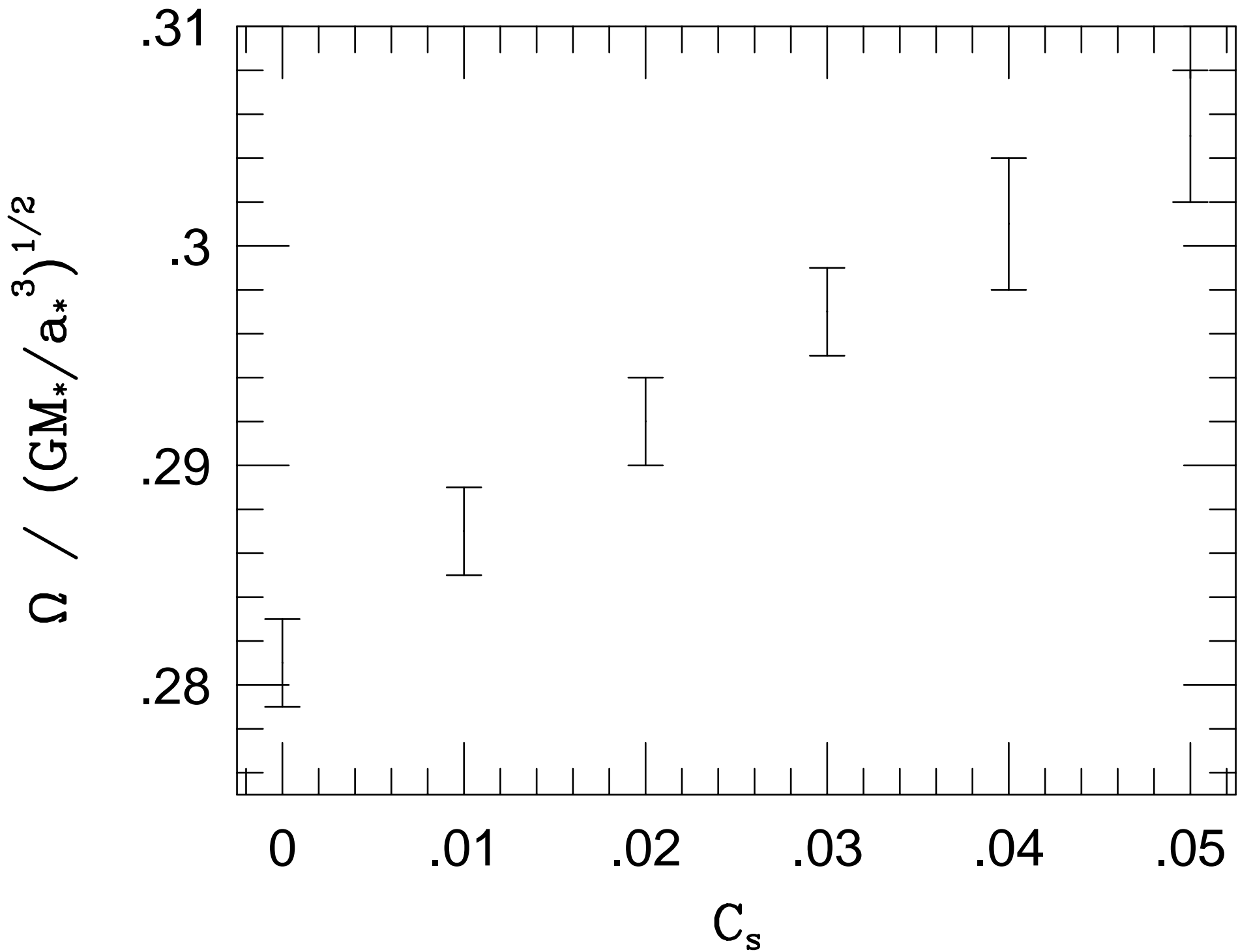
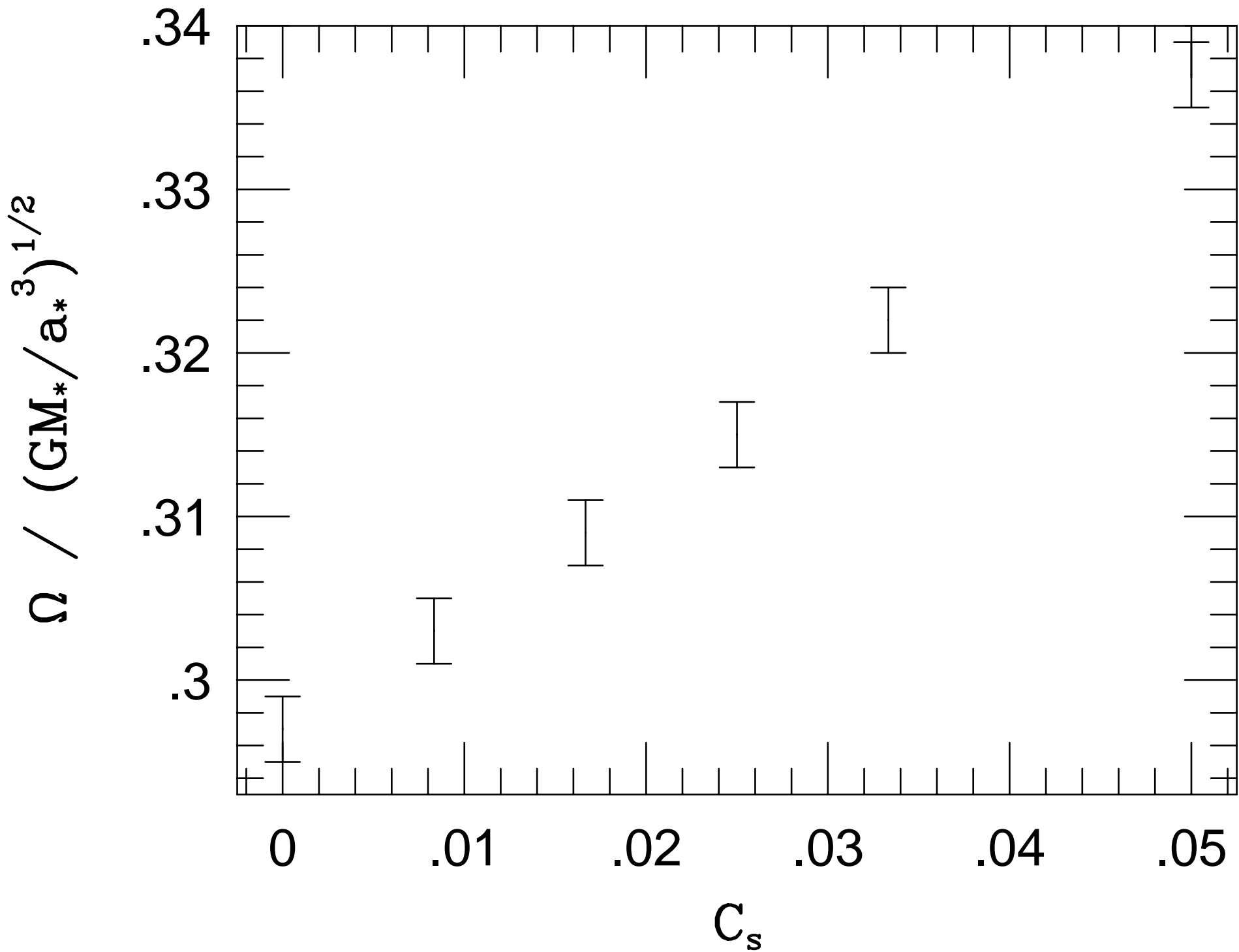
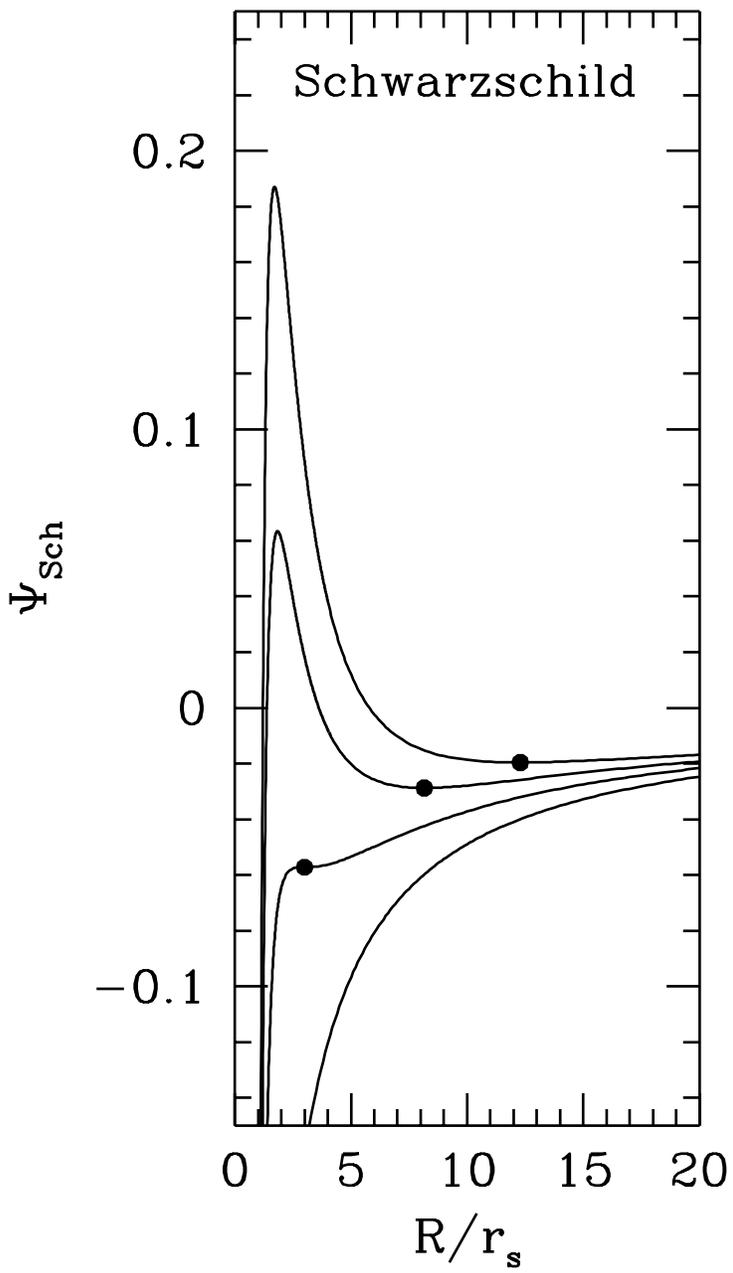
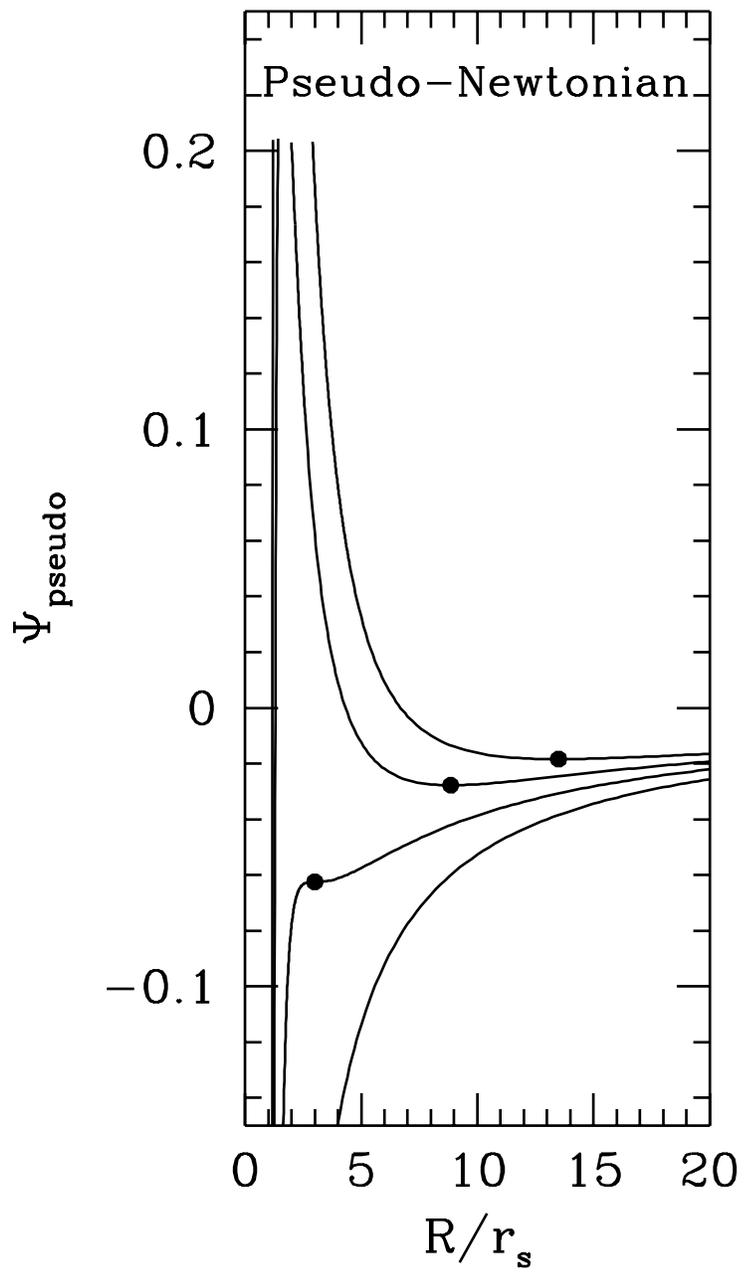
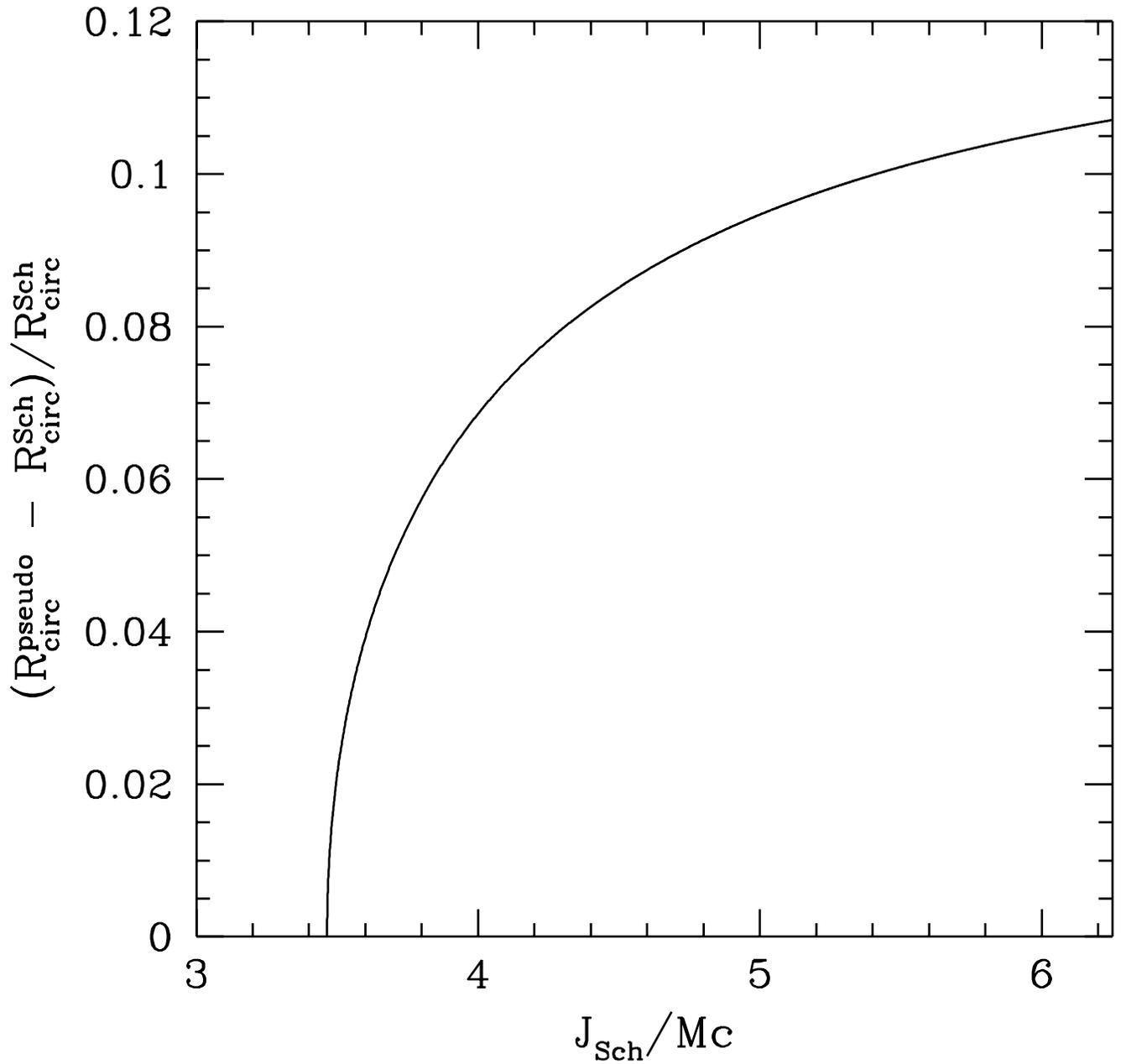
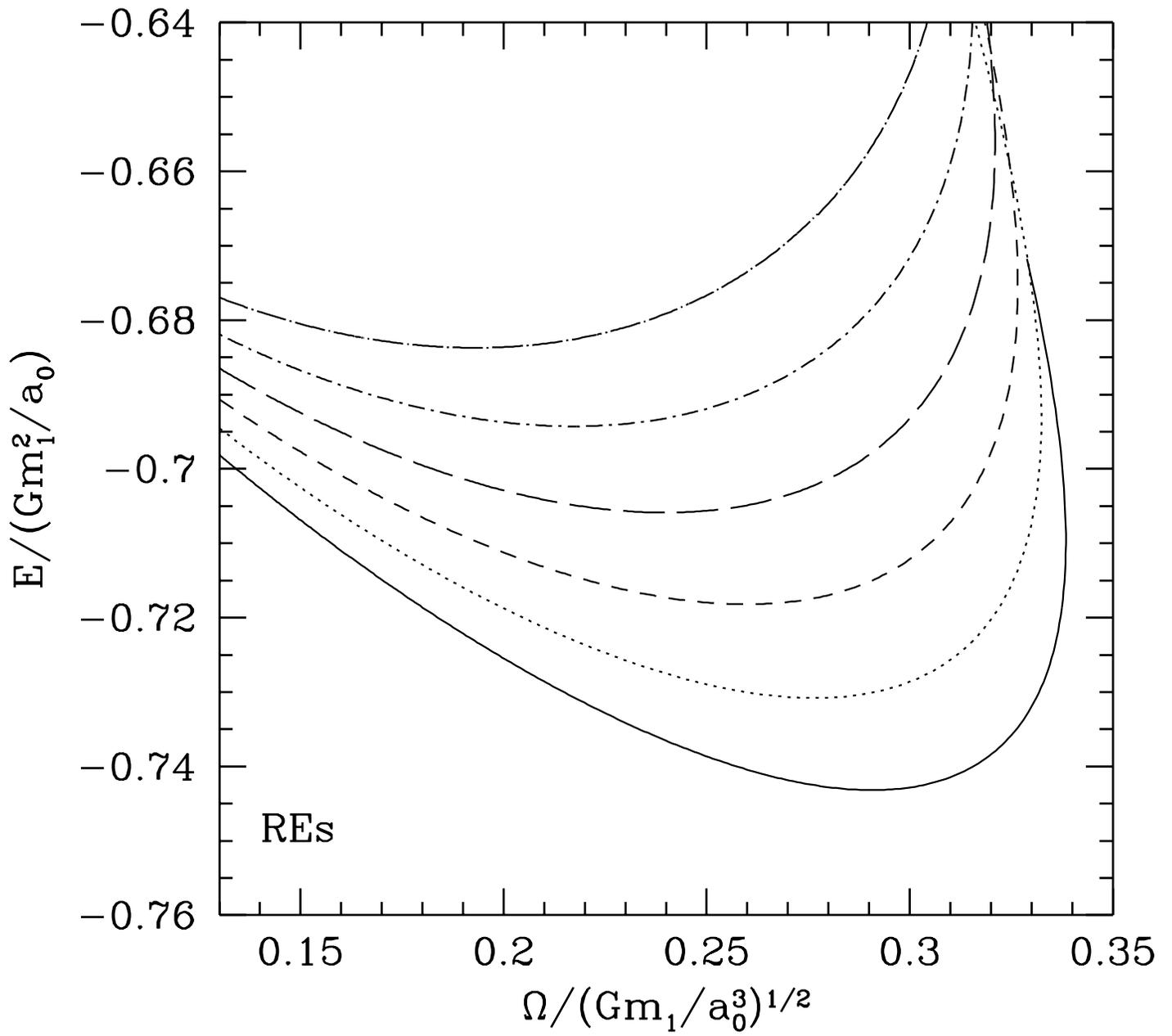
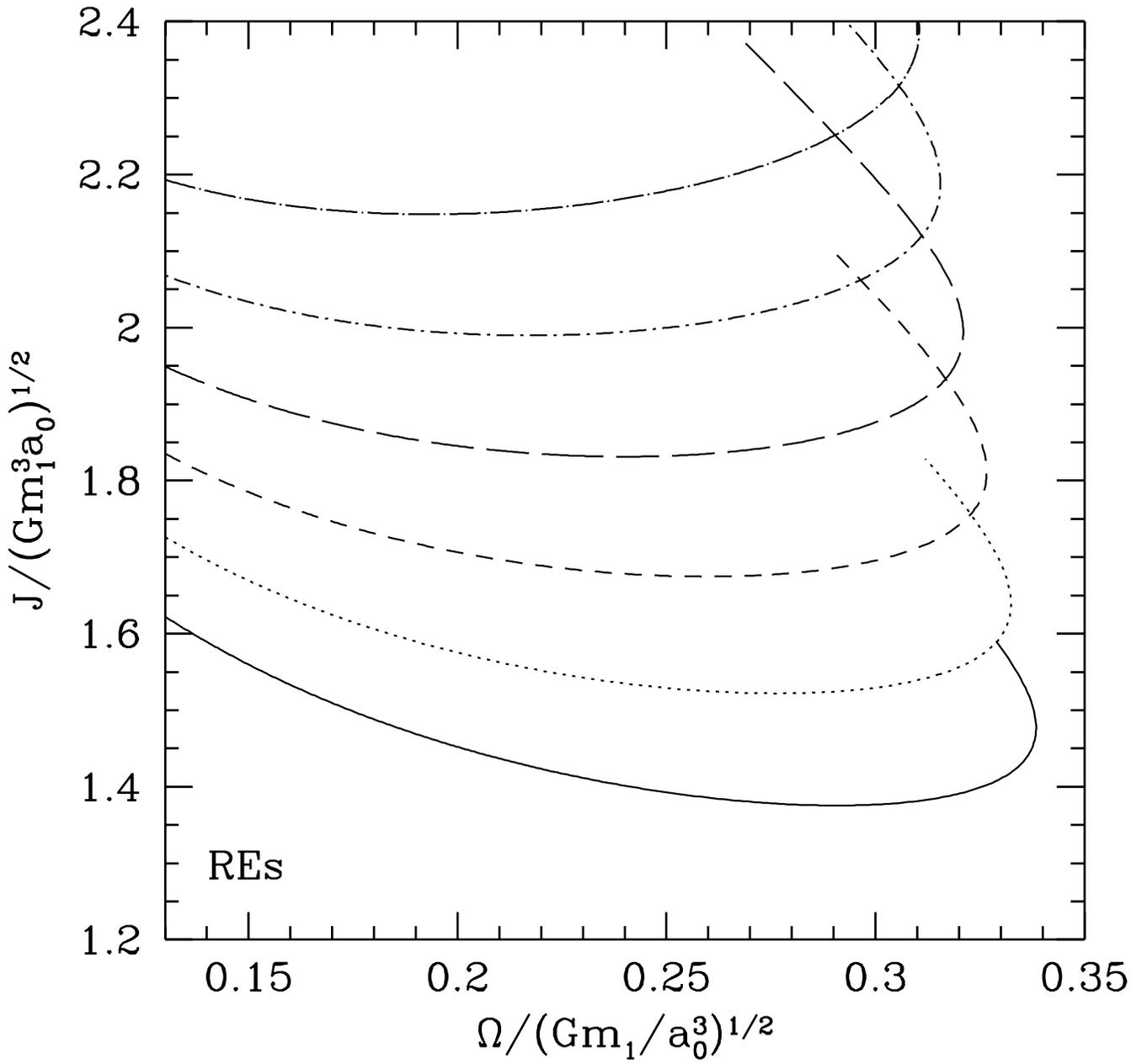
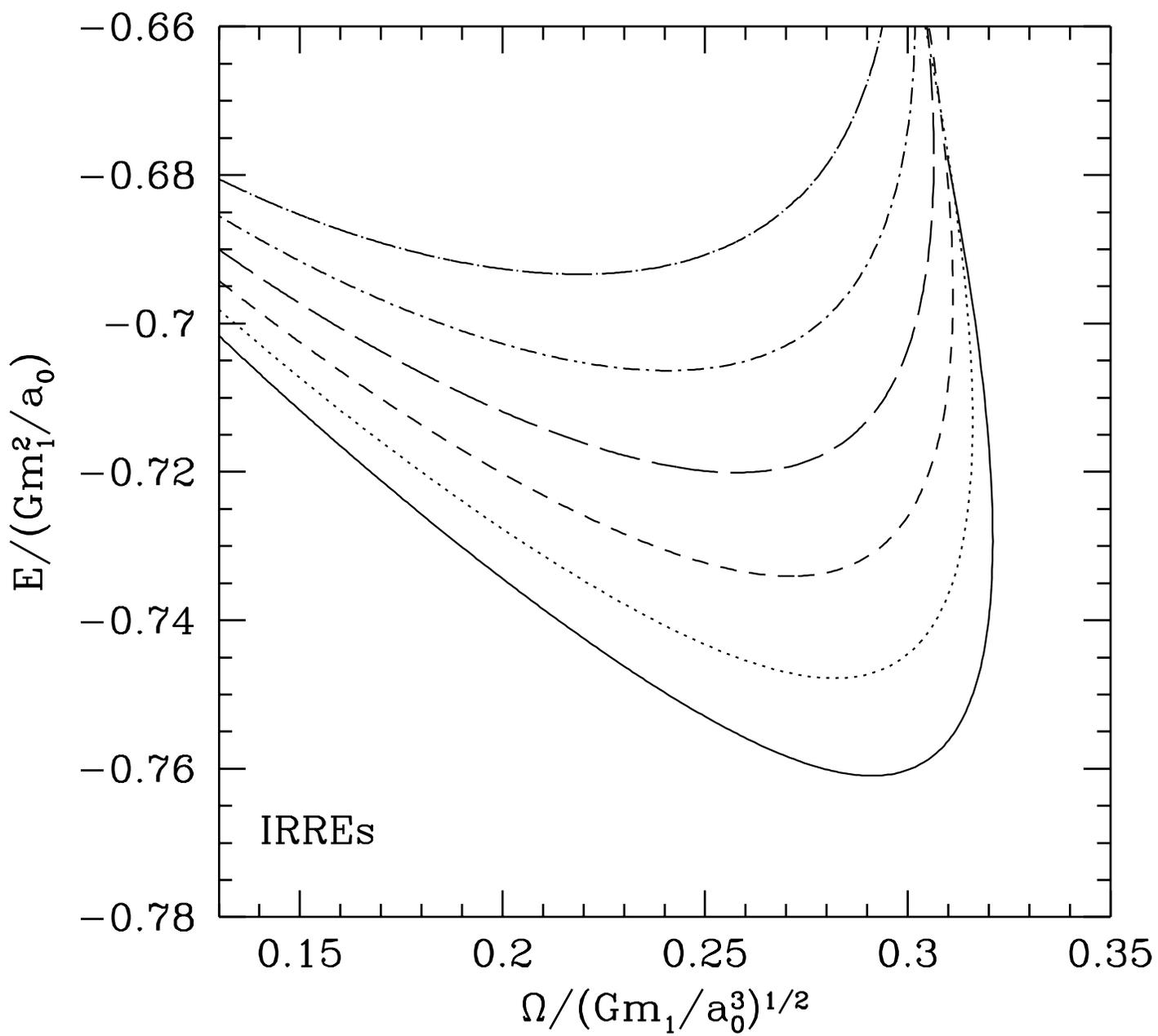
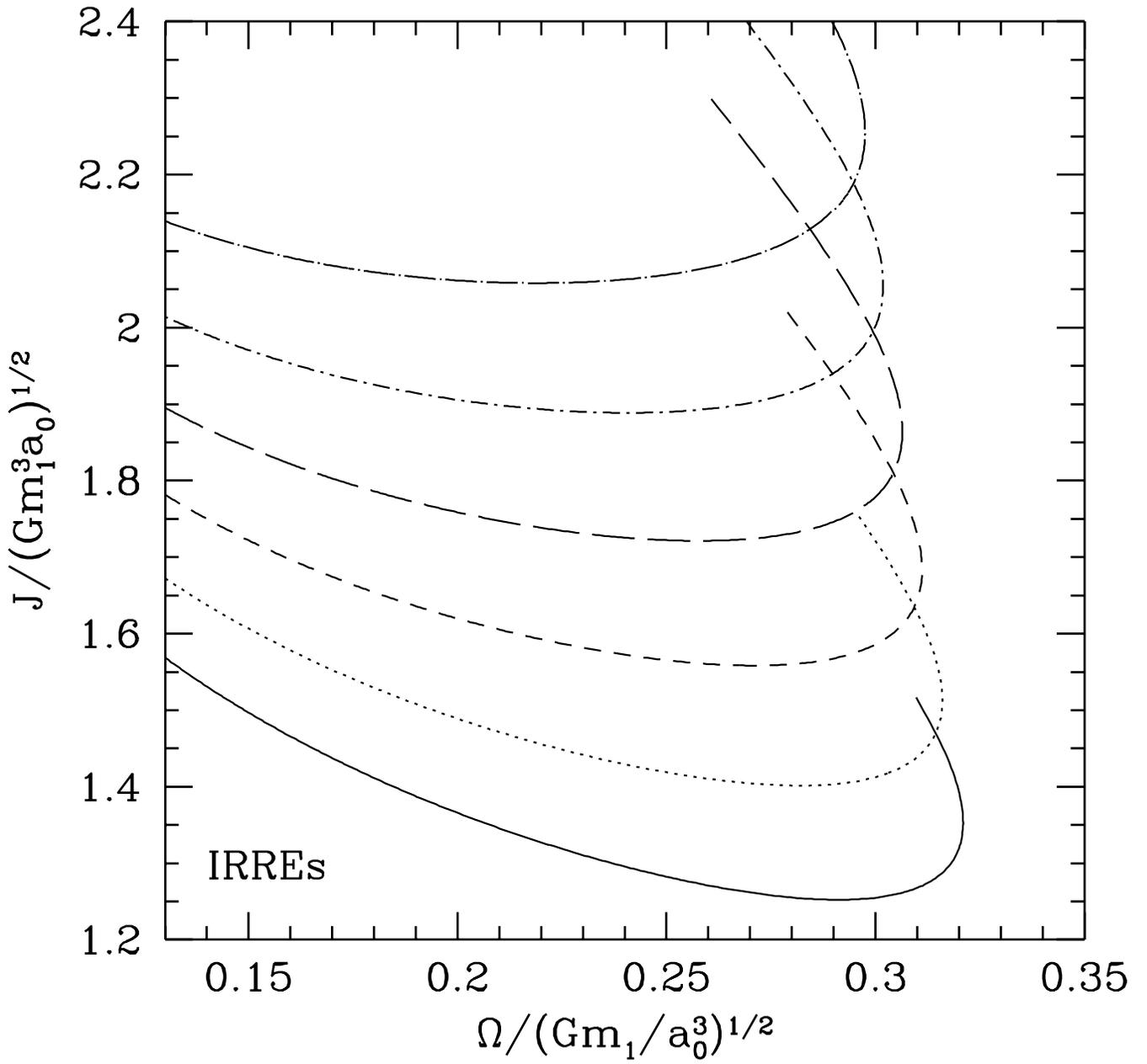
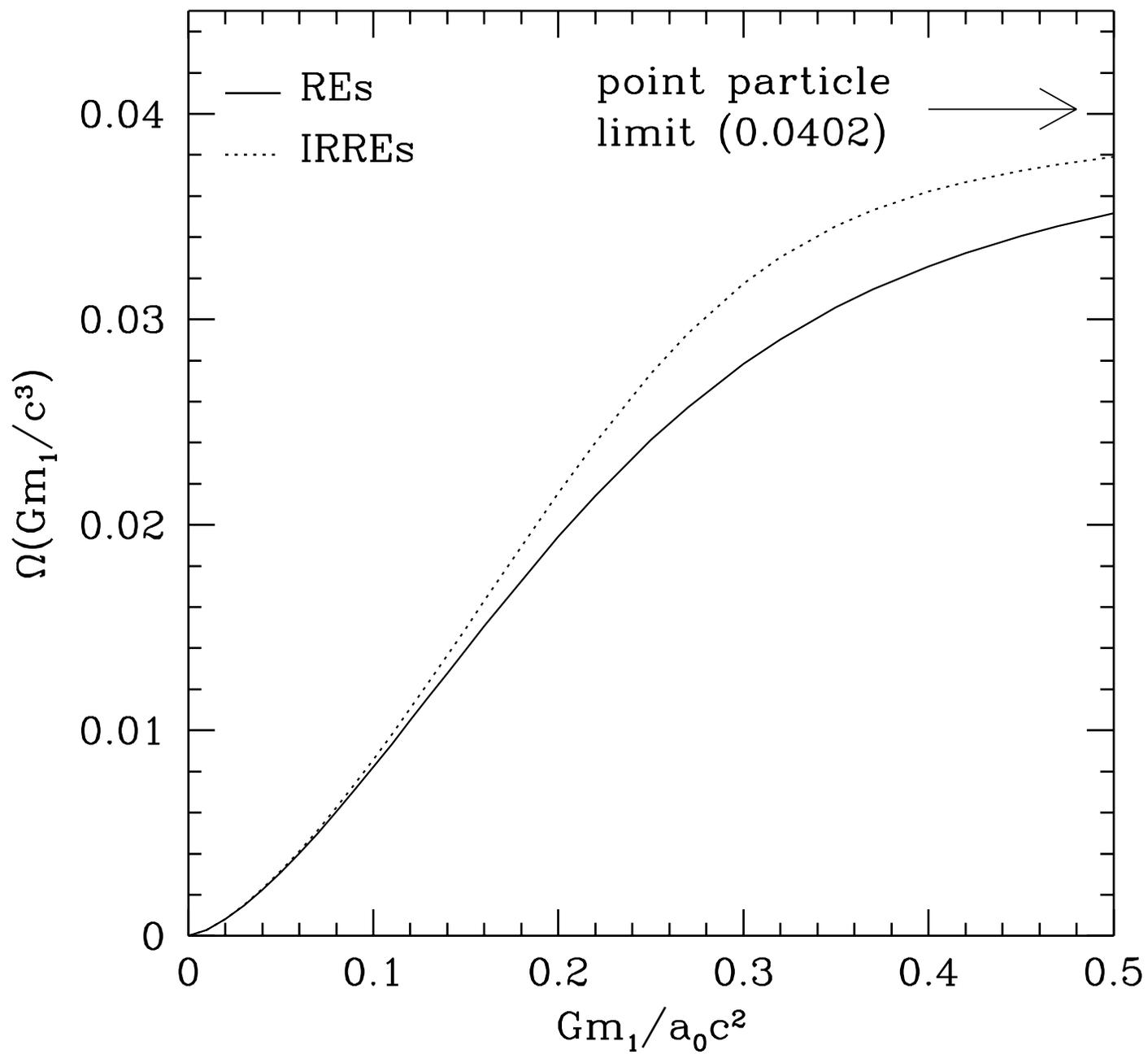
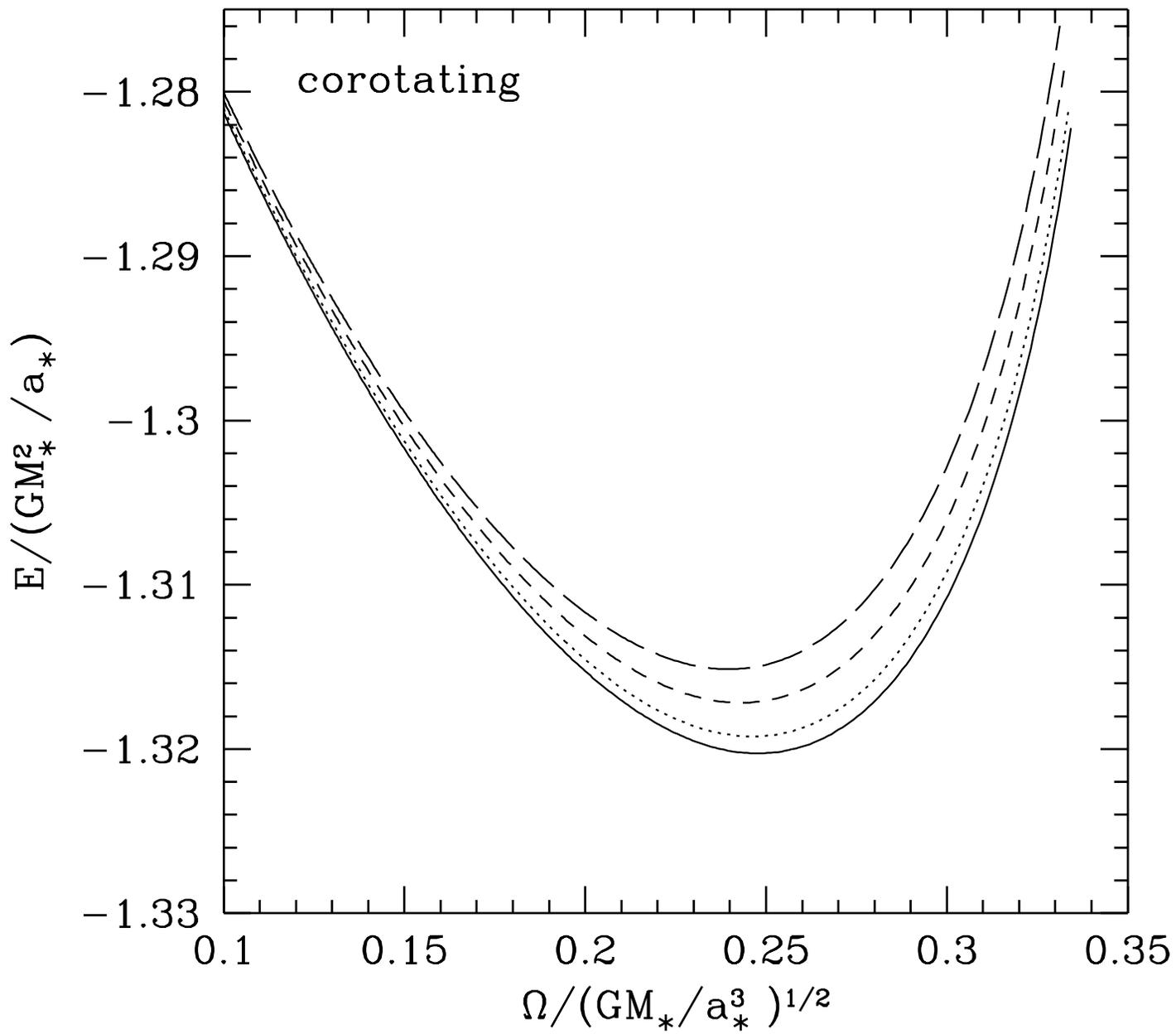
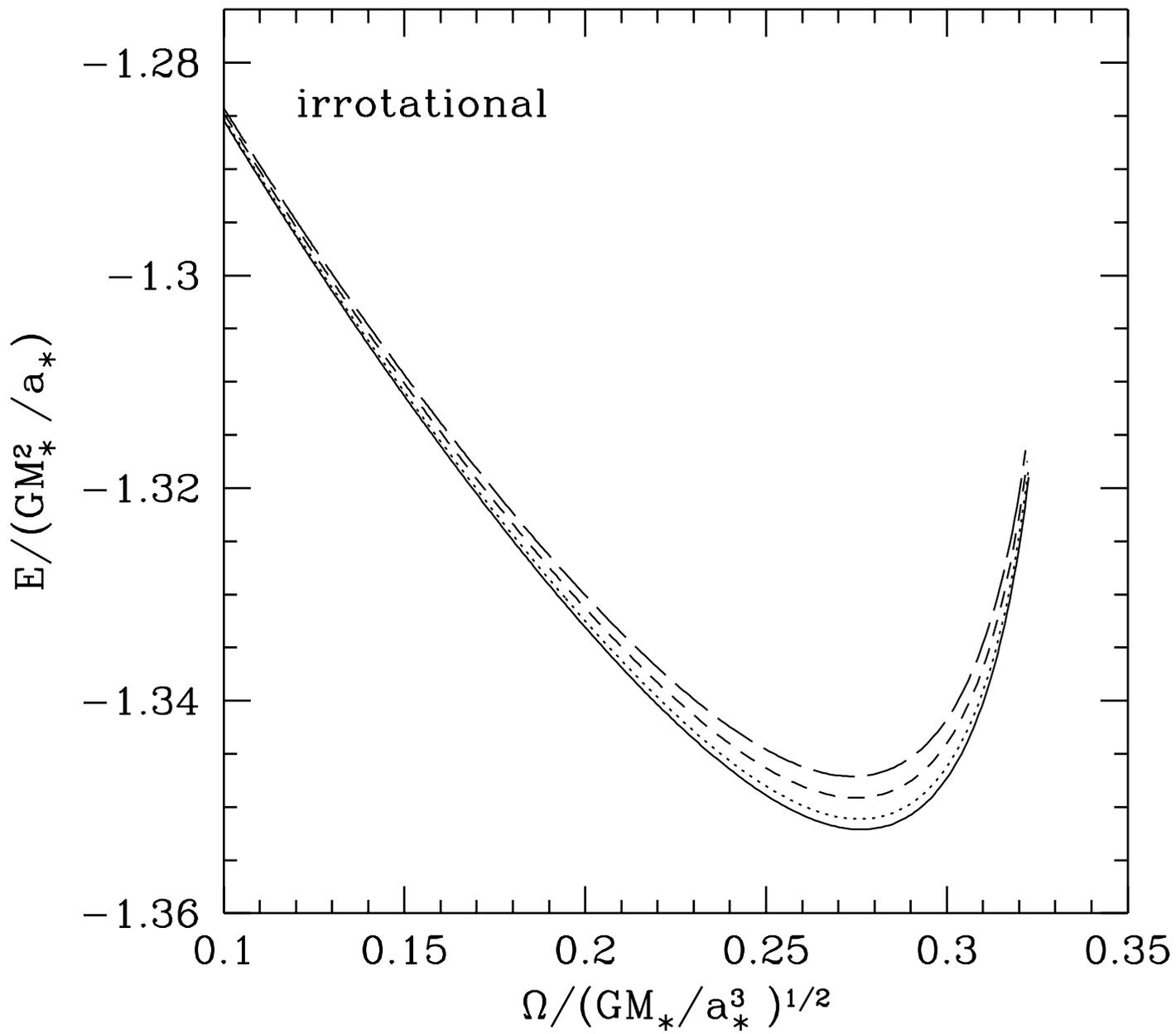
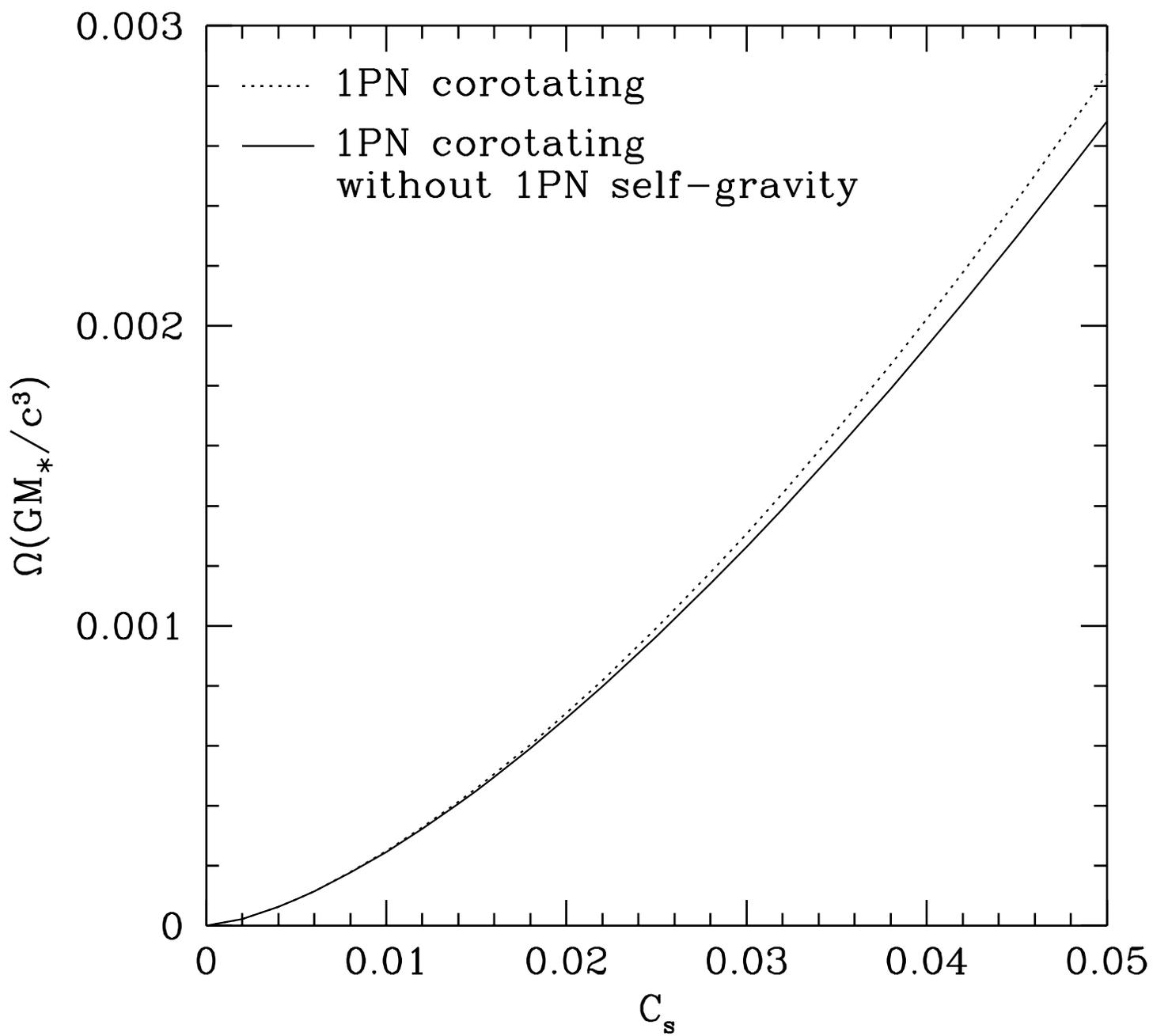
\end{document}